\documentclass[12pt,reqno]{amsart}

\usepackage{epsfig}
\usepackage{amsmath, amsthm, amsfonts}
\usepackage{graphicx}

\numberwithin{equation}{section}

\newcommand{\R}{{\mathbb R}}
\newcommand{\C}{{\mathbb C}}

\renewcommand{\Re}{{\operatorname{Re\,}}}
\renewcommand{\Im}{{\operatorname{Im\,}}}

\newcommand{\Tr}{{{\operatorname{Tr}}}}

\newcommand{\Ai}{{\operatorname{Ai}}}

\newcommand{\sign}{{\operatorname{sgn}\,}}
\newcommand{\Res}{{\operatorname{Res}\,}}
\newcommand{\al}{\alpha}
\newcommand{\be}{\beta}
\newcommand{\ga}{\gamma}
\newcommand{\Ga}{\Gamma}

\newcommand{\la}{\lambda}
\newcommand{\ep}{\varepsilon}
\newcommand{\de}{\delta}
\newcommand{\De}{\Delta}
\newcommand{\f}{\varphi}
\newcommand{\sg}{\sigma}
\newcommand{\om}{\omega}
\newcommand{\Om}{\Omega}
\newcommand{\z}{\zeta}

\newtheorem{theo}{{\sc \bf Theorem}}[section]

\newtheorem{lem}[theo]{{\sc \bf Lemma}}
\newtheorem{prop}[theo]{{\sc \bf Proposition}}

\begin{document}

\title[Exact Solution of the Six-Vertex Model]
{Exact Solution of the Six-Vertex Model with Domain Wall Boundary Conditions. Disordered Phase}

\author{Pavel M. Bleher}
\address{Department of Mathematical Sciences,
Indiana University-Purdue University Indianapolis,
402 N. Blackford St., Indianapolis, IN 46202, U.S.A.}
\email{bleher@math.iupui.edu}
\author{Vladimir V. Fokin}
\address{Department of Mathematical Sciences,
Indiana University-Purdue University Indianapolis,
402 N. Blackford St., Indianapolis, IN 46202, U.S.A.}
\email{vvf@math.iupui.edu}

\thanks{The first author is supported in part
by the National Science Foundation (NSF) Grant DMS-0354962.}

\date{\today}

\begin{abstract}
The six-vertex model, or the square ice model, with domain wall boundary conditions (DWBC)
has been introduced and solved for finite $N$ by Korepin and Izergin. The solution is based on
the Yang-Baxter equations and it represents the free energy in terms of an $N\times N$ Hankel 
determinant. Paul Zinn-Justin observed that the Izergin-Korepin formula can be re-expressed 
in terms of the partition function of a random matrix model with a nonpolynomial interaction. 
We use this observation to obtain the large $N$ asymptotics of the six-vertex model with DWBC
in the disordered phase.
The solution is based on the Riemann-Hilbert approach and the Deift-Zhou nonlinear steepest
descent method. As was noticed by Kuperberg, the problem of enumeration of alternating sign
matrices (the ASM problem) is a special case of the the six-vertex model. We compare the 
obtained exact solution of the six-vertex model with known exact results for the
1, 2, and 3 enumerations of ASMs, and also with the exact solution on the so-called free
fermion line. We prove the conjecture of Zinn-Justin that the partition function of
the six-vertex model with DWBC has the asymptotics, 
$Z_N\sim CN^\kappa e^{N^2f}$ as $N\to\infty$,
and we find the exact value of the exponent $\kappa$. 
\end{abstract}

\maketitle

\section{Introduction}
\vskip 1cm

The six-vertex model, or the model of two-dimensional ice, is stated on a square 
lattice with arrows on edges. The arrows obey the rule that at every vertex there 
are two arrows 
\begin{center}
 \begin{figure}[h]\label{arrows}
\begin{center}
   \scalebox{0.52}{\includegraphics{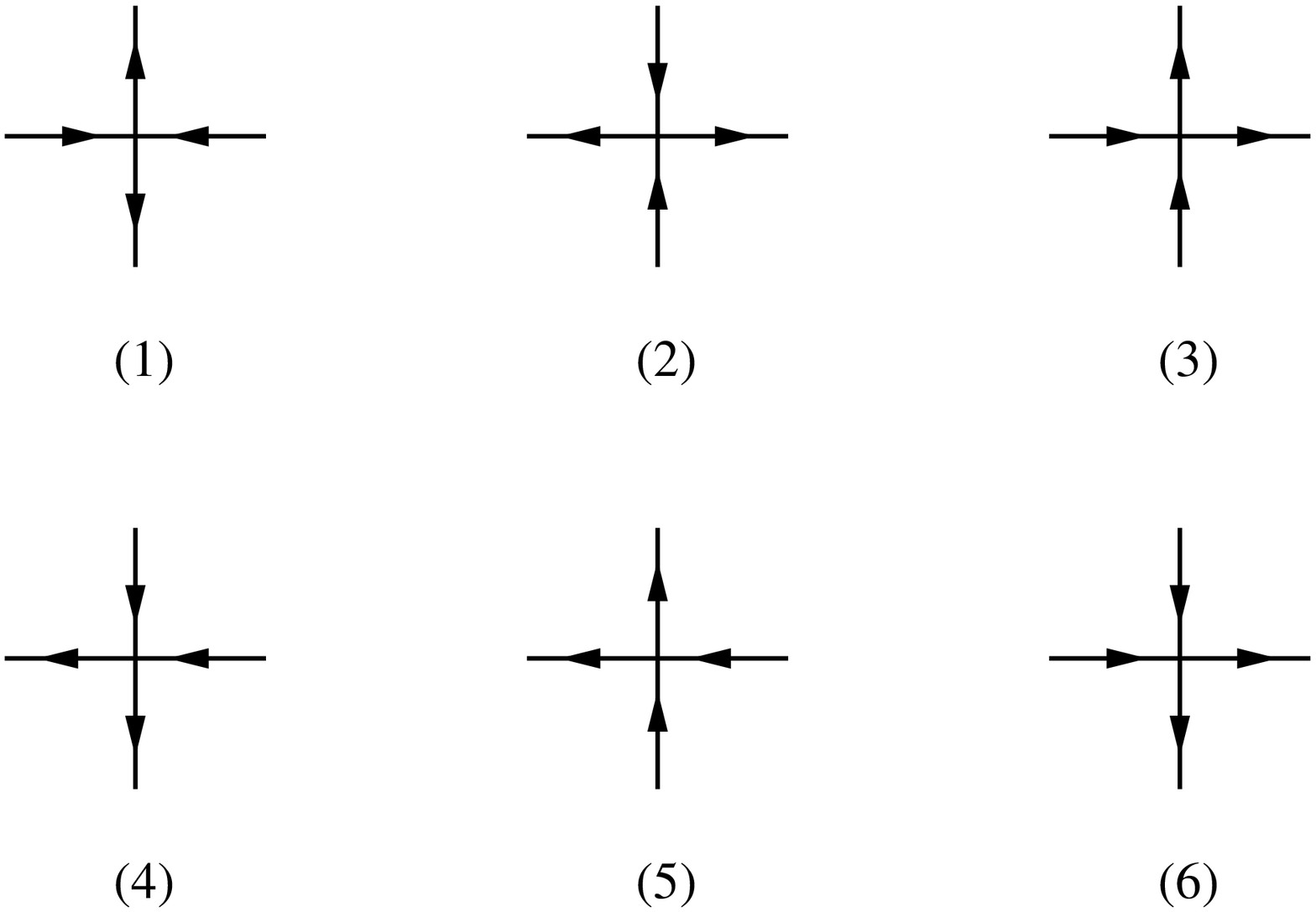}}
\end{center}
        \caption[The six arrow configurations allowed at a vertex]{The six arrow configurations allowed at a vertex.}
    \end{figure}
\end{center}
pointing in and two arrows pointing out. Such rule is sometimes 
called the ice-rule. There are only six possible configurations of arrows at each 
vertex, hence the name of the model, see Fig.~1.

We will consider the domain wall boundary conditions (DWBC), 
in which the arrows on the upper and lower boundaries point in the square, 
and the ones on the left and right boundaries point out. 
One possible configuration with DWBC on the $4\times 4$ lattice is shown on Fig.~2.

\begin{center}
 \begin{figure}[h]\label{example1}
\begin{center}
   \scalebox{0.52}{\includegraphics{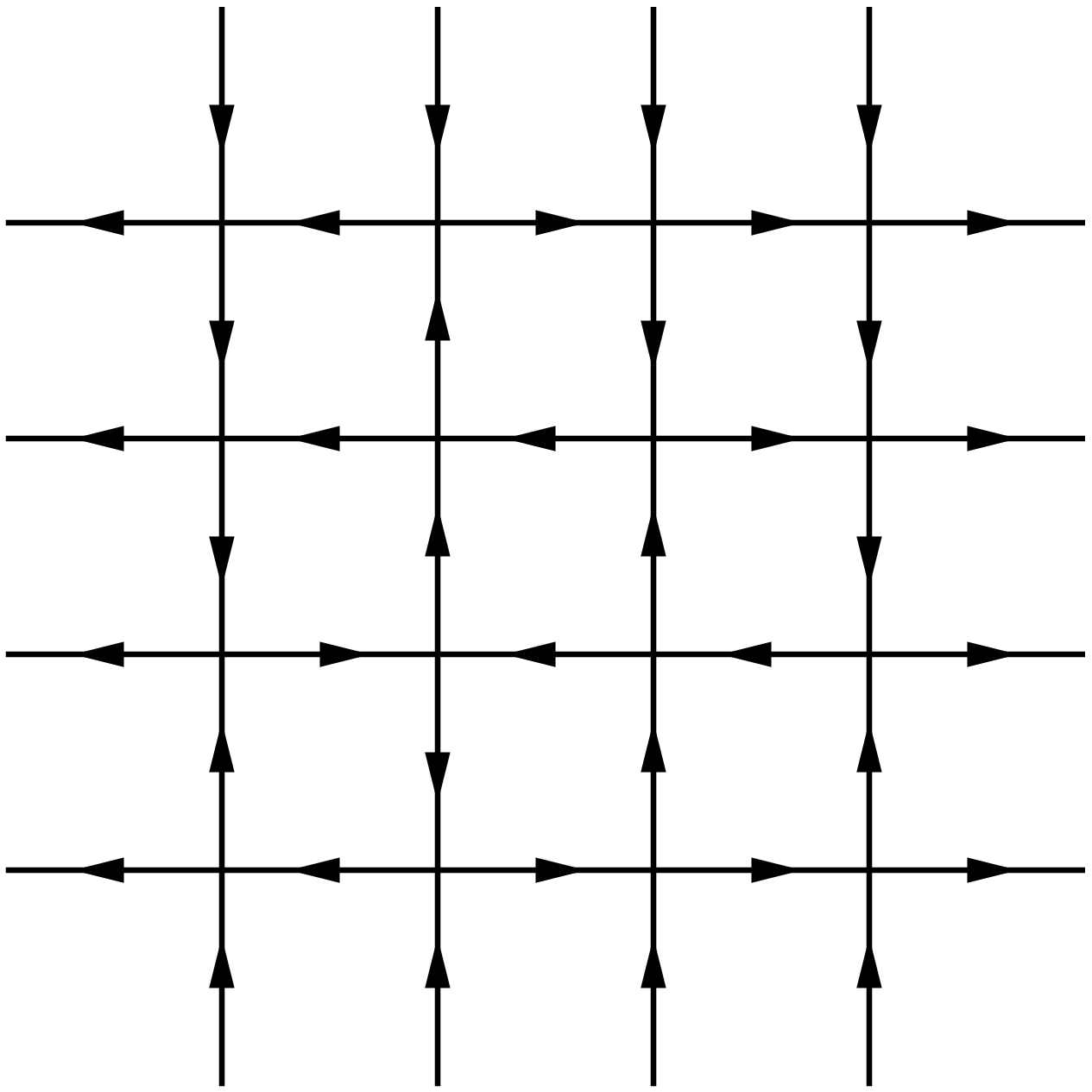}}
\end{center}
        \caption[An example of $4\times4$ configuration]{An example of $4\times4$ configuration.}
    \end{figure}
\end{center}

The name of the square ice comes from the two-dimensional arrangement 
of water molecules, $H_2O$, with oxygen atoms at the vertices of 
the lattice and one hydrogen atom between each pair of adjacent oxygen 
atoms. We place an arrow in the direction from  a hydrogen atom 
toward an oxygen atom if there is a bond between them. Thus, as 
we already noticed before, there are two inbound and two outbound 
arrows at each vertex.

\begin{center}
 \begin{figure}[h]\label{ice1}
\begin{center}
   \scalebox{0.52}{\includegraphics{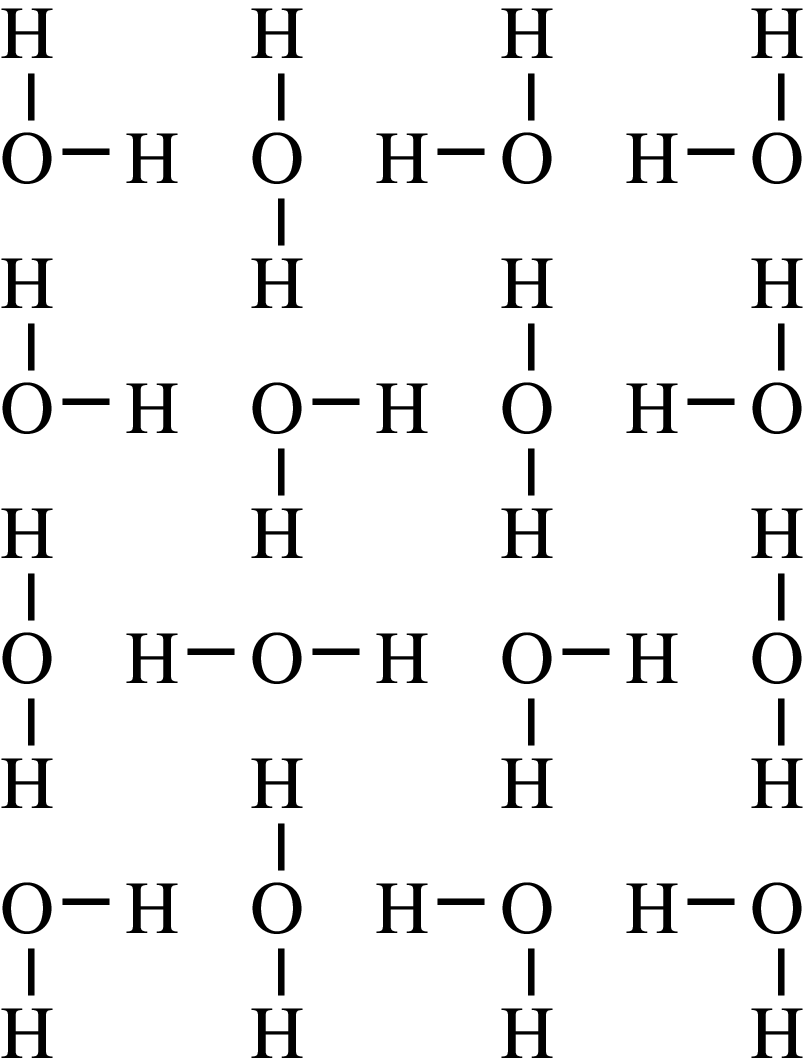}}
\end{center}
        \caption[The corresponding ice model]{The corresponding ice model.}
    \end{figure}
\end{center}

For each possible vertex state we assign a weight $w_i,\; i=1,\dots,6$, 
and define, as usual, the partition function, as a sum over all possible 
arrow configurations of the product of the vertex weights,
\begin{equation}\label{pf1}
Z_N=\sum_{\mbox{arrow configurations}}\prod_{i=1}^6w_i^{n_i},
\end{equation}
where $n_i$ is the number of vertices in the state $i$ in a given arrow configuration.
We will consider the case, when the weights are invariant under the 
simultaneous reversal of all arrows, i.e., 
\begin{equation}\label{pf2}
a:=w_3=w_4, \quad
b:=w_5=w_6, \quad
c:=w_1=w_2.
\end{equation}
Define the parameter $\Delta$ as
\begin{equation}\label{pf3}
\Delta=
\frac{a^2+b^2-c^2}{2ab}.
\end{equation}
There are three physical phases for the six-vertex model: the ferroelectric phase, $\Delta > 1$; the anti-ferroelectric phase, $\Delta<-1$; and, the disordered phase, $-1<\Delta<1$. The phase diagram of the model is shown on Fig.~4.

\begin{center}
 \begin{figure}[h]\label{PhaseDiagram1}
\begin{center}
   \scalebox{0.6}{\includegraphics{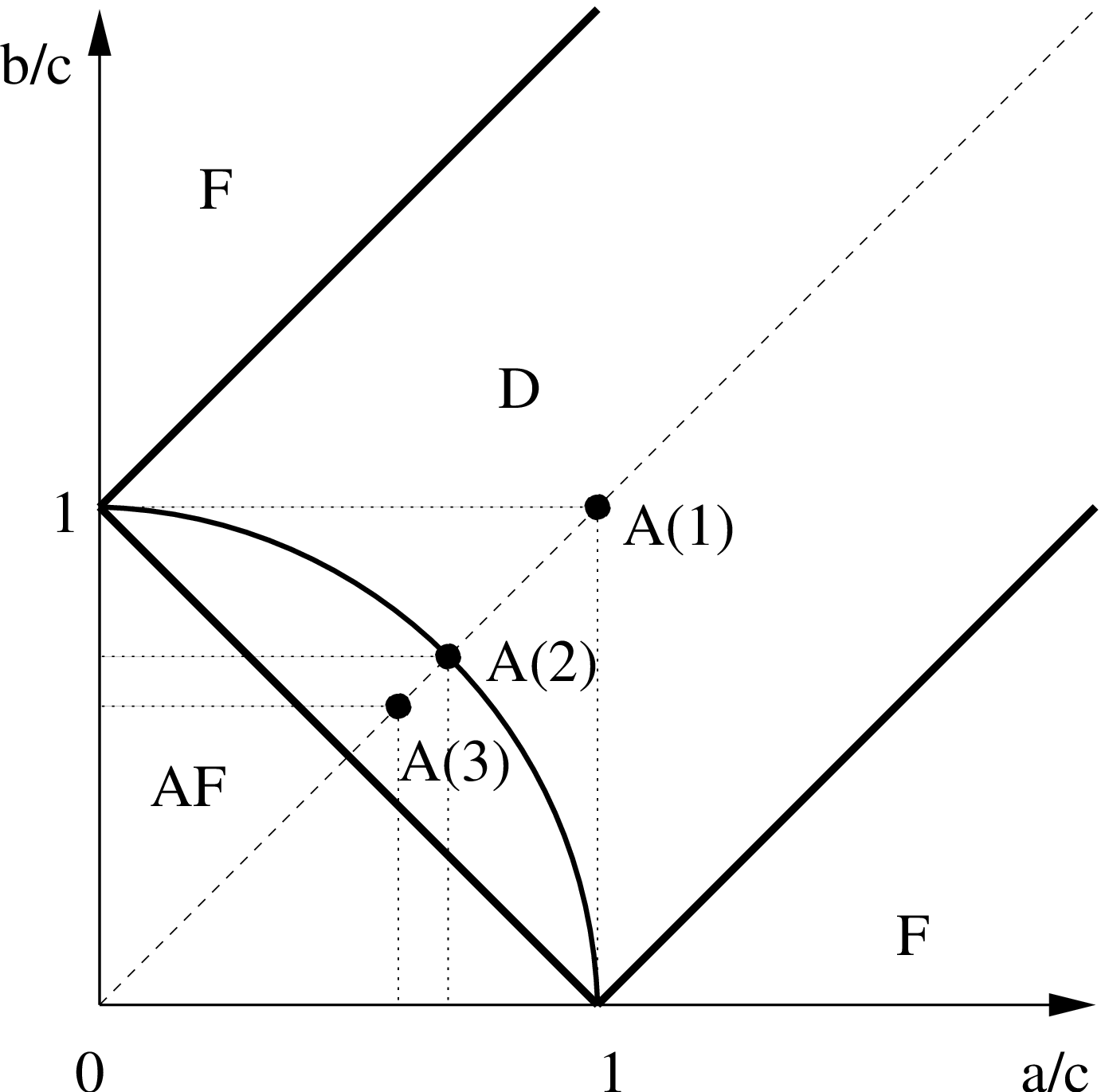}}
\end{center}
        \caption[The phase diagram of the model]{The phase diagram of the model,
 where {\bf F}, {\bf AF} and {\bf D} mark ferroelectric, antiferroelectric,  
and disordered  phases, respectively. The circular arc corresponds to the 
so-called "free fermion" line, where $\Delta=0$, and the three
dots correspond to 1-, 2-, and 3-enumeration of alternating sign matrices.}
    \end{figure}
\end{center}

In these phases we parametrize the weights in the standard way:
for the ferroelectric phase,
\begin{equation}\label{pf4}
a=\sinh(t-\ga), \quad
b=\sinh(t+\ga), \quad
c=\sinh(2\ga), \quad
|\ga|<t,
\end{equation}
for the anti-ferroelectric phase,
\begin{equation}\label{pf5}
a=\sinh(\ga-t), \quad
b=\sinh(\ga+t), \quad
c=\sinh(2\ga), \quad
|t|<\ga,
\end{equation}
and for the disordered phase,
\begin{equation}\label{pf6}
a=\sin(\ga-t), \quad
b=\sin(\ga+t), \quad
c=\sin(2\ga), \quad
|t|<\ga.
\end{equation}
Here we will discuss  the disordered phase, and we will use parametrization (\ref{pf6}).

A solution for the free energy of the six-vertex model with periodic 
boundary conditions (PBC) was found by Lieb
\cite{Lieb1}--\cite{Lieb4} by means of the Bethe Ansatz. In the most general form 
of the six-vertex model the Bethe Ansatz solution with PBC was obtained by
Sutherland \cite{Sut}. A detailed classification of
the phases of the model is given in the review paper of Lieb and Wu \cite
{LW}; see also the book of Baxter \cite{Bax}.
The six-vertex model with antiperiodic boundary conditions is solved
in \cite{BBOY}.

 The six-vertex model with DWBC was introduced 
by Korepin in \cite{Kor}, who derived an important 
recursion relation for the partition function of the model.
This lead to a beautiful determinantal formula of Izergin \cite{Ize},
for the partition function of the six-vertex model with DWBC.
A detailed proof of this formula and its generalizations are given in
the paper of Izergin, Coker, and Korepin \cite{ICK}. When the weights are 
parameterized according to (\ref{pf6}), the formula of Izergin is
\begin{equation} \label{pf7}
Z_N=\frac{[\sin(\ga+t)\sin(\ga-t)]^{N^2}}{\left(
\prod_{n=0}^{N-1}n!\right)^2}\,\tau_N\,,
\end{equation}
where $\tau_N$ is the Hankel determinant,
\begin{equation} \label{pf8}
\tau_N=\det\left(\frac{d^{i+k-2}\phi}{dt^{i+k-2}}\right)_{1\le i,k\le N},
\end{equation} 
and
\begin{equation} \label{pf9}
\phi(t)=\frac{\sin(2\ga)}{\sin(\ga+t)\sin(\ga-t)}\,.
\end{equation}
An elegant derivation of the 
Izergin determinantal formula from the Yang-Baxter equations is 
given in the papers of Korepin and Zinn-Justin \cite{KZ} and Kuperberg \cite {Kup}.
 
One of the applications of the determinantal formula is that it
implies that 
the partition function $\tau_N$ solves the Toda equation,
\begin{equation} \label{pf10}
\tau_N\tau''_N-{\tau'_N}^2=\tau_{N+1}\tau_{N-1},
\qquad N\ge 1,\qquad ({}')=\frac{\partial }{\partial t}\,,
\end{equation}
cf. \cite{Sog}. This was used by Korepin and Zinn-Justin \cite{KZ} to 
derive the free energy of the six-vertex model with DWBC, assuming
some Ansatz on the behavior of subdominant terms in the large $N$
asymptotics of the free energy.

Another application of the Izergin determinantal formula is that
$\tau_N$ can be expressed in terms of  a partition function of
a random matrix model. The relation to the random matrix model was
obtained and used by Zinn-Justin \cite{Z}. This relation will be
very important for us. It can be 
derived as follows.
For the evaluation of the Hankel determinant, it is convenient 
to use the integral representation of $\phi(t)$, namely, to write it 
in the form of the Laplace transform,
\begin{equation} \label{dph6}
\phi(t)=\int_{-\infty}^\infty e^{t\la}m(\la)d\la,
\end{equation} 
where
\begin{equation} \label{dph7}
m(\la)=\frac{\sinh\frac{\la}{2}(\pi-2\ga)}
{\sinh\frac{\la}{2}\pi}\,.
\end{equation} 
Then
\begin{equation} \label{dph8}
\frac{d^i\phi}{dt^i}=\int_{-\infty}^\infty \la^i
e^{t\la}m(\la)d\la,
\end{equation} 
and by substituting this into the Hankel determinant,
(\ref{pf8}), we obtain that
\begin{equation} \label{dph9}
\begin{aligned}
\tau_N&=\int \prod_{i=1}^N[e^{t\la_i}m(\la_i)d\la_i]
\det(\la_i^{i+k-2})_{1\le i,k\le N}\\
&=\int \prod_{i=1}^N[e^{t\la_i}m(\la_i)d\la_i]
\det(\la_i^{k-1})_{1\le i,k\le N}\prod_{i=1}^N\la_i^{i-1}.
\end{aligned}
\end{equation}
Consider any permutation $\sg\in S_N$ of variables $\la_i$. 
From the last equation we have that
\begin{equation} \label{dph10}
\tau_N=\int \prod_{i=1}^N[e^{t\la_i}m(\la_i)d\la_i]
(-1)^\sg\det(\la_i^{k-1})_{1\le i,k\le N}\prod_{i=1}^N\la_{\sg(i)}^{i-1}.
\end{equation}
By summing over $\sg\in S_N$, we obtain that
\begin{equation} \label{dph11}
\tau_N=\frac{1}{N!}\int \prod_{i=1}^N[e^{t\la_i}m(\la_i)d\la_i]
\De(\la)^2
\end{equation}
(see \cite{Z}), where $\De(\la)$ is the Vandermonde determinant,
\begin{equation} \label{dph12}
\De(\la)=\det(\la_i^{k-1})_{1\le i,k\le N}=\prod_{i<k}(\la_k-\la_i).
\end{equation}
Equation (\ref{dph11}) expresses $\tau_N$ in terms of a matrix model
integral. Namely, if $m(x)=e^{-V(x)}$, then
\begin{equation} \label{dph13}
\tau_N=\frac{\prod_{n=0}^{N-1}n!}{\pi^{N(N-1)/2}}\int dM e^{\Tr [tM-V(M)]},
\end{equation}
where the integration is over the space of $N\times N$ Hermitian matrices.
The matrix model integral can be solved, furthermore, in terms of
orthogonal polynomials.

Introduce monic polynomials $P_n(x)=x^n+\dots$ orthogonal 
on the line with respect to the weight $e^{tx}m(x)$, so that
\begin{equation} \label{dph14}
\int_{-\infty}^\infty P_n(x)P_m(x)e^{tx}m(x)dx=h_n\delta_{nm}.
\end{equation}
Then it follows from (\ref{dph11}) that
\begin{equation} \label{dph15}
\tau_N=\prod_{n=0}^{N-1}h_n. 
\end{equation}
The orthogonal polynomials satisfy 
the three term recurrent relation,
\begin{equation} \label{dph16}
xP_n(x)=P_{n+1}(x)+Q_nP_n(x)+R_nP_{n-1}(x),
\end{equation}
where $R_n$ can be found as
\begin{equation} \label{dph16a}
R_n=\frac{h_n}{h_{n-1}}\,,
\end{equation}
see, e.g., \cite{Sze}.
This gives that
\begin{equation} \label{dph17}
h_n=h_0\prod_{j=1}^n R_j,
\end{equation}
where
\begin{equation}\label{dph17a}
h_0=\int_{-\infty}^\infty e^{tx}m(x)dx
=\frac{\sin(2\ga)}{\sin(\ga+t)\sin(\ga-t)}\,.
\end{equation}
By substituting (\ref{dph17}) into (\ref{dph15}), we obtain that
\begin{equation} \label{dph18}
\tau_N=h_0^N\prod_{n=1}^{N-1}R_n^{N-n}. 
\end{equation}
We will prove the following asymptotics of the recurrent coefficients
$R_n$.

\begin{theo} \label{R_n}
As $n\to\infty$,
\begin{equation} \label{dph19}
R_n=\frac{n^2}{\ga^2}\left[R+\cos(n\om)\sum_{j:\;\kappa_j\le 2} 
c_j n^{-\kappa_j}+cn^{-2}+O(n^{-2-\ep})\right],\qquad \ep>0,
\end{equation}
where the sum is finite and it goes over $j=1,2,\dots$ such that $\kappa_j\le 2$,
\begin{equation} \label{dph20}
R=\left(\frac{\pi}{2\cos \frac{\pi \z}{2}}\right)^2,
\quad \z\equiv \frac{t}{\ga}\,;
\qquad\om=\pi(1+\z)\,; 
\qquad \kappa_j=1+\frac{2 j}{\frac{\pi}{2\ga}-1}\,,
\end{equation}
and
\begin{equation}\label{dph20a}
c_j=\frac{2\ga e^{\f(y_j)}}{\cos\frac{\pi\z}{2}}
(-1)^j\sin\frac{\pi j}{1-\frac{2\ga}{\pi}}\,,
\end{equation}
where
\begin{equation}\label{dph20b}
y_j=\frac{\pi j}{\frac{\pi}{2\ga}-1}\,,
\end{equation}
and
\begin{equation}\label{dph20c}
\f(y)=-\frac{2y}{\pi }\ln\left(2\pi\cos\frac{\pi\z}{2}\right)
+\frac{2}{\pi } \left[\int_0^\infty
\arg (\mu+iy) f(\mu)d\mu
+y\ln y -y\right],
\end{equation}
where
\begin{equation}\label{dph20d}
f(\mu)=\frac{\pi}{2\ga}\coth \mu\frac{\pi}{2\ga}
-\left(\frac{\pi}{2\ga}-1\right)
\coth \mu\left(\frac{\pi}{2\ga}-1\right)
-\sign \mu.
\end{equation}
Also,
\begin{equation}\label{dph20e}
c=\frac{\pi\ga^2}{6(\pi-2\ga)\cos^2\frac{\pi\zeta}{2}}
-\frac{\pi^2}{48\cos^2\frac{\pi\zeta}{2}}\,.
\end{equation}
The error term in (\ref{dph19}) is uniform on any compact
subset of the set
\begin{equation}\label{dph20f}
\left\{(t,\ga):\;|t|<\ga,\;0<\ga<\frac{\pi}{2}\right\}\,.
\end{equation}
\end{theo}

{\bf Remark.} The method of the proof allows an extension of 
formula (\ref{dph19}) to an asymptotic series in
negative powers of $n$. We stopped at terms of the order of $n^{-2}$,
because for higher order terms formula for $c_j$ becomes complex.

Denote 
\begin{equation} \label{dph21}
F_N=\frac{1}{N^2}\ln \frac{\tau_N}{\left(\prod_{n=0}^{N-1} n!\right)^2}.
\end{equation}
From Theorem \ref{R_n} we will derive the following result.

\begin{theo} \label{F_N}
As $N\to\infty$,
\begin{equation} \label{dph22}
F_N=F+O(N^{-1}),
\end{equation}
where
\begin{equation} \label{dph23}
F=\frac{1}{2}\ln \frac{R}{\ga^2}
=\ln\frac{\pi}{ 2\ga\cos \frac{\pi \z}{2}}\,.
\end{equation}
\end{theo} 

This coincides with the formula of work \cite{Z}, obtained
in the saddle-point approximation. Earlier it was derived 
in work \cite{KZ}, from some Ansatz for the free energy asymptotics. 
For the partition function $Z_N$ in (\ref{pf7}) we obtain 
from Theorem \ref{F_N} the formula, 
\begin{equation} \label{exact1}
\frac{1}{N^2}\ln Z_N
=f+O(N^{-1})\,\qquad
f=\ln\left(\frac{\pi[\cos(2t)-\cos(2\ga)]}{ 4\ga\cos \frac{\pi t}{2\ga}}\right).
\end{equation}
Let us compare this formula and asymptotics (\ref{dph19})  with known exact results.

{\bf The free fermion line, $\ga=\frac{\pi}{4}$, $|t|<\frac{\pi}{4}$.} 
In this case the exact result
is 
\begin{equation} \label{exact1a}
Z_N=1,
\end{equation}
see, e.g., \cite{CP}, which implies $f=0$. This agrees with formula (\ref{exact1}),
which also gives $f=0$ when $\ga=\frac{\pi}{4}$.
Moreover, the orthogonal polynomials in this case
are the Meixner-Pollaczek polynomials, for which
\begin{equation} \label{exact2}
R_n=\frac{4n^2}{\cos^2 2t}=\frac{n^2R}{\ga^2}\,,
\end{equation}
cf. \cite{CP}. Thus, formula (\ref{dph19}) is exact on the free
fermion line, with no error term. This agrees with Theorem \ref{R_n},
because from (\ref{dph20a}), (\ref{dph20e}), $c_j=c=0$ when $\ga=\frac{\pi}{4}$. 

{\bf The ASM (ice) point, $\ga=\frac{\pi}{3}\,$, $t=0$.} 
In this case we obtain from (\ref{pf6}) that
\begin{equation} \label{exact2a}
a=b=c=\frac{\sqrt 3}{2}\,,
\end{equation}
hence
\begin{equation} \label{exact3}
Z_N=\left(\frac{\sqrt 3}{2}\right)^{N^2}A(N),
\end{equation}
where $A(N)$ is the number of configurations in the six-vertex model
with DWBC. There is a one-to-one correspondence between the set of
configurations in the six-vertex model with DWBC and the set of
$N\times N$ alternating sign matrices. By definition, an
alternating sign matrix (ASM) is a matrix with the following properties:
\begin{itemize}
\item all entries of the matrix are $-1,0,1$;
\item if we look at the sequence of $(-1)$'s and 1's, they are
alternating along any row and any column;
\item the sum of entries is equal to 1 along any row and any column.
\end{itemize}
The above correspondence is established as follows: given a configuration 
of arrows on edges, we assign $(-1)$
to any vertex  of type (1) on Fig.~1, $1$ to any vertex  of type (2),
and 0 to any vertex of other types. Then the configuration on the
vertices gives rise to an ASM, and this correspondence is one-to-one.
For instance, Fig. 5 shows the ASM corresponding to the configuration
of arrows on Fig. 2.

\begin{center}
 \begin{figure}[h]\label{ASM}
\begin{center}
   \scalebox{0.45}{\includegraphics{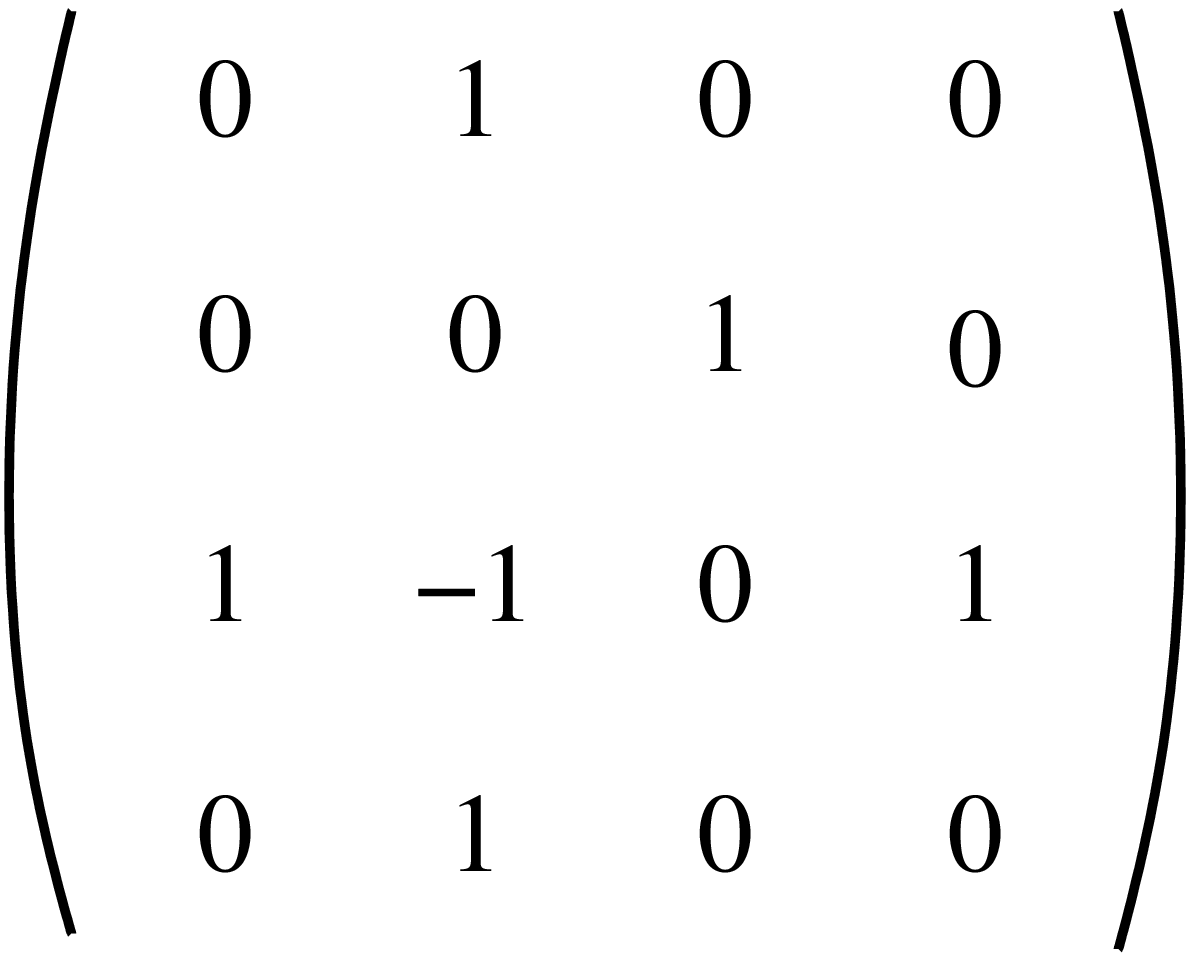}}
\end{center}
        \caption[ASM for the configuration of Fig. 2]{ASM for the configuration of Fig. 2.}
    \end{figure}
\end{center}

For the number of ASMs there is an exact formula:
\begin{equation} \label{exact4}
A(N)=\prod_{n=0}^{N-1}\frac{(3n+1)!n!}{(2n)!(2n+1)!}\,.
\end{equation}
This formula was  conjectured in \cite {MRR1}, \cite{MRR2}, and 
proved by Zeilberger \cite{Ze} by combinatorial arguments.
Another proof was given by Kuperberg \cite{Kup}, who used formula (\ref{pf7}).
The relation to classical orthogonal polynomials was found by Colomo and
Pronko \cite{CP}, who used this relation to give a new proof
of the ASM conjecture. The orthogonal polynomials in this case are the continuous
Hahn polynomials and from \cite{CP} we find that
\begin{equation} \label{exact5}
R_n=\frac{n^2(9n^2-1)}{4n^2-1}=\frac{9n^2}{4}+\frac{5}{16}+O(n^{-2}).
\end{equation}
Formula (\ref{dph19}) gives
\begin{equation} \label{exact6}
R_n=\frac{9n^2}{\pi^2}\left[\frac{\pi^2}{4}+\frac{5\pi^2}{144n^2}
+O(n^{-2-\ep})\right]\,,
\end{equation}
which agrees with (\ref{exact5}). From (\ref{exact4}) we find, see Appendix
\ref{AN}, that as $N\to\infty$,
\begin{equation} \label{exact6a}
A(N)=C\left(\frac{3\sqrt 3}{4}\right)^{N^2}N^{-\frac{5}{36}}
\left(1-\frac{115}{15552N^2}+O(N^{-3})\right),
\end{equation}
where $C>0$ is a constant, so that
\begin{equation} \label{exact6b}
Z_N=C\left(\frac{9}{8}\right)^{N^2}N^{-\frac{5}{36}}
\left(1-\frac{115}{15552N^2}+O(N^{-3})\right),\qquad N\to\infty.
\end{equation}
Formula (\ref{exact1}) gives $f=\ln \frac{9}{8}\,$, which agrees with 
the last formula.

{\bf The $x=3$ ASM  point, $\ga=\frac{\pi}{6}\,$, $t=0$.} Here the exact result 
is 
\begin{equation} \label{exact7}
Z_N=\frac{ 3^{N/2}}{2^{N^2}}A(N;3),
\end{equation}
where
\begin{equation} \label{exact8}
\left\{\begin{aligned}
A(2m+1;3)&=3^{m(m+1)}\prod_{k=1}^m\left[\frac{(3k-1)!}{(m+k)!}\right]^2,\\
A(2m+2;3)&=3^m\frac{(3m+2)!m!}{[(2m+1)!]^2}A(2m+1;3)\,.
\end{aligned}\right.
\end{equation}
In this case $A(N;3)$ counts the number of alternating sign matrices with weight $3^k$,
where $k$ is the number of $(-1)$ entries. Formula (\ref{exact8}) for $A(N;3)$
was conjectured in \cite {MRR1}, \cite{MRR2} and proved in \cite{Kup}. 
The relation to classical orthogonal polynomials was again found by Colomo and
Pronko \cite{CP}, who used it to give a new proof of formula (\ref{exact8})
for the 3-enumeration of ASMs. The orthogonal polynomials in this case are expressed
in terms of the continuous dual
Hahn polynomials and from \cite{CP} we find that
\begin{equation} \label{exact9}
\begin{aligned}
R_{2m}=36m^2,\qquad
R_{2m+1}=4(3m+1)(3m+2)\,.
\end{aligned}
\end{equation}
In this case the subdominant term in the asymptotics of $R_n$ exhibits a
period 2 oscillation. Namely, we obtain from the last formula that
\begin{equation} \label{exact9a}
R_n=9n^2+\frac{-1+(-1)^n}{2}\,.
\end{equation}
This perfectly fits to the frequency value $\om=\pi$ for $\z=0$ in (\ref{dph20}).
Moreover, formula (\ref{dph19}) gives
\begin{equation} \label{exact10}
R_n=\frac{36n^2}{\pi^2}\left[\frac{\pi^2}{4}+\frac{(-1)^nc_1}{n^{2}}-\frac{\pi^2}{72n^2}
+O(n^{-2-\ep})\right]\,,
\end{equation}
which agrees with (\ref{exact9a}) and it provides with the value of $c_1=\frac{\pi^2}{72}$.

From (\ref{exact8}) we find, see Appendix
\ref{AN}, that as $m\to\infty$,
\begin{equation} \label{exact11}
A(2m;3)=C_3\left(\frac{3}{2}\right)^{4m^2}3^{-m}
(2m)^{\frac{1}{18}}\left(1+\frac{77}{7776m^2}+O(N^{-3})\right),
\end{equation}
where $C_3>0$ is a constant, 
and
\begin{equation} \label{exact11a}
A(2m+1;3)=
C_3\left(\frac{3}{2}\right)^{(2m+1)^2}3^{-\frac{2m+1}{2}}
(2m+1)^{\frac{1}{18}}\left(1+\frac{131}{7776m^2}+O(m^{-3})\right).
\end{equation}
so that
\begin{equation} \label{exact11b}
A(N;3)=C_3\left(\frac{3}{2}\right)^{N^2}3^{-\frac{N}{2}}
N^{\frac{1}{18}}\left(1+\frac{104-27(-1)^N}{1944N^2}+O(N^{-3})\right),
\end{equation}
and
\begin{equation} \label{exact12}
Z_N=C_3\left(\frac{3}{4}\right)^{N^2}N^{\frac{1}{18}}
\left(1+\frac{104-27(-1)^N}{1944N^2}+O(N^{-3})\right),\qquad N\to\infty.
\end{equation}
Formula (\ref{exact1}) gives $f=\ln \frac{3}{4}\,$, which agrees with 
the last formula.

We have the identity,
\begin{equation} \label{dph24}
\frac{\partial^2 F_N}{\partial t^2}=\frac{R_N}{N^2}\,,
\end{equation}
see, e.g., \cite{BI3}, which is equivalent to the Toda equation
(\ref{pf10}).
By using identity (\ref{dph24}), we obtain from Theorem
\ref{R_n} the following asymptotics.

\begin{theo} \label{F_N_t} As $N\to\infty$,
\begin{equation} \label{dph25}
\frac{\partial^2(F_N-F)}{\partial t^2}=\frac{1}{\ga^2}
\cos(N\om)\sum_{j:\;\kappa_j\le 2}  c_j N^{-\kappa_j}+cN^{-2}+O(N^{-2-\ep}).
\end{equation}
\end{theo} 

This gives a quasiperiodic asymptotics, as $N\to\infty$, of the second derivative
of the subdominant terms. 

{\bfseries Zinn-Justin's conjecture.}
Paul Zinn-Justin conjectured in \cite{Z} that
\begin{equation} \label{dph26}
Z_N\sim CN^{\kappa}e^{N^2f}\,,
\end{equation}
i.e.,
\begin{equation} \label{dph26a}
\lim_{N\to\infty}\frac{Z_N}{CN^{\kappa}e^{N^2f}}=1\,.
\end{equation}
Formulae (\ref{exact1a}), (\ref{exact6b}), and (\ref{exact12}) confirm this
conjecture, with the value of $\kappa$ given as
\begin{equation} \label{dph27}
\kappa=
\left\{
\begin{aligned}
&0,\quad \ga=\frac{\pi}{4},\quad |t|<\frac{\pi}{4}\,;\\
&-\frac{5}{36},\quad \ga=\frac{\pi}{3},\quad t=0;\\
&\frac{1}{18},\quad \ga=\frac{\pi}{6},\quad t=0.
\end{aligned}\right.
\end{equation}
Bogoliubov, Kitaev and Zvonarev obtained in \cite {BKZ} the asymptotics of $Z_N$
on the line $\frac{a}{c}+\frac{b}{c}=1$, separating the
disordered and antiferroelectric phases. This corresponds to the
value $\ga=0$. They found that in this case formula (\ref{dph26})
holds with $\kappa=\frac{1}{12}\,$.

With the help of Theorem \ref{R_n} we will prove the following result.

\begin{theo} \label{kappa} We have that
\begin{equation} \label{dph27b}
Z_N= CN^{\kappa}e^{N^2f}\left(1+O(N^{-\ep})\right),\qquad \ep>0,
\end{equation}
where
\begin{equation} \label{dph27c}
\kappa=\frac{1}{12}-\frac{2\ga^2}{3\pi(\pi-2\ga)}\,,
\end{equation}
and $C>0$ is a constant.
\end{theo} 

This proves the conjecture of Zinn-Justin, and it gives the exact 
value of the exponent $\kappa$. Let us remark, that the presence of the power-like factor
$N^\kappa$ in the asymptotics of $Z_N$ in (\ref{dph27b})
is rather unusual from the point of view of random matrix models.
As was proven rigorously by Ercolani and McLaughlin \cite{EM},
in the one-matrix model with an independent of $N$ analytic interaction $V(M)=M^2+tV_1(M)$,
where $t>0$ is small, $N^{-2}\ln \frac{Z_N(t)}{Z_N(0)}$ is expanded into an 
asymptotic series in powers of $N^{-2}$.

The set-up for the remainder of the paper is the following:
\begin{itemize}
\item  
In Section
\ref{Rescaling} we describe a rescaling of the weight, which was
introduced by Zinn-Justin \cite{Z}, and which is convenient in the
subsequent calculations. The rescaled random matrix model 
is described by a potential $V_N(x)$ such that as $N\to\infty$,
it approaches a limiting potential $V(x)$. 
\item 
In Sections \ref{EM}-\ref{EMVN}
we evaluate the equilibrium measures for the random matrix models,
first for the limiting one, corresponding to $V(x)$, and then for the
random matrix model, which corresponds to  $V_N(x)$.
\item 
In Section \ref{RHP} we remind the Riemann-Hilbert problem for orthogonal polynomials,
and in Sections \ref{TRHP}-\ref{ASRHP} we carry out the large $N$ 
asymptotic analysis of the Riemann-Hilbert
problem, via a sequence of transformations and the Deift-Zhou nonlinear steepest descent
method. 
\item
We use the results of this analysis in
Section \ref{asympt}, where we obtain the large $N$ asymptotics of the recurrent 
coefficients and
prove Theorem \ref{R_n}. The central point in the derivation of the subdominant 
asymptotics of the recurrent coefficient is a deformation of the lenses boundary,
see Section \ref{R_n}. During this deformation, 
the lenses boundary crosses poles of the function
$e^{-NG_N(z)}$, and every time it crosses a pole, a new subdominant term 
arises in the asymptotics of the recurrent coefficient. Section \ref{Proofs}
gives a proof to Theorems \ref{F_N}-\ref{kappa}. 
\item
Finally,
there are several  Appendices to the paper, where some auxiliary results
are proved and some exact large $N$ asymptotics are obtained.
\end{itemize}

\section{Rescaling of the Weight}\label{Rescaling}

Following \cite{Z}, let us 
substitute $\la_i=\frac{N\mu_i}{\ga}$ in (\ref{dph11}). This reduces $\tau_N$ to
\begin{equation} \label{resc0}
\tau_N=\frac{N^{N^2}\tilde\tau_N}{N!\ga^{N^2}},
\end{equation}
where 
\begin{equation} \label{resc1}
\tilde\tau_N=\int 
\prod_{i=1}^N
\left[e^{N\zeta\mu_i}m\left(\frac{N\mu_i}{\ga}\right)d\mu_i\right]
\De(\mu)^2,
\end{equation}
and
\begin{equation} \label{resc2}
\zeta=\frac{t}{\ga},\qquad -1<\zeta<1.
\end{equation}
The polynomials
\begin{equation} \label{resc3}
P_{Nn}(x)\equiv \left(\frac{\ga}{N}\right)^n P_n\left(\frac{Nx}{\ga}\right)
\end{equation}
are monic polynomials orthogonal with respect to the weight
$e^{N\z x}m\left(\frac{Nx}{\ga}\right)$, so that
\begin{equation} \label{resc4}
\int_{-\infty}^\infty P_{Nn}(x)P_{Nm}(x)e^{N\z x}m\left(\frac{Nx}{\ga}\right)dx
=h_{Nn}\delta_{nm},
\end{equation}
where
\begin{equation} \label{resc5}
h_{Nn}=\left(\frac{\ga}{N}\right)^{2n+1} h_n.
\end{equation}
It follows from (\ref{resc1}) that
\begin{equation} \label{resc6}
\tilde\tau_N=\prod_{n=0}^{N-1}h_{Nn}. 
\end{equation}
The polynomials $P_{Nn}(x)$ satisfy the three term recurrent
relation,
\begin{equation} \label{resc7}
xP_{Nn}(x)=P_{N,n+1}(x)+Q_{Nn}P_{Nn}(x)+R_{Nn}P_{N,n-1}(x),
\end{equation}
where
\begin{equation} \label{resc8}
R_{Nn}=\left(\frac{\ga}{N}\right)^2R_n,\qquad Q_{Nn}=\frac{\ga}{N}Q_n.
\end{equation}
In what follows we will evaluate the asymptotics of $R_{NN}$ and $Q_{NN}$
as $N\to\infty$. In particular, for $R_{NN}$ we will obtain the formula
\begin{equation} \label{resc8a}
R_{NN}=R+\cos(N\om)\sum_{j:\;\kappa_j\le 2} 
c_j N^{-\kappa_j}+O(N^{-2}).
\end{equation}
Then (\ref{resc8}) will provide us with the needed asymptotics
of $R_n$ as $n\to\infty$.

Define
\begin{equation} \label{resc9}
V_N(\mu)=
-\zeta\mu-\frac{1}{N}\ln\left[\frac{\sinh N\mu(\frac{\pi}{2\ga}-1)}
{\sinh N\mu\frac{\pi}{2\ga}}\right]\,.
\end{equation}
Then 
\begin{equation} \label{resc10}
e^{-NV_N(\mu)}=e^{N\zeta\mu}\frac{\sinh N\mu(\frac{\pi}{2\ga}-1)}
{\sinh N\mu\frac{\pi}{2\ga}}
=e^{N\zeta\mu}m\left(\frac{N\mu}{\ga}\right),
\end{equation}
hence
\begin{equation} \label{resc11}
\tilde\tau_N=
\int \prod_{i=1}^N[e^{-NV_N(\mu_i)} d\mu_i]
\De(\mu)^2,
\end{equation}
Observe that as $N\to\infty$,
\begin{equation} \label{resc12}
 V_N(\mu)\to V(\mu)\equiv
-\zeta\mu+|\mu|.
\end{equation}
We will evaluate the equilibrium measures, first for $V$
and then for $V_N$. But before we discuss
some general formulae for equilibrium measures.

\section { Equilibrium Measure} \label{EM}

In this section we remind some facts concerning equilibrium measures,
see \cite{DKM}, \cite{DKMVZ}. 
Let $V(x)$ be a real analytic function such that 
\begin{equation} \label{em_a}
 \lim_{x\to\pm\infty}\frac {V(x)}{\ln|x|}=\infty.
\end{equation}
The equilibrium measure, $\nu^{\rm{eq}}=\nu^{\rm{eq}}_V$, for $V$ is defined as 
a minimizer of the functional
\begin{equation}\label {em_b}
I_V(\nu)=-\iint_{\R^2} \ln|x-y|d\nu(x)d\nu(y)+\int_{\R^1} V(x)d\nu(x),
\end{equation}
over all probability measures $\nu$ on $\R^1$. The minimizer 
exists and it is unique.  The equilibrium measure
has the following properties:
\begin{itemize}
\item It is absolutely continuous with respect to the Lebesgue measure,
$d\nu^{\rm{eq}}(x)=\rho(x)dx$.
\item It is supported by a finite number of disjoint intervals,
$S=\cup_{l=1}^q[\al_l,\be_l]$.
\item On $S$,
\begin{equation}\label {em_c}
\rho(x)=\frac{1}{2\pi i}h(x)\sqrt{R(x+i0)},\qquad R(z)\equiv \prod_{l=1}^q
(z-\al_l)(z-\be_l),
\end{equation}
where $h(x)$ is a real
analytic function on the real line,
and $\sqrt{R(z)}$ is taken on the principal sheet.
\end{itemize}
The function $h(x)$ is expressed by the contour integral,
\begin{equation}\label {em_d}
h(x)=\frac{1}{2\pi i}\oint_{\Gamma}\frac{V'(s)ds}{(x-s)\sqrt{R(s)}},
\qquad x\in S,
\end{equation}
over any closed contour $\Gamma$,
with $S$ in its interior. For the equilibrium measure,
consider its resolvent, 
\begin{equation}\label {EM1}
\om(z)=\int_{S}\frac{\rho(\mu)d\mu}{z-\mu},\quad
z\in\C\setminus S,
\end{equation}
and the $g$-function,
\begin{equation}\label {EM2}
g(z)=\int_S\rho(\mu)\log(z-\mu)d\mu, \qquad z\in\C\setminus(-\infty,\be_q],
\end{equation}
where for $\log z$ the principal branch is taken. Then
\begin{equation}\label {EM3}
g'(z)=\om(z),
\end{equation}
and, by the jump property of the Cauchy integral, 
\begin{equation}\label {EM3a}
\om(\mu+i0)-\om(\mu-i0)=-2\pi i\rho(\mu),\qquad \mu\in S.
\end{equation}
As $z\to\infty$,
\begin{equation}\label {EM4}
\om(z)=z^{-1}+O(z^{-2}),\qquad g(z)=\log z +O(z^{-1}).
\end{equation}
The equilibrium measure is uniquely determined by the condition that
there exists a real constant $l$ such that
\begin{itemize}
\item For any $\mu\in S$,
\begin{equation}\label {g1}
g(\mu+i0)+g(\mu-i0)-V(\mu)-l=0,\qquad \mu\in S.
\end{equation}
\item For any $\mu\in \R\setminus S$,
\begin{equation}\label {g2}
g(\mu+i0)+g(\mu-i0)-V(\mu)-l\le 0,\qquad \mu\in \R\setminus S,
\end{equation}
\end{itemize}
see \cite{DKM}, \cite {DKMVZ}. It implies the equation,
\begin{equation}
\om(\mu+i0)+\om(\mu-i0)=V'(\mu),\qquad \mu\in S.
\label {g3}
\end{equation}
A solution to this equation can be found as
\begin{equation}
\om(z)=\frac{\sqrt{R(z)}}{2\pi i}\int_S\frac{V'(x)dx}{(z-x)\sqrt{R(x+i0)}}.
\label {g3a}
\end{equation}
From (\ref{EM3a}), 
\begin{equation}\label {g4}
g(\mu+i0)-g(\mu-i0)=\left\{
\begin{aligned}
&2\pi i,\qquad \mu\le \al_1,\\
&2\pi i\int_{\mu}^{\be_q}\rho(s)\chi_S(s)ds,\qquad \al_1\le\mu\le\be_q,\\
&0,\;\;\qquad \mu\ge \be_q,
\end{aligned}
\right.
\end{equation}
where $\chi_S$ is the characteristic function of $S$. 

We will be interested in the case when $V$ is convex. In this
case the equilibrium measure is supported by one interval, say, $[\al,\be]$,
and (\ref{g4}) reduces to 
\begin{equation}\label {g5}
g(\mu+i0)-g(\mu-i0)=\left\{
\begin{aligned}
&2\pi i,\qquad \mu\le \al,\\
&2\pi i\int_{\mu}^{\be}\rho(s)ds,\qquad \al\le\mu\le\be,\\
&0,\;\;\qquad \mu\ge \be.
\end{aligned}
\right.
\end{equation}
For $z_0\in\C$ and $r>0$, we will us the standard notation for the disk, 
\begin{equation}\label {g6}
D(z_0,r)=\{z\in\C:\;|z-z_0|<r\}.
\end{equation}
From (\ref{em_c}), (\ref{g1}) and (\ref{g5}), one obtains that
there exists $r>0$ such that
\begin{equation}\label {g7}
-2g(z)+V(z)+l=\int_{\be}^z h(s)\sqrt{(s-\al)(s-\be)}ds,\qquad
z\in D(\be,r)\setminus [\al,\be], 
\end{equation}
and
\begin{equation}\label {g7a}
-2g(z)+V(z)+l=-2\pi i\,\sign (\Im z)+\int_z^{\al} h(s)\sqrt{(\al-s)(\be-s)}ds,\qquad
z\in D(\al,r)\setminus \R. 
\end{equation}
Finally, it follows from (\ref{EM4}) that 
the function $e^{Ng(z)}$ is analytic on $\C\setminus [\al,\be]$, and
\begin{equation}\label{g8}
e^{Ng(z)}=z^N+O(z^{N-1}),\qquad z\to\infty.
\end{equation}

\section { Equilibrium Measure for $V$} \label{EMV}

In this section we consider the equilibrium measure for the potential
\begin{equation} \label{EMV1}
 V(\mu)= -\zeta\mu+|\mu|,\qquad |\zeta|<1.
\end{equation}
It is obviously a convex function, and 
\begin{equation} \label{EMV2}
 V'(\mu)= -\zeta+\sign(\mu).
\end{equation}
Integral (\ref{g3a}) is explicitly evaluated in this case as
\begin{equation}\label{V3}
\om(z)=\frac{1-\z}{2}+\frac{2}{i\pi}\log\left[\frac{\sqrt{\be(z-\al)}
-i\sqrt{-\al(z-\be)}}{\sqrt{z(\be-\al)}}\right],
\end{equation}
and from the asymptotics,
\begin{equation}\label{EMV2a}
\om(z)=z^{-1}+O(z^{-2}),\qquad z\to\infty,
\end{equation}
one finds that
\begin{equation}
\al=-\pi\tan\frac{\pi}{4}(1-\z),\quad
\be=\pi\tan\frac{\pi}{4}(1+\z),
\label {V4}
\end{equation} 
see \cite{Z}. Observe that
\begin{equation}
\be+\al=2\pi\tan\frac{\pi\z}{2}\,,\qquad
\be-\al=\frac{2\pi}{\cos\frac{\pi\z}{2}}\,,\qquad (-\al)\be=\pi^2.
\label {V4a}
\end{equation} 
For the square root in (\ref{V3}) we take the principal branch,
with a cut on the negative half-axis.
The function $\om(z)$ is analytic on $\C\setminus [\al,\be]$.
On $[\al,\be]$,
\begin{equation}
\om(\mu\pm i0)=\frac{-\z+\sign(\mu)}{2}
\pm\frac{2}{i\pi}\log\left[\frac{\sqrt{\be(\mu-\al)}
+\sqrt{-\al(\be-\mu)}}{\sqrt{|\mu|(\be-\al)}}\right],\quad \al<\mu<\be,
\label {V5}
\end{equation}
so that the density function $\rho(\mu)$ 
is equal to
\begin{equation}
\rho(\mu)= 
\frac{2}{\pi^2}\log\left[\frac{\sqrt{\be(\mu-\al)}
+\sqrt{-\al(\be-\mu)}}{\sqrt{|\mu|(\be-\al)}}\right],\quad \al<\mu<\be.
\label {V7}
\end{equation}
Observe that $\rho(\mu)$ has a logarithmic singularity at the origin.
From (\ref{V5}),
\begin{equation}
\om(\al)=\frac{-\z-1}{2}\,,\quad \om(\be)=\frac{-\z+1}{2}\,.
\label {V8}
\end{equation}
From (\ref{V3}), (\ref{V4}) we obtain that
\begin{equation}\label{V11}
g(z)=z\om(z)+
2\log\left(\sqrt{z-\al}+\sqrt{z-\be}\right)-(1+2\log 2)
\end{equation}
(see Appendix \ref{A} below). This implies that
\begin{equation}\label{V12}
\int_{\mu}^{\be}\rho(s)ds=-\mu\rho(\mu)+\frac{2}{\pi}
\arctan\sqrt{\frac{\be-\mu}{\mu-\al}},\qquad \al\le\mu\le\be.
\end{equation}
In particular, 
\begin{equation}\label{V12a}
\int_0^{\be}\rho(s)ds=\frac{1+\z}{2}\,.
\end{equation}
In addition, we have that
\begin{equation}\label{V13}
g(\mu+i0)+g(\mu-i0)-V(\mu)-l=0,
\quad \al\le\mu\le \be.
\end{equation}
By (\ref{V8}), $\om(\be)=\frac{1-\z}{2}$, hence
\begin{equation}\label{V14}
l= 2g(\be)-V(\be)=2\ln(\be-\al)-2-4\ln 2.
\end{equation}
Define an analytic continuation of the potential $V(\mu)=-\z\mu+|\mu|$
from $\R$ to $\C$ as
\begin{equation}\label {V15}
V(z)=\left\{
\begin{aligned}
{}&(1-\z)z,\quad \Re z\ge 0,\\
{}&(-1-\z)z,\quad \Re z\le 0
\end{aligned}
\right.
\end{equation}
(the function $V(z)$ is two-valued on $\Re z=0$).
In what follows we will use the function $h(z)$ defined by the formulae,
\begin{equation}\label {V16}
h(z)=\frac{4i}{\pi \sqrt{(z-\al)(z-\be)}}
\log\left[\frac{\sqrt{\be(z-\al)}
-i\sqrt{(-\al)(z-\be)}}{\sqrt{ z(\be-\al)}}\right],\quad \Re z\ge 0.
\end{equation}
and
\begin{equation}\label {V16a}
h(z)=-\frac{4i}{\pi \sqrt{(\al-z)(\be-z)}}
\log\left[\frac{\sqrt{(-\al)(\be-z)}+i\sqrt{\be(\al-z)}}
{\sqrt{(-z)(\be-\al)}}\right],\quad \Re z\le 0,
\end{equation}
where the square root and logarithm are taken on the principal sheet.
The function $h(z)$ has the following properties:
\begin{enumerate}
\item [(i)] {$h(z)$ is analytic in $\C\setminus i\R$ and
\begin{equation}\label {V17}
h(\al)=\frac{4}{(-\al)\,(\be-\al)}>0,\quad
h(\be)=\frac{4}{\be\,(\be-\al)}>0;
\end{equation}
\begin{equation}
h'(\al)=\frac{4(\be-3\al)}{3\al^2\,(\be-\al)^2},\quad
h'(\be)=\frac{4(\al-3\be)}{3\be^2\,(\be-\al)^2};
\end{equation}}
\item [(ii)] {by (\ref{V7}),
\begin{equation}\label {V18a}
\rho(\mu)=\frac{1}{2\pi } h(\mu)\sqrt {(\mu-\al)(\be-\mu)},\quad \al<\mu<\be;
\end{equation}}
\item [(iii)] {by (\ref{V3}),
\begin{equation}\label {V19a}
\om(z)=\frac{V'(z)}{2}-\frac{h(z)\sqrt{(z-\al)(z-\be)}}{2}
\end{equation}}
\end{enumerate}
Asymptotic formulae for orthogonal polynomials with weight
(\ref{EMV1}) with $\zeta=0$ were obtained, via the Riemann-Hilbert approach,
in paper \cite{KM} by Kriecherbauer and McLaughlin. In fact, they studied
a more general case, of
the Freud potentials of the form $V(\mu)=|\mu|^{\al}$. 

\section { Equilibrium Measure for $V_N$}\label{EMVN}

From (\ref{resc9}) we obtain that
\begin{equation}\label {VN1}
V'_N(\mu)=-\left(\frac{\pi}{2\ga}-1\right)
\coth N\mu\left(\frac{\pi}{2\ga}-1\right)
+\frac{\pi}{2\ga}\coth N\mu\frac{\pi}{2\ga}
-\zeta.
\end{equation}
The function $V'_N(\mu)$
is increasing, hence $V_N(\mu)$ is convex. Its equilibrium measure,
$\rho_N(\mu) d\mu$, is supported by one interval $[\al_N,\be_N]$.
As $N\to\infty$, the equilibrium measure for $V_N$ converges
to the one for $V$. In this section we will derive some
asymptotic formulas for $\al_N$, $\be_N$ and $\rho_N(\mu)$ as $N\to\infty$.
Consider the resolvent,
\begin{equation}\label {VN2}
\om_N(z)=\int_{\al_N}^{\be_N}\frac{d\mu\,\rho_N(\mu)}{z-\mu},\quad
z\in\C\setminus[\al_N,\be_N].
\end{equation}
Then
\begin{equation}\label {VN3a}
\rho_N(\mu)=-\frac{1}{2\pi i}[\om_N(\mu+i0)-\om_N(\mu-i0)],
\qquad \al_N< \mu< \be_N,
\end{equation}
and
\begin{equation}\label {VN3aa}
\om_N(\mu\pm i0)=\mp \pi i \rho_N(\mu)+ P.V. \int_{\al_N}^{\be_N}
\frac{\rho_N(x)dx}{\mu-x},
\qquad \al_N< \mu< \be_N,
\end{equation}
where $P.V.\int$ is the principal value of the integral.
The resolvent solves the equation,
\begin{equation}
\om_N(\mu+i0)+\om_N(\mu-i0)=V'_N(\mu),\qquad \al_N< \mu< \be_N.
\label {VN3}
\end{equation}
The solution to this equation is
\begin{equation}\label{VN4}
\om_N(z)=-\frac{\sqrt{R_N(z)}}{2\pi i}\int_{\al_N}^{\be_N}
\frac{V'_N(x)}{(z-x)\sqrt{R_N(x)}_+}dx,\quad z\in\C\setminus[\al_N,\be_N], 
\end{equation}
where
\begin{equation}\label{VN4a}
R_N(z)=(z-\al_N)(z-\be_N),
\end{equation}
and $\sqrt{R_N(z)}$ is taken on the principal sheet, with a cut on $[\al_N,\be_N]$.
As usual,
\[
\sqrt{R_N(x)}_+=\lim_{\ep\to+0}\sqrt{R_N(x+i\ep)}.
\]

{\bfseries Evaluation of the end-points.}
From (\ref{VN2}) we have that $\om_N(z)\sim\frac{1}{z}$ as $z\to \infty$.
By evaluating the large $z$ asymptotics of the integral on the right
in (\ref{VN4}), we obtain the equations,
\begin{equation}\label{VN5}
\frac{1}{2\pi }\int_{\al_N}^{\be_N}\frac{V'_N(x)}{\sqrt{(x-\al_N)(\be_N-x)}}dx=0,
\end{equation}
and
\begin{equation}\label{VN6}
\frac{1}{2\pi }\int_{\al_N}^{\be_N}\frac{xV'_N(x)}
{\sqrt{(x-\al_N)(\be_N-x)}}dx=1.
\end{equation}
From these two equations we obtain the following asymptotics 
of $\al_N$, $\be_N$ as $N\to\infty$.

\begin{prop}\label{alN-beN} As $N\to\infty$,
\begin{equation}\label{VN7}
\begin{aligned} 
&\al_N=\al+N^{-2}\frac{\ga^2\left(2\sin\frac{\pi\zeta}{2}-1\right)}{3(\pi-2\ga)\cos\frac{\pi\zeta}{2}}
+O(N^{-3}),\\
&\be_N=\be+N^{-2}\frac{\ga^2\left(2\sin\frac{\pi\zeta}{2}+1\right)}{3(\pi-2\ga)\cos\frac{\pi\zeta}{2}}
+O(N^{-3}),
\end{aligned}
\end{equation}
where $\al$, $\be$ are given in (\ref{V4}).
\end{prop}

Proof of Proposition \ref{alN-beN} is given in Appendix \ref{AppD} below.

{\bfseries Evaluation of the density.}
Consider now the asymptotics of the density function $\rho_N(x)$. As
$N\to\infty$, it approaches the density function $\rho(x)$ given
in (\ref{V7}). The density $\rho(x)$ has a logarithmic
singularity at $x=0$. For $\rho_N$ the singularity is smoothed out
and we are interested in the large $N$ asymptotics of $\rho_N$ near
the origin. 
From (\ref{VN3a}) and (\ref{VN4}) we obtain that
\begin{equation}\label{VN4aa}
\rho_N(\mu)=-\frac{\sqrt{r_N(\mu)}}{2\pi^2}P.V.\int_{\al_N}^{\be_N}
\frac{V'_N(x)dx}{(\mu-x)\sqrt{r_N(x)}}\,,\qquad \al_N<\mu<\be_N.
\end{equation}
where
\begin{equation}\label{VN4ab}
r_N(x)=(x-\al_N)(\be_N-x).
\end{equation}
Observe that $\rho_N(\mu)$ is analytic for $\al_N<\mu<\be_N.$
From (\ref{VN1}) we have that
\begin{equation}\label {VNa}
V'_N(x)=\sign x-\zeta+f(Nx),
\end{equation}
where
\begin{equation}\label {VNb}
f(x)=\frac{\pi}{2\ga}\coth x\frac{\pi}{2\ga}
-\left(\frac{\pi}{2\ga}-1\right)
\coth x\left(\frac{\pi}{2\ga}-1\right)
-\sign x,
\end{equation}
hence
\begin{equation}\label {VNc}
\rho_N(\mu)=\rho_N^0(\mu)+\rho_N^1(\mu),
\end{equation}
where
\begin{equation}\label{VNd}
\rho_N^0(\mu)=-\frac{\sqrt{r_N(\mu)}}{2\pi^2}P.V.\int_{\al_N}^{\be_N}
\frac{(\sign x-\zeta)dx}{(\mu-x)\sqrt{r_N(x)}}\,,\qquad \al_N<\mu<\be_N,
\end{equation}
and
\begin{equation}\label{VNd1}
\rho_N^1(\mu)=-\frac{\sqrt{r_N(\mu)}}{2\pi^2}P.V.\int_{\al_N}^{\be_N}
\frac{f(Nx)dx}{(\mu-x)\sqrt{r_N(x)}}\,,\qquad \al_N<\mu<\be_N.
\end{equation}
The function $\rho_N^0(\mu)$ is evaluated explicitly as
\begin{equation}\label {VNe}
\rho_N^0(\mu)= 
\frac{2}{\pi^2}\log\left[\frac{\sqrt{\be_N(\mu-\al_N)}
+\sqrt{-\al_N(\be_N-\mu)}}{\sqrt{|\mu|(\be_N-\al_N)}}\right],\quad \al_N<\mu<\be_N.
\end{equation}
[cf. (\ref{V7})]. Set
\begin{equation}\label {VNe1}
\om_N^0(z)=\int_{\al_N}^{\be_N}\frac{d\mu\,\rho_N^0(\mu)}{z-\mu},\quad
z\in\C\setminus[\al_N,\be_N].
\end{equation}
Then
\begin{equation}\label{VNe2}
\om_N^0(z)=\frac{1-\z}{2}+\frac{2}{i\pi}\log\left[\frac{\sqrt{\be_N(z-\al_N)}
-i\sqrt{-\al_N(z-\be_N)}}{\sqrt{z(\be_N-\al_N)}}\right],
\end{equation}
[cf. (\ref{V3})].

To describe the large $N$ asymptotics of $\rho_N^1(\mu)$,
introduce the function,
\begin{equation}\label {VNf}
k(\mu)=P.V.\int_{-\infty}^{\infty}
\frac{f(x)dx}{\mu-x}\,.
\end{equation}
From expilicit formula (\ref{VNb}), we have the following properties of
the function $f(x)$:
\begin{itemize}
\item
 $f(x)$ satisfies the estimate,
\begin{equation}\label {VNg}
|f(x)|\le C_0e^{-C|x|},
\end{equation}
with some $C_0,C>0$;
\item $f(x)$ is an odd function;
\item the function $f(x)+\sign x$ is real analytic.
\end{itemize}
These properties of $f(x)$ imply the following
properties of the function $k(\mu)$:
\begin{itemize}
\item
the function $k(\mu)$ is even and
\begin{equation}\label {kmu1}
k(\mu)=-2\ln|\mu|+k_0(\mu),
\end{equation}
where $k_0(\mu)$ is real analytic;
\item
 as $\mu\to\infty$,
\begin{equation}\label {kmu2}
k(\mu)=\frac{C}{\mu^2}+O(\mu^{-4}),\qquad \mu\to\infty,
\end{equation}
where
\begin{equation}\label {kmu3}
C=\int_{-\infty}^\infty xf(x)dx=-\frac{2\pi\ga^2}{3(\pi-2\ga)}\,.
\end{equation}
\end{itemize}
We use the properties of $k(\mu)$ to prove the following asymptotics 
of the function $\rho_N^1(\mu)$. 

\begin{prop} \label{rhoN1} As $N\to\infty$,
uniformly in the interval $\al_N\le \mu\le \be_N$,
\begin{equation}\label{rhoN1_1}
\rho_N^1(\mu)=-\frac{1}{2\pi^2}\,k(N\mu)+O(N^{-2}).
\end{equation}
In addition, there exists a family of complex domains
$\{\Om_r,\;r>0\}$ such that $[\al_N+r, \be_N-r]\subset\Om_r$ and
$\Om_r\subset\Om_{r'}$
whenever $r>r'$, and such that the function 
\begin{equation}\label{rhoN1_2}
\ep_N(\mu)\equiv \rho_N^1(\mu)+\frac{1}{2\pi^2}\,k(N\mu).
\end{equation}
can be analytically continued to $\Om_r$ and as $N\to\infty$,
\begin{equation}\label{rhoN1_3}
\sup_{z\in\Om_r}|\ep_N(z)|=O(N^{-2}).
\end{equation}
\end{prop}

The proof of Proposition \ref{rhoN1} is given below in Appendix \ref{E}.

Proposition \ref{rhoN1} implies that
\begin{equation}\label{rh_0}
\rho_N(\mu)=\rho_N^0(\mu)-\frac{1}{2\pi^2}\,k(N\mu)+O(N^{-2}),\qquad
\mu\in[\al_N,\be_N],
\end{equation}
and this equation can be extended to the complex domain $\Om_r$, $r>0$.
This can be further specified as follows. Let $r>0$ be an arbitrary
fixed number such that $r\le \frac 12\min\{-\al,\be\}$. Then
\begin{itemize}
\item For $\mu\in[\al_N+r,\be_N-r]$,
\begin{equation}\label{rh_1}
\rho_N(\mu)=\rho(\mu)-\frac{1}{2\pi^2}\,k(N\mu)+O(N^{-2}),
\end{equation}
where $\rho(\mu)$ is given in (\ref{V7}).
\item
For $\mu\in[\al_N,\al_N+r]\cup [\be_N-r,\be_N]$,
\begin{equation}\label{rh_2}
\rho_N(\mu)=\rho_N^0(\mu)+O(N^{-2}).
\end{equation}
\end{itemize}
Observe that (\ref{rh_1}) implies that
\begin{equation}\label{rh_3}
\rho_N(\mu)=\frac{1}{\pi^2}\ln N+a(\mu)-\frac{1}{2\pi^2}\,k_0(N\mu)+O(N^{-2}),
\qquad \mu\in[\al_N+r,\be_N-r],
\end{equation}
where
\begin{equation}\label{rh_4}
a(\mu)= 
\frac{2}{\pi^2}\log\left[\frac{\sqrt{\be(\mu-\al)}
+\sqrt{-\al(\be-\mu)}}{\sqrt{\be-\al}}\right],\quad \al<\mu<\be.
\end{equation}
From (\ref{rh_0}) we obtain the following result:

\begin{prop} \label{int_rho}
We have that
\begin{equation}\label{rh_5}
\int_0^{\be_N}\rho_N(\mu)d\mu
=\frac{1+\zeta}{2}+O(N^{-2}).
\end{equation}
\end{prop}

{\it Proof.} By an explicit integration of (\ref{VNe}) we have that 
\begin{equation}\label{rh_6}
\int_\mu^{\be_N}\rho_N^0(x)dx
=-\mu\rho_N^0(\mu)+\frac{2}{\pi}
\arctan\sqrt{\frac{\be_N-\mu}{\mu-\al_N}},\qquad \al_N\le\mu\le\be_N.
\end{equation}
In particular,
\begin{equation}\label{rh_7}
\int_{\al_N}^{\be_N}\rho_N^0(x)dx=1,
\end{equation}
and
\begin{equation}\label{rh_7a}
\int_0^{\be_N}\rho_N^0(x)dx=\frac{1+\zeta}{2}+O(N^{-2}).
\end{equation}
Since 
\begin{equation}\label{rh_8}
\int_{\al_N}^{\be_N}\rho_N(x)dx=1,
\end{equation}
we obtain from (\ref{rh_0}) that
\begin{equation}\label{rh_9}
\int_{-\infty}^\infty k(x)dx=0.
\end{equation}
Since $k(x)$ is even, this implies that
\begin{equation}\label{rh_10}
\int_0^\infty k(x)dx=0.
\end{equation}
Therefore,
\begin{equation}\label{rh_11}
\int_0^{\be_N}\rho_N(\mu)d\mu
=\int_0^{\be_N}\rho_N^0(\mu)d\mu
-\frac{1}{2\pi^2}\int_0^{\be_N}k(N\mu)d\mu+O(N^{-2})
=\frac{1+\zeta}{2}+O(N^{-2}).
\end{equation}
Proposition \ref{int_rho} is proved.

{\bfseries Evaluation of the resolvent.}
The large $N$ asymptotics of the function $\om_N(z)$
can be obtained as follows. By (\ref{VNc}) and (\ref{rhoN1_2}),
\begin{equation}\label{rh_12}
\rho_N(\mu)=\rho_N^0(\mu)-\frac{1}{2\pi^2}\,k(N\mu)+\ep_N(\mu),
\end{equation}
hence
\begin{equation}\label{om_1}
\om_N(z)=\om_N^0(z)-\frac{1}{2\pi^2}\,m(Nz)+\xi_N(z),
\end{equation}
where
\begin{equation}\label{om_2}
m(z)=\int_{-\infty}^\infty \frac {k(\mu)d\mu}{z-\mu},\qquad
\xi_N(z)=\int_{-\infty}^\infty \frac {\ep_N(\mu)d\mu}{z-\mu}.
\end{equation}
Observe that $\xi_N(z)$ is an analytic function in $(\Om_r\setminus \R)$, $r>0$. 
Consider 
a complex domain $U_r$ such that the closure of $U_r$ belongs 
to $\Om_r$ and $[\al_N+r,\be_N-r]\subset U_r$. 
Then from (\ref{rhoN1_3}) and the analyticity of $\ep_N(z)$ in $\Om_r$ 
we obtain that
\begin{equation}\label{om_3}
\sup_{z\in U_r\setminus \R}|\xi_N(z)|=O(N^{-2}).
\end{equation}
We have that
\begin{equation}\label{om_5}
m(z)=\sign(\Im z)\pi i\int_{-\infty}^\infty \frac {f(\mu)d\mu}{z-\mu},
\end{equation}
Indeed, if we introduce the Fourier transform,
\begin{equation}\label{om_6}
\tilde k(\tau)=\frac{1}{2\pi}\int_{-\infty}^\infty e^{-i\tau \mu} k(\mu)d\mu,
\end{equation}
then (\ref{om_2}) implies that
\begin{equation}\label{om_7}
\tilde m_{\pm}(\tau)=\mp 2\pi i\theta(\pm \tau)\tilde k(\tau),
\end{equation}
where $m_{\pm}( \mu)=m(\mu\pm i0)$,
and $\theta(\tau)=1$ for $\tau\ge 0$ and $\theta(\tau)=0$ for $\tau< 0$.
Also, from (\ref{VNf}),
\begin{equation}\label{om_8}
\tilde k(\tau)=-\pi i\,\sign(\tau)\tilde f(\tau),
\end{equation}
hence
\begin{equation}\label{om_9}
\tilde m_{\pm}(\tau)=- 2\pi^2 \theta(\pm \tau)\tilde f(\tau).
\end{equation}
By taking the inverse Fourier transforms, we obtain that
\begin{equation}\label{om_10}
 m(\mu\pm i0)=\pm \pi i
\int_{-\infty}^\infty \frac {f(x)dx}{\mu\pm i0-x}\,,
\end{equation}
which implies, by means of analytic continuation, (\ref{om_5}).

By using the listed above properties of $f(x)$, we obtain from (\ref{om_5})
the following properties of $m(z)$, $z\in(\C\setminus\R)$:
\begin{itemize}
\item
The symmetry conditions,
\begin{equation}\label{om_11}
m(-z)=-m(z), \qquad m(\bar z)=\overline {m(z)}. 
\end{equation}
\item The representation,
\begin{equation}\label{om_12}
 m(z)=\pm 2\pi i \log z+m_0(z),\qquad \pm \Im z>0,
\end{equation}
with $\log z$ on the principal sheet,
where $m_0(z)$ is analytic in the closed half-planes $\{\pm \Im z\ge 0\}$,
and
\begin{equation}\label{om_13}
m_0(-z)=-m_0(z), \qquad m_0(\bar z)=\overline {m_0(z)}. 
\end{equation}
\item 
As $z\to\infty$,
\begin{equation}\label{om_14}
m(z)=\frac{C\sign(\Im z)}{z^2}+O(z^{-4}).
\end{equation}
\end{itemize}

We summarize the properties of $\om_N(z)$ in the following proposition.

\begin{prop}
For any $r>0$ there exists an
independent of $N$ complex neighborhood $U_r$
of the interval $\al_N+r\le \mu\le \be_N-r$ such that
for $z\in U_r$, equation (\ref{om_1}) holds, in which
$\om_N^0(z)$ is given by (\ref{VNe2}), $m(z),$ by (\ref{om_5}),
and $\xi_N(z)$ satisfies estimate (\ref{om_3}).
In addition, for $z\in U_r$,
\begin{equation}\label{om_15}
 \om_N(z)=\mp \frac{i\ln N}{\pi}+b(z)-\frac{1}{2\pi^2}m_0(Nz)+O(N^{-2}),
\qquad \pm\Im z>0,
\end{equation}
where
\begin{equation}\label{om_16}
b(z)=\frac{1-\z}{2}+\frac{2}{i\pi}\log\left[\frac{\sqrt{\be(z-\al)}
-i\sqrt{-\al(z-\be)}}{\sqrt{\be-\al}}\right],
\end{equation}
with a cut on $[\al,\be]$, and
\begin{equation}\label{om_17}
m_0(z)=\pm\pi i\left[\int_{-\infty}^\infty \frac {f(\mu)d\mu}{z-\mu}
-2\log z\right]\,,\qquad \pm\Im z>0,  
\end{equation}
with $\log z$ on the principal sheet.
\end{prop}

Observe that both $b(z)$ and $m_0(Nz)$ have a jump across $[\al,\be]$,
and 
\begin{equation}\label{om_18}
b(\bar z)=\overline{b(z)},\qquad
m_0(\bar z)=\overline{m_0(z)}.
\end{equation}
By using (\ref{V4a}), we find that
\begin{equation}\label{om_19}
b(+i0)=\frac{1-\z}{2}+\frac{1}{i\pi}\ln\left(2\pi\cos\frac{\pi\z}{2}\right).
\end{equation}

{\bfseries Evaluation of the constant of integration.} Let us evaluate $l_N$.
By (\ref{g1}), for any $\mu\in [\al_N,\be_N]$,
\begin{equation}\label{lN1}
l_N=V_N(\mu)-g_{N-}(\mu)-g_{N+}(\mu).
\end{equation}
Take $\mu=\frac{\be}{2}$. By (\ref{resc9}), there exists $c>0$ such that
\begin{equation}\label{lN2}
V_N\left(\frac{\be}{2}\right)=V\left(\frac{\be}{2}\right)+O(e^{-cN})\,.
\end{equation}
Also, by (\ref{EM2}),
\begin{equation}\label{lN3}
g_{N-}\left(\frac{\be}{2}\right)+g_{N+}\left(\frac{\be}{2}\right)
=2\int_{\al_N}^{\be_N}\rho_N(x)\ln\left|\frac{\be}{2}-x\right|\,dx
\end{equation}
By (\ref{rh_0}) and (\ref{VN7}), we can reduce this to
\begin{equation}\label{lN4}
\begin{aligned}
g_{N-}\left(\frac{\be}{2}\right)+g_{N+}\left(\frac{\be}{2}\right)
&=2\int_{\al}^{\be}\rho(x)\ln\left|\frac{\be}{2}-x\right|\,dx\\
&-\frac{1}{\pi^2}\int_{\al_N}^{\be_N} k(Nx)\ln\left|\frac{\be}{2}-x\right|\,dx+O(N^{-2}).
\end{aligned}
\end{equation}
From (\ref{kmu2}) and (\ref{rh_9}) we obtain that
\begin{equation}\label{lN5}
\int_{\al_N}^{\be_N} k(Nx)\ln\left|\frac{\be}{2}-x\right|\,dx=O(N^{-2}),
\end{equation}
hence
\begin{equation}\label{lN5a}
g_{N-}\left(\frac{\be}{2}\right)+g_{N+}\left(\frac{\be}{2}\right)
=2\int_{\al}^{\be}\rho(x)\ln\left|\frac{\be}{2}-x\right|\,dx
+O(N^{-2}).
\end{equation}
Thus,
\begin{equation}\label{lN6}
\begin{aligned}
l_N&=V_N\left(\frac{\be}{2}\right)-g_{N-}\left(\frac{\be}{2}\right)
-g_{N+}\left(\frac{\be}{2}\right)\\
&=V\left(\frac{\be}{2}\right)-g_{-}\left(\frac{\be}{2}\right)
-g_{+}\left(\frac{\be}{2}\right)+O(N^{-2})=l+O(N^{-2}),
\end{aligned}
\end{equation}
where by (\ref{V14}), $l=2\ln(\be-\al)-2-4\ln 2$.

\section {Riemann-Hilbert Problem} \label {RHP}

The Riemann-Hilbert (RH) problem for orthogonal polynomials with respect to
the weight $w(\mu)$ is the following:
\begin{enumerate}
\item [(i)] {(\it analyticity)} $Y(z)=\left( Y_{ij}(z)\right)_{i,j=1,2}$ is a matrix
valued analytic 
function on $\C\setminus \R$ which has limits on the real
line, $Y_{\pm}(\mu)$, so that for all $A>0$, 
\begin{equation}
\lim_{\ep\to 0^+} 
\max_{-A\le \mu\le A} |Y(\mu\pm i\ep)-Y_{\pm}(\mu)|=0.
\label {2.1}
\end{equation} 
\item
[(ii)] {\it (jump condition)}
\begin{equation}
Y_+(\mu)=Y_-(\mu)
\begin{pmatrix}
1 & w(\mu) \\
0 & 1
\end{pmatrix}.
\label {2.2}
\end{equation}
\item
[(iii)] {\it (asymptotics at infinity)}
\begin{equation}
Y(z)= \left[I+O\left(|z|^{-1}\right)\right]
\begin{pmatrix}
z^n & 0 \\
0 & z^{-n}
\end{pmatrix},\quad |z|\to\infty.
\label {2.3}
\end{equation}
\end{enumerate}
 
\begin{prop}\label {Y} The RH problem (i)-(iii) has a unique
solution given by 
\begin{equation}
Y(z)=\begin{pmatrix}
\pi_n(z) & \int_{\R} \frac {\pi_n(\mu)w(\mu)d\mu}{(\mu-z)2\pi i} \\
-\frac{2\pi i\pi_{n-1}(z)}{h_{n-1}}
& \int_{\R} \frac {-\pi_{n-1}(\mu)w(\mu)d\mu}{(\mu-z)h_{n-1}}
\end{pmatrix}
\label {2.4}
\end{equation}
where $\pi_n(\mu)=\mu^n+\dots$ denotes the $n$-th monic
orthogonal polynomial with respect to the measure $w(\mu)d\mu$
and $h_n=\int_{\R} \pi_n(\mu)^2w(\mu)d\mu$. Furthermore,
there exist $2\times 2$ matrices $Y_j,\; j=1,2,\dots$,
so that for all $m\ge 1$,
\begin{equation}
Y(z)
\begin{pmatrix}
z^{-n} & 0 \\
0 & z^n
\end{pmatrix}=I+\frac{Y_1}{z}+\frac{Y_2}{z^2}+\dots+\frac{Y_m}{z^m}
+O\left(|z|^{-m-1}\right),\quad
|z|\to\infty, 
\label {2.5}
\end{equation}
and
\begin{equation}
\begin{aligned}
&h_n=-2\pi i (Y_1)_{12}, \quad h_{n-1}=-\frac{2\pi i}{(Y_1)_{21}}\\
&R_n=(Y_1)_{21}(Y_1)_{12}, \\
&Q_n=\frac{(Y_2)_{21}}{(Y_1)_{21}}+(Y_1)_{11},
\end{aligned}
\label {2.6}
\end{equation}
where $Q_n,\;R_n$ are the recurrence coefficients associated 
to the orthogonal polynomials,
\begin{equation}
z\pi_n(z)=\pi_{n+1}(z)+Q_n\pi_n(z)+R_n\pi_{n-1}(z).
\label {2.7}
\end{equation}
\end{prop}

RH problem (i)-(iii) and Proposition 1 hold for a general weight $w(\mu)$ (see 
[2] for conditions on $w(\mu)$). In our case 
\begin{equation}
w(\mu)=e^{-NV_N(\mu)},
\label {2.8}
\end{equation}
and (\ref{2.6}) reads
\begin{equation}
\begin{aligned}
&h_{Nn}=-2\pi i (Y_1)_{12}, \quad h_{N,n-1}=-\frac{2\pi i}{(Y_1)_{21}}\\
&R_{Nn}=(Y_1)_{21}(Y_1)_{12}, \\
&Q_{Nn}=\frac{(Y_2)_{21}}{(Y_1)_{21}}+(Y_1)_{11},
\end{aligned}
\label {2.9}
\end{equation}

\section {Transformations of the RH Problem} \label {TRHP}

We will follow \cite{DKMVZ} to find the asymptotics of the solution
$Y(z)$ to the Riemann-Hilbert problem (i)-(iii) in the case when
$n=N$ and $N\to\infty$.

{\bf Transformation of the RH problem (\ref{2.1})-(\ref{2.3}).} Set
\begin{equation}
T(z)\equiv e^{-N\frac{l_N}{2} \sg_3}Y(z)e^{-N\left(g_N(z)-\frac{l_N}{2}\right) \sg_3},
\quad z\in \C\setminus\R;\qquad\sg_3=\begin{pmatrix} 1 & 0 \\ 0 & -1 \end{pmatrix}.
\label {t1}
\end{equation}
where
\begin{equation}
g_N(z)=\int_{\al_N}^{\be_N}\rho_N(\mu)\log(z-\mu)d\mu.
\label {t1a}
\end{equation}
Then  $T(z)$ solves
the following RH problem: 
\begin{enumerate}
\item
[(i)] {(\it analyticity)} $T(z)=\left( T_{ij}(z)\right)_{i,j=1,2}$ is a matrix
valued analytic 
function on $\C\setminus \R$ which has limits on the real
line, $T_{\pm}(\mu)$, so that for all $A>0$, 
\begin{equation}
\lim_{\ep\to 0^+} 
\max_{-A\le \mu\le A} |T(\mu\pm i\ep)-T_{\pm}(\mu)|=0.
\label {t2}
\end{equation} 
\item
[(ii)] {\it (jump condition)}
\begin{equation}
T_+(\mu)=T_-(\mu)J_T(\mu),
\label {t3}
\end{equation}
where
\begin{equation}
J_T(\mu)=\begin{pmatrix}
e^{-N(g_{N+}(\mu)-g_{N-}(\mu))} & 1 \\
0 & e^{N(g_{N+}(\mu)-g_{N-}(\mu))}
\end{pmatrix},\quad \mu\in[\al_N,\be_N],
\label {t3a}
\end{equation} 
and
\begin{equation}
J_T(\mu)=\begin{pmatrix}
1 & e^{N[g_{N+}(\mu)+g_{N-}(\mu)-V_N(\mu)-l_N]}  \\
0 & 1
\end{pmatrix},\quad \mu\in\R\setminus [\al_N,\be_N].
\label {t3b}
\end{equation} 
\item
[(iii)] {\it (asymptotics at infinity)}
\begin{equation}
T(z)= I+O\left(|z|^{-1}\right),\quad |z|\to\infty.
\label {t4}
\end{equation}
\end{enumerate}
The key point here is that the (21) element of the matrix $J_T(\mu)$ on 
$[\al_N,\be_N]$ is equal to 1, due to equation (\ref{g1}).
For convenience, let us rewrite the recurrent coefficients $Q_{NN}$, $R_{NN}$ 
in the new terms:
\begin{equation}
\begin{aligned}
&h_{NN}=-2\pi i e^{Nl_N}(T_1)_{12}, \quad 
h_{N,N-1}=-\frac{2\pi ie^{Nl_N}}{(T_1)_{21}}\\
&R_{NN}=(T_1)_{21}(T_1)_{12}, \\
&Q_{NN}=\frac{(T_2)_{21}}{(T_1)_{21}}+(T_1)_{11}.
\end{aligned}
\label {t5}
\end{equation}

{\bf Jump matrix factorization.} Denote for the sake of brevity
\begin{equation}
G_N(\mu)=g_{N+}(\mu)-g_{N-}(\mu).
\label {jm1}
\end{equation}
There is the following factorization
of the jump matrix $J_T$ on $[\al_N,\be_N]$:
\begin{equation}
\begin{pmatrix}
e^{-NG_N(\mu)} & 1 \\
0 & e^{NG_N(\mu)}
\end{pmatrix}
=\begin{pmatrix}
1 & 0 \\
e^{NG_N(\mu)} & 1 
\end{pmatrix}
\begin{pmatrix}
0 & 1 \\
-1 & 0
\end{pmatrix}
\begin{pmatrix}
1 & 0 \\
e^{-NG_N(\mu)} & 1 
\end{pmatrix}\equiv v_-v_0v_+.
\label {jm2}
\end{equation}
Substituting this factorization into (\ref{t3}) for $\mu\in[\al_N,\be_N]$,
we obtain that
\begin{equation}
T_+(\mu)=T_-(\mu)v_-(\mu)v_0v_+(\mu),
\label {jm3}
\end{equation}
or
\begin{equation}
\left[T_+(\mu)v_+^{-1}(\mu)\right]=\left[T_-(\mu)v_-(\mu)\right]v_0,
\quad \mu\in[\al_N,\be_N].
\label {jm4}
\end{equation}

{\bf Lenses.}
By using the factorization of jump matrix (\ref{jm2}) above,
we can transform the RH problem for $T$ in the following way. Consider the contours
$\Sigma_N^+$ and $\Sigma_N^-$ on the complex plane from $\al_N$
to $\be_N$, as shown on Figure 6. 

\begin{center}
 \begin{figure}[h]\label{figure6}
\begin{center}
   \scalebox{0.5}{\includegraphics{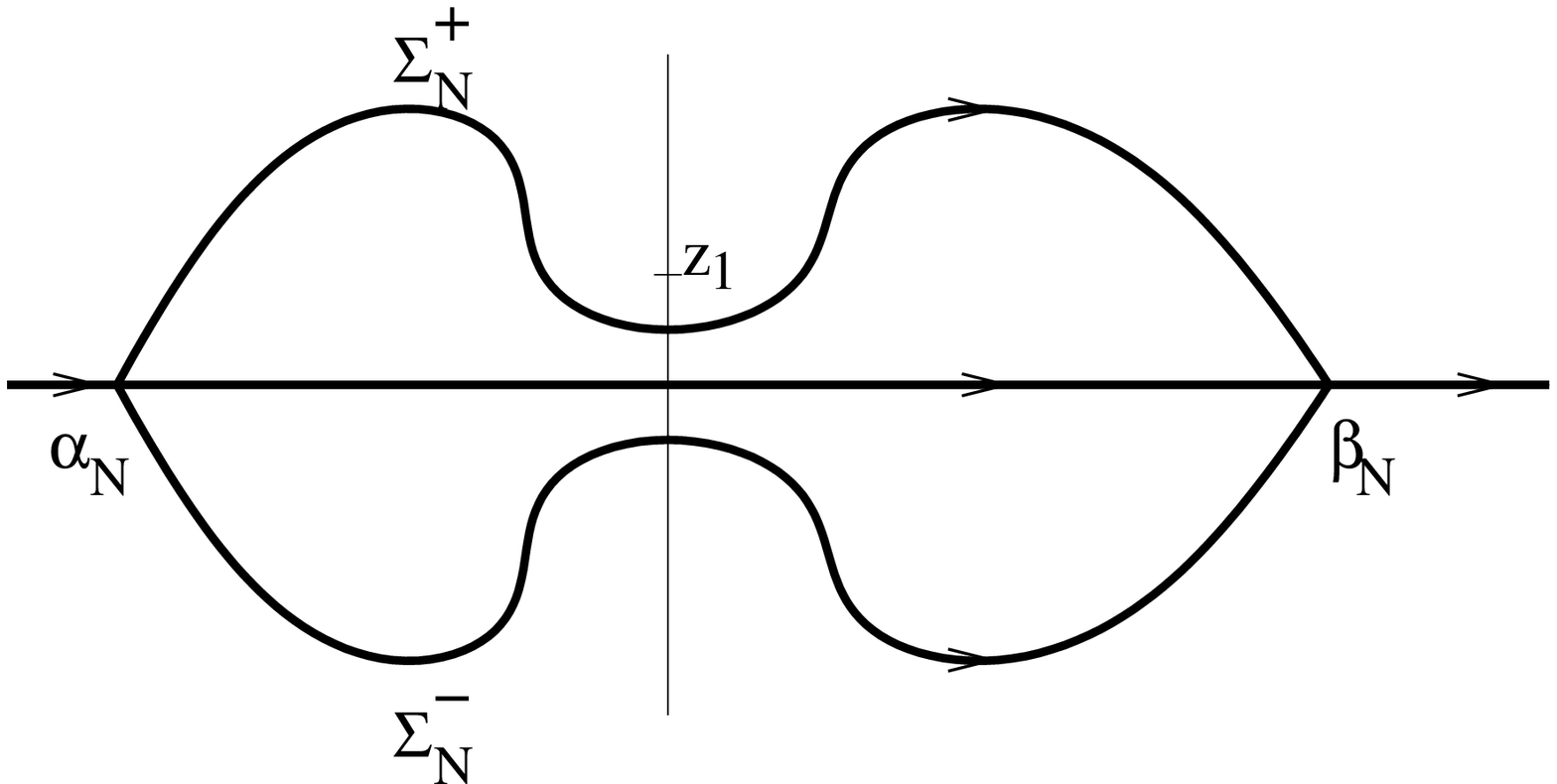}}
\end{center}
        \caption{The lenses.}
    \end{figure}
\end{center}

The contours $\Sigma_N^{\pm}$ go
closer and closer to the origin as $N\to\infty$. Namely, by (\ref{resc10}),
\begin{equation} \label{defcona}
e^{NV_N(z)}=e^{-N\zeta z}\frac
{\sinh Nz\frac{\pi}{2\ga}}{\sinh Nz(\frac{\pi}{2\ga}-1)},
\end{equation}
so that the function $e^{NV_N(z)}$ has poles on the imaginary axis.
Consider the first pole in the upper half-plane,
\begin{equation} \label{defconb}
z_1=\frac{iN^{-1}\pi}{\frac{\pi}{2\ga}-1}\,.
\end{equation}
The contour $\Sigma_N^+$ should be in the upper half-plane and
it should cross the imaginary axis below $z_1$, say, at $\frac{1}{2}z_1$. We take 
$\Sigma_N^-=\overline{\Sigma_N^+}$.
We call the region between $\Sigma_N^+$ (respectively,
$\Sigma_N^-$) and
$[\al_N,\be_N]$ the upper (respectively, lower) lens. Let 
\begin{equation}
S(z)=\left\{
\begin{aligned}
&T(z),\quad \mbox{outside of the lenses}, \\
&T(z)[v_+(z)]^{-1},\quad \mbox{in the upper lens}, \\
&T(z)v_-(z),\quad \mbox{in the lower lens}.
\end{aligned}
\right.
\label {defcon1}
\end{equation}
Then $S(z)$ solves the following RH problem:
\begin{enumerate}
\item [(i)] {(\it analyticity)} $S(z)$ is analytic on $\C\setminus (\R \cup \Sigma)$.
\item [(ii)] {\it (jump condition)}
\begin{equation}\label {defcon2}
S_+(z)=S_-(z)J_S(z),
\end{equation}
where 
\begin{equation}
J_S(z)=\left\{
\begin{aligned}
&\begin{pmatrix}
1 & e^{N[g_{N+}(z)+g_{N-}(z)-V_N(z)-l_N]} \\
0 & 1
\end{pmatrix},\quad z\in \R\setminus [\al_N,\be_N], \\
&\begin{pmatrix}
0 & 1 \\
-1 & 0
\end{pmatrix},\quad z\in [\al_N,\be_N], \\
&\begin{pmatrix}
1 & 0 \\
e^{-NG_N(z)} & 1
\end{pmatrix},\quad z\in \Sigma_N^+, \\
&\begin{pmatrix}
1 & 0 \\
e^{NG_N(z)} & 1
\end{pmatrix},\quad z\in \Sigma_N^-.
\end{aligned}
\right.
\label {defcon3}
\end{equation}
\item [(iii)] {\it (asymptotics at infinity)}
\begin{equation}\label {defcon4}
S(z)= I+O\left(|z|^{-1}\right),\quad |z|\to\infty.
\end{equation}
\end{enumerate}

{\bf Evaluation of the functions $e^{\pm NG_N(z)}$.} By (\ref{jm1}) and (\ref{g5}),
\begin{equation}\label {G_a}
G_N(\mu)=2\pi i\int_{\mu}^{\be_N}\rho_N(s)ds,\qquad \al_N\le \mu\le \be_N,
\end{equation}
hence, in particular, by (\ref{rh_5}),
\begin{equation}\label {G_b}
G_N(0)=2\pi i\int_0^{\be_N}\rho_N(s)ds
=\pi i(1+\z)+O(N^{-2}).
\end{equation}
Consider first $e^{-NG_N(z)}$
on $\Sigma_N^+$. From (\ref{g4}) we have that
\begin{equation}\label {G1}
g_{N-}(\mu)=V_N(\mu)+l_N-g_{N+}(\mu),\qquad \al_N\le\mu\le\be_N.
\end{equation}  
The RHS of this equation is extended to $\Im\mu>0$ and this gives us an analytic
continuation of $g_{N-}(\mu)$. By applying this continuation to (\ref{jm1}),
we obtain that for $z\in\Sigma_N^+$,
\begin{equation}\label {G2}
G_N(z)=2g_N(z)-V_N(z)-l_N,
\end{equation}
hence
\begin{equation}\label {G3}
e^{-NG_N(z)}=e^{-2Ng_N(z)+NV_N(z)+Nl_N}.
\end{equation}
By (\ref{resc10}),
\begin{equation}\label {G4}
e^{NV_N(z)}=\frac{\sinh Nz\frac{\pi}{2\ga}}
{\sinh Nz\left(\frac{\pi}{2\ga}-1\right)}e^{-N\z z}.
\end{equation}
In particular,
\begin{equation}\label {G5}
e^{NV_N(0)}=\frac{\pi}{\pi-2\ga}.
\end{equation}
Therefore,
\begin{equation}\label {G6}
\frac{e^{-NG_N(z)}}{e^{-NG_N(0)}}=\left(\frac{\pi-2\ga}{\pi}\right)
e^{-2N[g_N(z)-g_N(0)]+NV_N(z)},
\end{equation}
so that 
\begin{equation}\label {G7}
e^{-NG_N(z)}=C_N
e^{-2N[g_N(z)-g_N(0)]}
\left[\frac{\sinh Nz\frac{\pi}{2\ga}}
{\sinh Nz\left(\frac{\pi}{2\ga}-1\right)}e^{-N\z z}\right],
\end{equation}
where
\begin{equation}\label {G8}
C_N=\frac{\pi-2\ga}{\pi} e^{-NG_N(0)}
\end{equation}
By (\ref{G_b}),
\begin{equation}\label {G9}
C_N=\frac{\pi-2\ga}{\pi}e^{-N\pi i(1+\z)}(1+O(N^{-1})).
\end{equation}
Observe that if $\Im z>0$ then
\begin{equation}\label {G10}
g_N(z)-g_N(0)=\int_0^z \om_N(s)ds,
\end{equation}
where the integration is taken over the interval $[0,z]$.
From (\ref{om_15}) we obtain that 
\begin{equation}\label {G11}
g_N(z)-g_N(0)=-\frac{iz\ln N}{\pi}+\int_0^z b(s)ds-\frac{1}{2\pi^2}
\int_0^z m_0(Ns) ds +O(N^{-2}|z|).
\end{equation}
In particular, for $z=iN^{-1}y$, where $y>0$ is bounded, we obtain that
\begin{equation}\label {G12}
g_N(iN^{-1} y)-g_N(0)=\frac{yN^{-1}\ln N}{\pi }+ib(+i0)yN^{-1}-\frac{1}{2\pi^2}
M_0(iy)N^{-1}  +O(N^{-2}),
\end{equation}
where 
\begin{equation}\label {G13}
M_0(z)=\int_0^z m_0(s)ds.
\end{equation}
Thus, (\ref{G7}) gives that
\begin{equation}\label {G14}
e^{-NG_N(iN^{-1}y)}=e^{iN\om}k_N(y)
\frac{\sin y\frac{\pi}{2\ga}}
{\sin y\left(\frac{\pi}{2\ga}-1\right)}(1+O(N^{-1})),
\qquad y>0,
\end{equation}
where
\begin{equation}\label {G14a}
\om=-\pi(1+\z),
\end{equation}
and
\begin{equation}\label {G15}
k_N(y)=\frac{\pi-2\ga}{\pi}e^{\f(y)}N^{-\frac{2y}{\pi}},
\end{equation}
with
\begin{equation}\label {G15a}
\f(y)=-i2b(+i0)y+\frac{1}{\pi^2}M_0(iy)-i\z y.
\end{equation}
By using the value of $b(+i0)$ given in (\ref{om_19}),
we obtain that
\begin{equation}\label {G15b}
\f(y)=-iy-\frac{2y}{\pi}\ln\left(2\pi\cos\frac{\pi\z}{2}\right)+\frac{1}{\pi^2}M_0(iy).
\end{equation}
From (\ref{G13}) and (\ref{om_17}) we obtain that
\begin{equation}\label {G16}
\frac{1}{\pi^2}M_0(iy)=Q(iy)-Q(+i0),
\end{equation}
where 
\begin{equation}\label {G17}
Q(z):=\frac{i}{\pi} \left[\int_{-\infty}^\infty
\log(z-\mu) f(\mu)d\mu-2z\log z+2z\right].
\end{equation}
Since $f(\mu)$ is odd, we have that
\begin{equation}\label {G17a}
Q(iy)-Q(+i0)=\frac{2}{\pi } \left[\int_0^\infty
\arg (iy+\mu) f(\mu)d\mu
+y\ln y -y\right]+iy.
\end{equation}
Thus,
\begin{equation}\label {G17b}
\f(y)=-\frac{2y}{\pi }\ln\left(2\pi\cos\frac{\pi\z}{2}\right)
+\frac{2}{\pi } \left[\int_0^\infty
\arg (iy+\mu) f(\mu)d\mu
+y\ln y -y\right].
\end{equation}

Consider now $\Im z<0$. Similar to (\ref{G2}) we have that
$G_N(\mu)$ is analytically continued to $G_N(z)$ with $\Im z<0$ as
\begin{equation}\label {G18}
G_N(z)=-2g_N(z)+V_N(z)+l_N,\qquad \Im z<0.
\end{equation}
From (\ref{t1a}),
\begin{equation}\label {G19}
g_N(\bar z)=\overline{g_N(z)}.
\end{equation}
Also, $V_N(\bar z)=\overline{V_N(z)}$ and $l_N\in\R$, hence
\begin{equation}\label {G20}
G_N(\bar z)=-\overline{G_N(z)}.
\end{equation}
From (\ref{G14}) we obtain now that
\begin{equation}\label {G21}
e^{NG_N(-iN^{-1}y)}=e^{-N\overline{G_N(iN^{-1}y)}}=e^{-iN\om}k_N(y)
\frac{\sin y\frac{\pi}{2\ga}}
{\sin y\left(\frac{\pi}{2\ga}-1\right)}(1+O(N^{-1})),
\qquad y>0.
\end{equation}

{\bf Model RH problem.}
Note that the jump matrix $J_S(z)$ converges, as $N\to\infty$, to the 
identity matrix, except on the interval $[\al,\be] $ where it is 
constant. This leads to the following model RH problem.
\begin{enumerate}
\item[(i)] {$M(z)$ is analytic on $\C \setminus [\al,\be]$.}
\item[(ii)] {{\it (jump condition)}
\begin{equation}
M_{+}(z)=M_{-}(z)J_M, \quad z \in [\al_N,\be_N],
\end{equation}
where
\begin{equation}
J_M=\begin{pmatrix}
0 & 1 \\
-1 & 0
\end{pmatrix}.
\label {3.26}
\end{equation}}
\item[(iii)] {{\it (asymptotics at infinity)}
\begin{equation}
M(z)= I+O\left(|z|^{-1}\right),\quad |z|\to\infty.
\label {3.27}
\end{equation}}
\end{enumerate}

{\bf Solution to the model RH problem.}
The model RH problem can be solved explicitly. Namely, let us reduce it 
to a pair of scalar RH problems that are solved by
the Plemelj-Sohotski formula. By diagonalizing the matrix $J_M$, we have 
that
\begin{equation}
\begin{pmatrix}
0 & 1 \\
-1 & 0
\end{pmatrix}
=\frac {1}{2}
\begin{pmatrix}
1 & 1 \\
i & -i
\end{pmatrix}
\begin{pmatrix}
i & 0 \\
0 & -i
\end{pmatrix}\begin{pmatrix}
1 & -i \\
1 & i
\end{pmatrix}
\end{equation}
Let
\begin{equation}
\tilde M(z)=
\begin{pmatrix}
1 & -i \\
1 & i
\end{pmatrix}
M(z)
\begin{pmatrix}
1 & 1 \\
i & -i
\end{pmatrix}.
\end{equation}
Then, clearly
\begin{enumerate}
\item[(i)] {$\tilde M(z)$ is analytic on $\C \setminus [\al_N,\be_N] $.}
\item[(ii)] {
\begin{equation}
\tilde M_+(z)=\tilde M_-(z)
\begin{pmatrix}
i & 0 \\
0 & -i
\end{pmatrix}, \quad z \in [\al_N,\be_N].
\end{equation}}
\item[(iii)] {
\begin{equation}
\tilde M(z)= I+O\left(|z|^{-1}\right),\quad |z|\to\infty.
\end{equation}}
\end{enumerate}
 
Thus,
\begin{equation}
\begin{aligned}
\tilde M(z)&=
\begin{pmatrix}
e^{\frac{1}{2\pi i}\int_{\al_N}^{\be_N}\frac{\log i}{s-z}ds} & 0 \\
0 & e^{\frac{1}{2\pi i}\int_{\al_N}^{\be_N}\frac{\log (-i)}{s-z}ds}
\end{pmatrix} \\
&=\begin{pmatrix}
e^{\frac{1}{4}\log\frac{\be_N-z}{\al_N-z}} & 0 \\
0 & e^{-\frac{1}{4}\log\frac{\be_N-z}{\al_N-z}} 
\end{pmatrix} \\
&=\begin{pmatrix}
\ga_N^{-1} & 0 \\
0 & \ga_N
\end{pmatrix},
\end{aligned}
\end{equation}
where
\begin{equation}
\ga_N(z)=\left(\frac{z-\al_N}{z-\be_N}\right)^{1/4}
\label {3.28}
\end{equation}
with cut on $[\al_N,\be_N]$ and the branch such that $\ga_N(\infty)=1$.
Then
\begin{equation}
\begin{aligned}
M(z)&=
\begin{pmatrix}
1 & 1 \\
i & -i
\end{pmatrix}
\ga_N^{\sg_3}
\begin{pmatrix}
1 & 1 \\
i & -i
\end{pmatrix}^{-1} \\
&=\begin{pmatrix}
\frac{\ga_N(z)+\ga_N^{-1}(z)}{2} & \frac{\ga_N(z)-\ga_N^{-1}(z)}{(-2i)} \\
\frac{\ga_N(z)-\ga_N^{-1}(z)}{2i} & \frac{\ga_N(z)+\ga_N^{-1}(z)}{2}
\end{pmatrix},\quad \det M(z)=1.
\end{aligned}
\label {3.29}
\end{equation} 
At infinity we have that
\begin{equation}
M(z)=I+\frac{1}{z}
\begin{pmatrix}
0 & \frac{\be_N-\al_N}{-4i} \\
\frac{\be_N-\al_N}{4i}  & 0
\end{pmatrix}
+\frac{1}{z^2}
\begin{pmatrix}
\frac{(\be_N-\al_N)^2}{32} & \frac{\be_N^2-\al_N^2}{-8i} \\
\frac{\be_N^2-\al_N^2}{8i} & \frac{(\be_N-\al_N)^2}{32} 
\end{pmatrix}
+O(|z|^{-3}).
\label {3.30}
\end{equation}
At the origin,
\begin{equation}
\ga_N(+i0)=\left(\frac{-\al_N}{+i0-\be_N}\right)^{1/4}
=\sqrt{\tan\frac{\pi}{4}(1-\z)}\,e^{-\pi i/4}+O(N^{-2}),
\label {3.31}
\end{equation}
hence
\begin{equation}
M(+i0)=
\begin{pmatrix}
p+iq & p-iq \\
-p+iq & p+iq 
\end{pmatrix}+O(N^{-2}),
\label {3.32}
\end{equation}
where
\begin{equation}
p,q=\frac{\sqrt 2}{4}\left[\sqrt{\tan\frac\pi 4(1+\z)}\pm 
\sqrt{\tan\frac\pi 4(1-\z)}\right].
\label {3.33}
\end{equation}
We have the conjugation condition,
\begin{equation}
M(\bar z)=\sg_3 \overline{M(z)}\sg_3,
\label {3.34}
\end{equation}
hence
\begin{equation}
M(-i0)=
\begin{pmatrix}
p-iq & -p-iq \\
p+iq & p-iq 
\end{pmatrix}+O(N^{-2}).
\label {3.35}
\end{equation}

\section{ Parametrix at the edge points}\label{Edge_Points}

We consider small disks $D(\be_N,r)$,   $D(\al_N,r)$ of radius $r>0$, 
centered at the edge points,  and we look for a local parametrix {\it P} 
defined on  $D(\be_N,r)\cup D(\al_N,r)$ such that 
\begin{enumerate}
\item[(i)] {$ P(z)$ is analytic on $(D(\be_N,r)\cup D(\al_N,r)) \setminus (\R\cup\Sigma_N) $,
where $\Sigma_N=\Sigma_N^+\cup\Sigma_N^-$ is the boundary of the lenses, see Figure 6.}
\item[(ii)] {
$
 P_+(z)=P_-(z)J_S(z) , \quad z \in (D(\be_N,r)\cup D(\al_N,r)) \cap (\R\cup\Sigma).
$}
\item[(iii)] {
$
P(z)=\left( I+O\left(N^{-1}\right)\right)M(z),\quad z\in\partial D(\be_N,r)
\cup \partial D(\al_N,r),\quad N\to\infty.
$}
\end{enumerate}
We consider the right edge point $\be_N$ in detail.
Note that by (\ref{g7}),  we have that for $z\in D(\be_N,r)$,
\begin{equation}\label{ep1}
\begin{aligned}
-g_N(z)+\frac{V_N(z)}{2}+\frac{l_N}{2}&=\frac{1}{2}\int_{\be_N}^z
h_N(\mu)\sqrt{(\mu-\al_N)(\mu-\be_N)}\,d\mu\\
&=\frac{2}{3}a_N(z)(z-\be_N)^{3/2}, \quad z \in D(\be_N,r) \setminus [\al_N,\be_N],
\end{aligned}
\end{equation}
where $a_N(z)$ is an analytic function in $D(\be_N,r)$ such that
\begin{equation}\label{ep1a}
a_N(\be_N)=\frac{1}{2}\,h_N(\be_N)\sqrt{\be_N-\al_N}=\frac{2}{\be\sqrt{\be-\al}}
+O(N^{-2})\,>0.
\end{equation}
Define the analytic function,
\begin{equation}\label{ep1b}
\la_N(z)=\left[\frac{3}{2}\left(-g_N(z)+\frac{V_N(z)}{2}+\frac{l_N}{2}\right)\right]^{2/3}
=a_N(z)^{2/3}(z-\be_N),
\end{equation}
so that $\la'_N(\be_N)=a_N(\be_N)^{2/3}>0$,
and consider the conformal mapping, 
\begin{equation}\label{ep1c}
\la_N:\; D(\be_N,r)\to\C.
\end{equation}
We will assume that the contours $\Sigma_N^{\pm}$ are chosen in $D(\be_N,r)$ in
such a way that
\begin{equation}\label{ep1d}
\la_N:\; \Sigma_N^{\pm}\to\left\{ z:\;\arg z=\pm\frac{2\pi}{3}\right\}\,.
\end{equation}
Let us transform the RH problem on the matrix $S(z)$ in $D(\be_N,r)$. Let 
\begin{equation}\label{ep1e}
\Phi(z)=S(z)e^{N\left( g_N(z)-\frac{V_N(z)}{2}-\frac{l_N}{2}\right)\sigma_3}.
\end{equation}

\begin{lem}\label{J_Phi}
$\Phi(z)$ satisfies the jump condition 
\begin{equation}\label{ep2}
\Phi_+(z)=\Phi_-(z)J_{\Phi},
\end{equation}
where
\begin{equation}\label{ep2a}
J_{\Phi}=
\left\{
\begin{aligned}
{}&\begin{pmatrix}
1 & 1 \\
0 & 1 
\end{pmatrix},\quad \mbox{\rm for $\arg z = 0$},\\
{}&\begin{pmatrix}
1 & 0 \\
1 & 1 
\end{pmatrix}, \quad \mbox{\rm for $ z\in\Sigma_N^+$}, \\
{}&\begin{pmatrix}
0 & 1 \\
-1 & 0 
\end{pmatrix}, \quad \mbox{\rm for $ \arg z = \pi$}, \\
{}&\begin{pmatrix}
1 & 0 \\
1 & 1  
\end{pmatrix},\quad \mbox{\rm for $z\in\Sigma_N^-$}.
\end{aligned}\right.
\end{equation}
\end{lem}

We will use a model solution to (\ref{ep2}), which is constructed explicitly in
a standard way out of the Airy functions. The Airy function $\Ai(z)$
solves the equation $y'' = zy$ and for any $\varepsilon >0$, in the
sector $\pi + \varepsilon \leq \arg z \leq \pi - \varepsilon$, it has
the asymptotics as $z \to \infty$,
\begin{equation}\label{ep3}
\begin{aligned}
\Ai(z)&=\frac{1}{2\sqrt\pi }z^{-1/4}e^{-\frac{2}{3}z^{3/2}}
\left(1-\frac{5}{48}z^{-3/2}+\frac{385}{4608}z^{-3}+O(z^{-9/2})\right),\\
\Ai'(z)&=-\frac{1}{2\sqrt\pi }z^{1/4}e^{-\frac{2}{3}z^{3/2}}
\left(1+\frac{7}{48}z^{-3/2}-\frac{455}{4608}z^{-3}+O(z^{-9/2})\right).
\end{aligned}
\end{equation}
The functions $\Ai(\om z)$, $\Ai(\om^2 z)$, where
$\om=e^{\frac{2\pi i}{3}}$, also solve the equation $y''=zy$, and we
have the linear relation,
\begin{equation}\label{ep4}
\Ai(z)+\om\Ai(\om z)+\om^2\Ai(\om^2 z)=0.
\end{equation}
Write
\begin{equation}\label{ep5}
y_0(z)=\Ai(z), \quad y_1(z)=\om\Ai(\om z),
\quad y_2(z)=\om^2\Ai(\om^2 z),
\end{equation}
and we use these functions to define
\begin{equation}\label{ep6}
\Phi(z)=
\left\{
\begin{aligned}
{}&\begin{pmatrix}
y_0(z) & -y_2(z) \\
y_0'(z) & -y_2'(z) 
\end{pmatrix},\quad \mbox{for $0 < \arg z < 2\pi/3$},\\
{}&\begin{pmatrix}
-y_1(z) & - y_2(z) \\
-y_1'(z) & -y_2'(z) 
\end{pmatrix}, \quad \mbox{for $2\pi/3 < \arg z < \pi$}, \\
{}&\begin{pmatrix}
-y_2(z) & y_1(z) \\
-y_2'(z) & y_1'(z) 
\end{pmatrix}, \quad \mbox{for $-\pi < \arg z < -2\pi/3$}, \\
{}&\begin{pmatrix}
y_0(z) & y_1(z) \\
y_0'(z) & y_1'(z)  
\end{pmatrix},\quad \mbox{for $-2\pi/3 < \arg z < 0$}.
\end{aligned}\right.
\end{equation}
Then in the sector $0 < \arg z < 2\pi/3$,
\begin{equation}\label{ep6.1}
\Phi(z)=
\begin{pmatrix}
\Phi_{11}(z) & \Phi_{12}(z) \\
\Phi_{21}(z) & \Phi_{22}(z) 
\end{pmatrix},
\end{equation}
where
\begin{equation}\label{ep6.2}
\begin{aligned}
\Phi_{11}(z)&=\frac{1}{2\sqrt\pi}
z^{-1/4}e^{-\frac{2}{3}z^{3/2}}\left(1-\frac{5}{48}z^{-3/2}
+O(z^{-3})\right),\\
\Phi_{12}(z)&=\frac{1}{2\sqrt\pi}(-\om^2)
(\om^2z)^{-1/4}e^{-\frac{2}{3}(\om^2z)^{3/2}}\left(1-\frac{5}{48}(\om^2z)^{-3/2}
+O(z^{-3})\right),\\
\Phi_{21}(z)&=-\frac{1}{2\sqrt\pi}
z^{-1/4}e^{-\frac{2}{3}z^{3/2}}\left(1+\frac{7}{48}z^{-3/2}
+O(z^{-3})\right),\\
\Phi_{22}(z)&=\frac{1}{2\sqrt\pi}\om
(\om^2z)^{1/4}e^{-\frac{2}{3}(\om^2z)^{3/2}}\left(1+\frac{7}{48}(\om^2z)^{-3/2}
+O(z^{-3})\right).
\end{aligned}
\end{equation}
where for $z^{-1/4}$, $z^{1/4}$, and $z^{3/2}$ the principal branches are taken, with the cut on $(-\infty,0)$. Since $\om^2=e^{\frac{4\pi i}{3}}$ and $0<\arg z<\frac{2\pi}{3}$, we have that $\arg \om^2z=\arg z -\frac{2\pi}{3}$, hence $(\om^2z)^{1/4}=e^{-\frac{\pi i}{6}}z^{1/4}$, 
$(\om^2z)^{-1/4}=e^{\frac{\pi i}{6}}z^{-1/4}$, 
$(\om^2z)^{3/2}=-z^{3/2}$, and
$(\om^2z)^{-3/2}=-z^{-3/2}$.
Substituting these expressions into (\ref {ep6.1}), we obtain that 
\begin{equation}\label{ep6.3}
\begin{aligned}
\Phi(z) &=\frac{1}{2\sqrt\pi}z^{-\sigma_3/4}
\left[\begin{pmatrix}
1 & i \\
-1 & i
\end{pmatrix}
+\frac{1}{48}
\begin{pmatrix}
-5 & 5i \\
-7 & -7i
\end{pmatrix}
z^{-3/2}+O(z^{-3})\right]
e^{-\frac{2}{3}z^{3/2}\sigma_3}
\end{aligned}
\end{equation}
Note that $\Phi(z)$ satisfies the jump condition 
$\Phi_+(z)=\Phi_-(z)J_{\Phi}$. 
Define
\begin{equation}\label{ep7}
P(z)=E(z)N^{\frac{1}{6}\sigma_3}\Phi(N^{2/3}\la_N(z))
e^{N\left(-g_N(z)+\frac{V_N(z)}{2}+\frac{l_N}{2}\right)\sigma_3},
\end{equation}
where $E(z)$ is an analytic prefactor that has to be chosen 
to satisfy the matching condition 
$P(z)=\left( I+O\left(N^{-1}\right)\right)M(z)$ on the boundary 
of $D(\be_N,r)$. 
Then
\begin{equation}\label{ep9}
\begin{aligned}
E(z)&=\sqrt\pi M(z)
\begin{pmatrix}
1 & -1 \\
-i & -i  
\end{pmatrix}
(\la_N(z))^{\sigma_3/4}\\
&=\sqrt\pi
\begin{pmatrix}
1 & -1 \\
-i & -i  
\end{pmatrix}
\begin{pmatrix}
\ga_N(z)\la_N^{1/4}(z) & 0 \\
0& \ga_N^{-1}(z)\la_N^{-1/4}(z) 
\end{pmatrix}.
\end{aligned}
\end{equation}
Recall the definition of $\ga_N(z)=\left(\frac{z-\al_N}{z-\be_N}\right)^{1/4}$ and note that 
\[
\ga_N(z)\la_N^{1/4}(z)=(z-\al_N)^{1/4}(a_N(z))^{1/6}.
\]
  Therefore $E(z)$ is indeed an analytic function in $D(\be_N,r)$.

A similar construction works for a parametrix $P$
around the other edge point. Namely, by (\ref{g7a}),  we have that for $z\in D(\al_N,r)$,
\begin{equation}\label{ep1_1}
\begin{aligned}
-g_N(z)+\frac{V_N(z)}{2}+\frac{l_N}{2}&+\pi i\,\sign(\Im z)=\frac{1}{2}\int_z^{\al_N}
h_N(\mu)\sqrt{(\al_N-\mu)(\be_N-\mu)}\,d\mu\\
&=\frac{2}{3}a_N(z)(\al_N-z)^{3/2}, \quad z \in D(\al_N,r) \setminus [\al_N,\be_N],
\end{aligned}
\end{equation}
where $a_N(z)$ is an analytic function in $D(\al_N,r)$ such that
\begin{equation}\label{ep1a_1}
a_N(\al_N)=\frac{1}{2}\,h(\al_N)\sqrt{\be_N-\al_N}=\frac{2}{(-\al)\sqrt{\be-\al}}
+O(N^{-2})\,>0.
\end{equation}
Define the analytic function,
\begin{equation}\label{ep1b_1}
\la_N(z)=\left[\frac{3}{2}\left(-g_N(z)+\frac{V_N(z)}{2}+\frac{l_N}{2}
+\pi i\,\sign(\Im z)\right)\right]^{2/3}
=a_N(z)^{2/3}(\al_N-z),
\end{equation}
so that $\la'_N(\al_N)=-a_N(\al_N)^{2/3}<0$, and then define $P(z)$ by the formula,
\begin{equation}\label{ep7_1}
P(z)=\sg_3 E(z)N^{\frac{1}{6}\sigma_3}\Phi(N^{2/3}\la_N(z))
e^{N\left(-g_N(z)+\frac{V_N(z)}{2}+\frac{l_N}{2}\right)\sigma_3}\sg_3,
\end{equation}
where
\begin{equation}\label{ep9_1}
\begin{aligned}
E(z)&=\sqrt\pi \sg_3 M(z)\sg_3
\begin{pmatrix}
1 & -1 \\
-i & -i  
\end{pmatrix}
(\la_N(z))^{\sigma_3/4}\\
&=\sqrt\pi
\begin{pmatrix}
1 & -1 \\
-i & -i  
\end{pmatrix}
\begin{pmatrix}
\ga^{-1}_N(z)\la_N^{1/4}(z) & 0 \\
0& \ga_N(z)\la_N^{-1/4}(z) 
\end{pmatrix}.
\end{aligned}
\end{equation}
Observe that the function 
\[
\ga^{-1}_N(z)\la_N^{1/4}(z)=(\be_N-z)^{1/4}(a_N(z))^{1/6}.
\]
is analytic in $D(\al_N,r)$, hence $E(z)$ is analytic as well.

\section { Approximate solution to the RH problem} \label{ASRHP}

Define
\begin{equation}\label{ap1}
R(z)=
\left\{
\begin{aligned}
{}&S(z)P^{-1}(z),\quad \mbox{\rm if}\; z\in D(\al_N,r)\cup D(\be_N,r), \\
{}&S(z)M^{-1}(z), \quad \mbox{otherwise}.
\end{aligned}\right.
\end{equation}
Then, in $D(\al_N,r)\cup D(\be_N,r)$ we have that
\begin{equation}
\begin{aligned}
R_+(z)&=S_+(z) P_+^{-1}(z)=S_-(z) J_S(z) J_S^{-1}(z) P_-^{-1}(z)=S_-(z) P_-^{-1}(z)\\
&=R_-(z),
\end{aligned}
\end{equation}
on $\Sigma_N^+ \cup \Sigma_N^- \cup (\R \setminus [\al_N-r,\be_N+r])$,
\begin{equation}
\begin{aligned}
R_+(z)&=S_+(z) M^{-1}(z)=S_-(z) J_S(z) M^{-1}(z) =S_-(z) M^{-1}(z) M(z) J_S(z) M^{-1}(z) \\
&=R_-(z)M(z) J_S(z) M^{-1}(z),
\end{aligned}
\end{equation}
on $[\al_N+r,\be_N-r]$,
\begin{equation}
\begin{aligned}
R_+(z)&=S_+(z) M_+^{-1}(z)=S_-(z)
\begin{pmatrix}
0 & 1 \\
-1 & 0  
\end{pmatrix} 
\begin{pmatrix}
0 & 1 \\
-1 & 0  
\end{pmatrix} ^{-1}
M_-^{-1}(z)=S_-(z) M_-^{-1}(z)\\
&=R_-(z),
\end{aligned}
\end{equation}
and on $\partial D(\al_N,r)\cup \partial D(\be_N,r)$ the jump matrix is
\begin{equation}
J_R(z)=R_-^{-1}(z) R_+(z)
=P(z) S^{-1}(z) S(z) M^{-1}(z)
=P(z) M^{-1}(z).
\end{equation}
Introduce the contour $\Sigma_R$, which consists of the six arcs,
\begin{equation}\label{J_a}
\Sigma_R=(-\infty,\al_N-r)\cup\Sigma_R^{\al}\cup\Sigma_R^+
\cup\Sigma_R^-\cup\Sigma_R^{\be}\cup(\be_N+r,\infty),
\end{equation}
where 
\begin{equation}\label{J_b}
\Sigma_R^{\al}=\partial D(\al_N,r),\qquad
\Sigma_R^{\be}=\partial D(\be_N,r),\qquad
\Sigma_R^{\pm}=\Sigma^{\pm}_N\setminus [D(\al_N,r)\cup  D(\be_N,r)].
\end{equation}
see Fig. 7. The orientation of the arcs is shown on Fig. 7. 

\begin{center}
 \begin{figure}[h]\label{figure7}
\begin{center}
   \scalebox{0.5}{\includegraphics{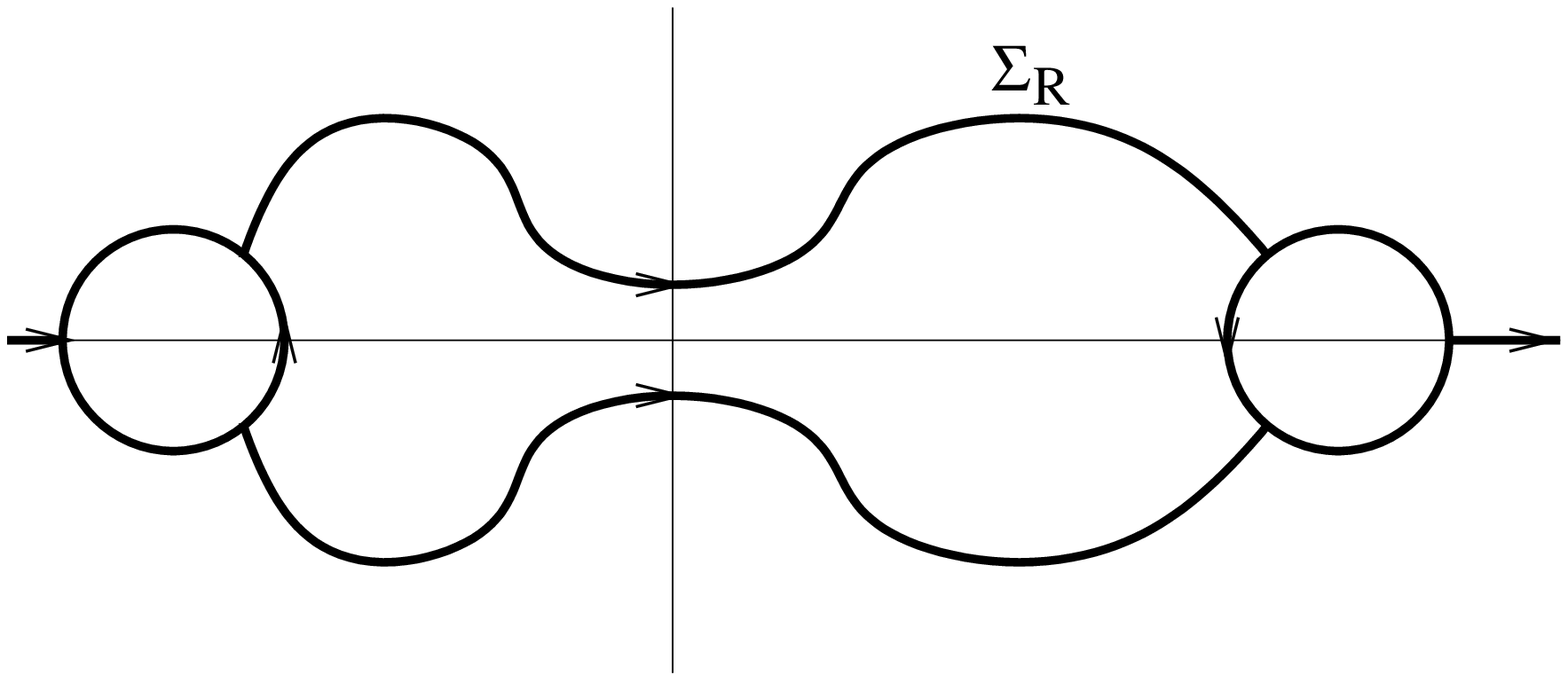}}
\end{center}
        \caption{The contour $\Sigma_R$.}
    \end{figure}
\end{center}

For the sake of brevity we will denote
\begin{equation}\label{J_c}
\Sigma_R^{\infty}=(-\infty,\al_N-r)\cup(\be_N+r,\infty).
\end{equation}
We have the following 

\begin{lem}\label{J_R}
$S(z)$ is a solution of the Riemann-Hilbert problem (\ref{defcon2}) - (\ref{defcon4}) if and only if 
$R(z)$ is a solution of the following RH problem:
\begin{enumerate}
\item[(i)] $ R(z)$ is analytic on $\C \setminus \Sigma_R$,
\item[(ii)] {
$R_+(z)=R_-(z)J_R(z),$ $z\in\Sigma_R$, where
\begin{equation}\label{J1}
J_R(z)=
\left\{
\begin{aligned}
{}&M(z)J_S(z)M^{-1}(z),\quad \mbox{on} \quad \Sigma_R \setminus 
\left(\partial D(\al_N,r)\cup \partial D(\be_N,r)\right) ,\\
{}&P(z)M^{-1}(z), \quad \mbox{ on}\quad \partial D(\al_N,r)\cup \partial D(\be_N,r).
\end{aligned}\right.
\end{equation}
}
\item[(iii)] {
$
R(z)= I+O\left(z^{-1}\right),\quad z\to\infty.
$}
\end{enumerate}
\end{lem}

We evaluate the jump matrix $J_R$ on different pieces of $\Sigma_R$.

{\bfseries Jump matrix $J_R(z)$ on $\Sigma_R^{\be}$.}
We have on $\partial D(\be_N,r)$ that
\begin{equation}\label{J2}
\begin{aligned}
J_R(z)&=P(z) M^{-1}(z)
=E(z)N^{\frac{1}{6}\sigma_3}\Phi(N^{2/3}\la_N(z))
e^{N(-g_N(z)+\frac{V_N(z)}{2}+\frac{l_N}{2})\sigma_3} M^{-1}(z)\\
&=\sqrt{\pi}M(z)
\begin{pmatrix}
1 & -1 \\
-i & -i
\end{pmatrix}
\la_N^{\frac{\sigma_3}{4}}(z) N^{\frac{1}{6}}
\frac{1}{2\sqrt{\pi}} N^{-\frac{1}{6}} \la_N^{-\frac{\sigma_3}{4}}(z)
\left[
\begin{pmatrix}
1 & i \\
-i & i
\end{pmatrix}+N^{-1}
\frac{1}{48}
\right.\\
&\left.
\times\begin{pmatrix}
-5 & 5i \\
-7 & -7i
\end{pmatrix}
\la_N^{-3/2}(z)+O\left(N^{-2}\right)\right]
e^{-\frac{2}{3}N\la_N^{3/2}(z)\sigma_3}
e^{N(-g_N(z)+\frac{V_N(z)}{2}+\frac{l_N}{2})\sigma_3} M^{-1}(z)\\
&=M(z)
\left[I
+\frac{1}{48}N^{-1}
\begin{pmatrix}
1 & 6i \\
6i & -1
\end{pmatrix}
\la_N^{-3/2}(z)+O\left(N^{-2}\right)
\right]M^{-1}(z)\\
&=I+N^{-1}J_R^1(z)+O\left(N^{-2}\right),
\end{aligned}
\end{equation}
where
\begin{equation}\label{J2a}
J_R^1(z)=
\frac{1}{96(z-\be)^2(z-\al)^{1/2}a(z)}
\begin{pmatrix}
-5(z-\al)+7(z-\be) & i[5(z-\al)+7(z-\be)] \\
i[5(z-\al)+7(z-\be)] & 5(z-\al)-7(z-\be)
\end{pmatrix},
\end{equation}
and
\begin{equation}\label{J3}
a(z)=\frac{3}{4(z-\be)^{3/2}}\int_\be^z  h(s)\sqrt{(s-\al)(s-\be)}\,ds,
\end{equation}
where $h(s)$ is defined in (\ref{V16}).

{\bf Jump matrix $J_R(z)$ on $\Sigma_R^{\al}$.}
Similarly, on $\partial D(\al_N,r)$,
\begin{equation}\label{J4}
\begin{aligned}
J_R(z)&=M(z)
\left[I
+\frac{1}{48}N^{-1}
\begin{pmatrix}
1 & -6i \\
-6i & -1
\end{pmatrix}
\la_N^{-3/2}(z)+O\left(N^{-2}\right)
\right]M^{-1}(z)\\
&=I+N^{-1}J_R^1(z)+O\left(N^{-2}\right),
\end{aligned}
\end{equation}
where
\begin{equation}\label{J5}
J_R^1(z)=
\frac{1}{96(\al-z)^2(\be-z)^{1/2}a(z)}
\begin{pmatrix}
7(\al-z)-5(\be-z) & i[-7(\al-z)-5(\be-z)] \\
i[-7(\al-z)-5(\be-z)] & -7(\al-z)+5(\be-z)
\end{pmatrix},
\end{equation}
and
\begin{equation}\label{J6}
a(z)=\frac{3}{4(\al-z)^{3/2}}\int_z^\al  h(s)\sqrt{(\al-s)(\be-s)}\,ds,
\end{equation}
where $h(s)$ is defined in (\ref{V16a}).

{\bf Jump matrix $J_R(z)$ on $\Sigma_R^{\pm}$.} By (\ref{J1}) and 
(\ref{defcon3}), on $\Sigma_R^+$,
\begin{equation}\label{J7a}
J_R(z)=I+J_R^{\circ}(z),
\end{equation}
where
\begin{equation}\label{J8}
J_R^{\circ}(z)=e^{-NG_N(z)}M(z)\sg_+ M(z)^{-1},\qquad z\in\Sigma_R^+,
\end{equation}
and
\begin{equation}\label{J8a}
\sg_+=
\begin{pmatrix}
0 & 0 \\
1 & 0
\end{pmatrix},\qquad
\sg_-=
\begin{pmatrix}
0 & 1 \\
0 & 0
\end{pmatrix}.
\end{equation}
From (\ref{G6}), (\ref{G11}), we obtain that there exist
constants $C,\ga,c>0$ such that
\begin{equation}\label{J9}
\|J_R^{\circ}(z)\|\le C N^{-\ga}e^{-cN|\Im z|},
\end{equation}
where $\|J_R^{\circ}(z)\|$ is the sum of absolute values of the matrix
elements of $J_R^{\circ}(z)$.
On $\Sigma_R^-$ we also have equation (\ref{J7a}) with
 estimate (\ref{J9}).

{\bf Jump matrix $J_R(z)$ on $\Sigma_R^{\infty}$.}
By (\ref{defcon3}), 
$J_R(x)=I+J_R^{\circ}(x)$, where
\begin{equation}\label{J10}
J_R^{\circ}(z)=e^{N[g_{N+}(z)+g_{N-}(z)-V_N(z)-l_N]}M(z)\sg_-
M(z)^{-1},\qquad z\in\Sigma_R^{\infty}.
\end{equation}
In this case, there exist $C,c>0$ such that
\begin{equation}\label{J11}
\|J_R^{\circ}(z)\|\le Ce^{-cN|z|}.
\end{equation}

{\bf Solution of the RH problem for $R$ by perturbation theory.}
The estimates above show that $J_R^{\circ}(z)\to 0$ as $N\to \infty$.
We can apply the following general result.

\begin{prop}  \label{R}
Assume that $v(\la)$, $\la\in\Sigma_R$, solves the
equation
\begin{equation}
v(\la)=I-\frac{1}{2\pi i}\int_{\Sigma_R} 
\frac {v(\mu)J_R^{\circ}(\mu)}{\la_- -\mu}\,d\mu,\quad \la\in\Sigma_R,
\label {J12}
\end{equation}
where $\la_-$ means $\la-i0$, the value of the limit from the minus side,
and $J_R=I+J_R^{\circ}$.
Then
\begin{equation}
R(z)=I-\frac{1}{2\pi i}\int_{\Sigma_R} 
\frac {v(\mu)J_R^{\circ}(\mu)}{z-\mu}\,d\mu,\quad z\in\C\setminus\Sigma_R,
\label {J13}
\end{equation}
solves the following RH problem:
\begin{enumerate}
\item[(i)] $ R(z)$ is analytic on $\C \setminus \Sigma_R$,
\item[(ii)] {$R_+(\la)=R_-(\la)J_R(\la),$ $\la\in\Sigma_R$,}
\item[(iii)] {$R(z)= I+O\left(z^{-1}\right),\quad z\to\infty.$}
\end{enumerate}
\end{prop}

{\it Proof.} From (\ref{J12}), (\ref{J13}),
\begin{equation}
R_-(\la)=v(\la),\quad \la\in\Sigma_R.
\label {J14}
\end{equation}
By the jump property of the Cauchy transform,
\begin{equation}
R_+(\la)-R_-(\la)=v(\la)J_R^{\circ}(\la)=R_-(\la)J_R^{\circ}(\la),
\label {J15}
\end{equation}
hence $R_+(\la)=R_-(\la)J_R(\la)$. From (\ref{J13}), $R(z)=I+O(z^{-1})$.
Proposition \ref{R} is proved.

Equation (\ref{J12}) can be solved by perturbation theory, so that
\begin{equation}
v(\la)=I+\sum_{k=1}^\infty v_k(\la),
\label {J16}
\end{equation}
where for $k\ge 1$,
\begin{equation}
v_k(\la)=-\frac{1}{2\pi i}\int_{\Sigma_R} 
\frac {v_{k-1}(\mu)J_R^{\circ}(\mu)}{\la_- -\mu}\,d\mu,\quad \la\in\Sigma_R,
\label {J17}
\end{equation}
and $v_0(\la)=I$. Series (\ref{J16}) is estimated from above 
by a convergent geometric series, so it is absolutely convergent.
Observe that
\begin{equation}
v_1(\la)=-\frac{1}{2\pi i}\int_{\Sigma_R} 
\frac {J_R^{\circ}(\mu)}{\la_- -\mu}\,d\mu,\quad \la\in\Sigma_R,
\label {J17a}
\end{equation} 
The function $R(z)$ is given then as
\begin{equation}
R(z)=I+\sum_{k=1}^\infty R_k(z),
\label {J18}
\end{equation}
where
\begin{equation}
R_k(z)=-\frac{1}{2\pi i}\int_{\Sigma_R} 
\frac {v_{k-1}(\mu)J_R^{\circ}(\mu)}{z -\mu}\,d\mu.
\label {J19}
\end{equation}
In particular,
\begin{equation}
R_1(z)=-\frac{1}{2\pi i}\int_{\Sigma_R} 
\frac {J_R^{\circ}(\mu)}{z -\mu}\,d\mu.
\label {J20}
\end{equation}

\section{ Large $N$ asymptotics of the recurrent coefficients}\label{asympt}

From (\ref{2.9}), (\ref{t1}) and (\ref{defcon1}),
we obtain the formulae for the recurrent coefficients:
\begin{equation}\label{as1}
\begin{aligned}
&h_{NN}=-2\pi i e^{Nl_N}(S_1)_{12}, \\
&R_{NN}=(S_1)_{21}(S_1)_{12}, \\
&Q_{NN}=\frac{(S_2)_{21}}{(S_2)_{21}}+(S_1)_{11}.
\end{aligned}
\end{equation}
where
\begin{equation}\label{as2}
S(z)=I+ \frac{S_1}{z}+ \frac{S_2}{z^2}+O(z^{-3})\,,
\qquad z\to\infty.
\end{equation}
By (\ref{ap1}), $S(z)=R(z)M(z)$ for large $z$, hence
\begin{equation}\label{as3a}
S_1=M_1+R_1,
\end{equation}
where
\begin{equation}\label{as4}
M(z)=I+ \frac{M_1}{z}+O(z^{-2}),\qquad
R(z)=I+ \frac{R_1}{z}+O(z^{-2}).
\end{equation}
Therefore,
\begin{equation}\label{as3}
R_{NN}=(M_1+R_1)_{21}(M_1+R_1)_{12}.
\end{equation}
By (\ref{3.30}),
\begin{equation}\label{as5}
M_1=
\begin{pmatrix}
0 & \frac{\be_N-\al_N}{-4i} \\
\frac{\be_N-\al_N}{4i}  & 0
\end{pmatrix},
\end{equation}
hence
\begin{equation}\label{as5a}
R_{NN}=\left(\frac{\be_N-\al_N}{4}\right)^2+\frac{\be_N-\al_N}{4i}
[(R_1)_{12}-(R_1)_{21}]+(R_1)_{12}(R_1)_{21}.
\end{equation}
By (\ref{J13}),
\begin{equation}\label{as6}
\begin{aligned}
R_1&=-\frac{1}{2\pi i}\int_{\Sigma_R}v(\la)J_R^{\circ}(\la)d\la\\
   &=-\frac{1}{2\pi i}\int_{\Sigma_R}J_R^{\circ}(\la)d\la
    -\frac{1}{2\pi i}\int_{\Sigma_R}v_1(\la)J_R^{\circ}(\la)d\la
    -\dots
\end{aligned}
\end{equation}
We will call the first term on the right,
\begin{equation}\label{as6a}
R_1^{(1)}=-\frac{1}{2\pi i}\int_{\Sigma_R}J_R^{\circ}(\la)d\la,
\end{equation}
a linear term, the second one,
\begin{equation}\label{as6b}
R_1^{(2)}=-\frac{1}{2\pi i}\int_{\Sigma_R}v_1(\la)J_R^{\circ}(\la)d\la,
\end{equation}
a quadratic term, etc. By definition, we have that
\begin{equation}\label{as6c}
R_1=R_1^{(1)}+R_1^{(2)}+\dots
\end{equation}
First we evaluate the linear term.

{\bfseries Evaluation of the linear term.} Denote
\begin{equation}\label{as7}
\begin{aligned}
R_1^a&=-\frac{1}{2\pi i}\int_{\Sigma_R^a}J_R^{\circ}(\la)d\la,\qquad
a=\al,\be,+,-,\infty,
\end{aligned}
\end{equation}
so that
\begin{equation}\label{as7a}
R_1^{(1)}=R_1^{\al}+R_1^{\be}+R_1^+ +R_1^- +R_1^{\infty}.
\end{equation}
Let us evaluate $R_1^{\al}$, $R_1^{\be}$, $R_1^{\pm}$, and $R_1^{\infty}$.

{\bfseries Evaluation of $R_1^{\al}$, $R_1^{\be}$.} By (\ref{J2}),
\begin{equation}\label{as8}
\begin{aligned}
R_1^{\be}&=-\frac{N^{-1}}{2\pi i}\oint_{\partial D(\be_N,r)}
J_R^1(z)\,dz+O(N^{-2}),
\end{aligned}
\end{equation}
which can be evaluated by taking the residue at $z=\be$.
The result is
\begin{equation}\label{as9}
R_1^{\be}=-N^{-1}\frac{1}{192}
\begin{pmatrix}
3\be+\al & i(11\be-\al) \\
i(11\be-\al) & -3\be-\al
\end{pmatrix}
+O(N^{-2}).
\end{equation}
A similar expression holds
for $R_1^\al$. Namely,
\begin{equation}\label{as9a}
R_1^{\al}=-N^{-1}\frac{1}{192}
\begin{pmatrix}
-3\al-\be & i(11\al-\be) \\
i(11\al-\be) & 3\al+\be
\end{pmatrix}
+O(N^{-2}).
\end{equation}
By taking into account terms of the order of $N^{-2}$ in (\ref{J2}), we obtain
the error terms in (\ref{as9}), (\ref{as9a}) as $N^{-2}c_{\al,\be}+O(N^{-3})$,
where $c_{\al,\be}$ are some explicit matrices.

{\bf Evaluation of $R_1^{\pm}$.} In the usual case of a random matrix model
with an analytic potential $V(M)$ independent of $N$, the terms $R_1^{\pm}$,
which represent the integral over the lenses boundary, are exponentially
small as $N\to\infty$, see \cite{DKMVZ}. 
It is not the case in our situation because of the series of
poles of the function $e^{-NG_N(z)}$ on the imaginary axis. By (\ref{J8}),
\begin{equation}\label{as10}
R^+_1=-\frac{1}{2\pi i} \int_{\Sigma_R^+}J_R^{\circ}(\la)d\la
=-\frac{1}{2\pi i} \int_{\Sigma_R^+} e^{-NG_N(\la)}  M(\la)\sg_+ M(\la)^{-1}d\la.
\end{equation}
From (\ref{G7}) we obtain that
the function $e^{-NG_N(z)}$ has simple poles at the points
\begin{equation}\label{as11}
z=z_j=iN^{-1}y_j,\qquad y_j=\frac{j\pi}{\frac{\pi}{2\ga}-1}\,,
\qquad j=1,2,\dots,
\end{equation}
and by (\ref{G14}), the residue at $z_j$ is equal to
\begin{equation}\label{as12}
\underset{z=z_j}{\Res}\left[e^{-NG_N(z)}\right] 
=e^{iN\om}k_N(y_j)\frac{i(-1)^j\sin y_j\frac{\pi}{2\ga}}
{N\left(\frac{\pi}{2\ga}-1\right)}(1+O(N^{-1})),
\end{equation}
By using (\ref{G15}) we reduce this to
\begin{equation}\label{as13}
\underset{z=z_j}{\Res}\left[e^{-NG_N(z)}\right] 
=iC_je^{iN\om}N^{-\kappa_j}
(1+O(N^{-1})),
\end{equation}
where
\begin{equation}\label{as14}
\kappa_j=1+\frac{2j}{\frac{\pi}{2\ga}-1}\,,
\qquad
C_j=\frac{2\ga}{\pi}e^{\f(y_j)}(-1)^j\sin\left(\frac{\pi j}{1-\frac{2\ga}{\pi}}\right)\,.
\end{equation}
Observe that $\kappa_j>1$. From (\ref{G21}) we obtain that
\begin{equation}\label{as14a}
\underset{z=-z_j}{\Res}\left[e^{NG_N(z)}\right] 
=-iC_je^{-iN\om}N^{-\kappa_j}
(1+O(N^{-1})),
\end{equation}

Let us deform the contour $\Sigma_R^+$ up, crossing the poles.
Every time we cross a pole, the residue at the pole
appears on the right of (\ref{as10}), while the 
integral becomes  smaller  than the contribution from the pole.
This gives the asymptotic expansion as $N\to\infty$,
\begin{equation}\label{as15a}
R^+_1
\sim -\sum_{j=1}^{\infty} \underset{z=z_j}{\Res}\left[e^{-NG_N(z)}\right] M(z_j)
\sg_+ M(z_j)^{-1},
\end{equation}
where the $j$-th term is of the order of $N^{-\kappa_j}$. 
For our purposes it will be sufficient to consider 
terms with $\kappa_j\le 2$ only,
\begin{equation}\label{as15}
R^+_1
=-\sum_{j:\;\kappa_j\le 2} \underset{z=z_j}{\Res}\left[e^{-NG_N(z)}\right] M(z_j)
\sg_+ M(z_j)^{-1}+O(N^{-2-\ep}),
\end{equation}
where
\begin{equation}\label{as16}
j_0=\left[\frac{1}{2}\left(\frac{\pi}{2\ga}-1\right)\right]\,.
\end{equation}
In fact, since $z_j=O(N^{-1})$, we can replace $M(z_j)$ by $M(+i0)$,
\begin{equation}\label{as17}
R^+_1
=-\sum_{j:\;\kappa_j\le 2} \underset{z=z_j}{\Res}\left[e^{-NG_N(z)}\right] M(+i0)
\sg_+ M(+i0)^{-1}+O(N^{-2-\ep}).
\end{equation}
Let us rewrite this in terms of the matrix elements,
\begin{equation}\label{as17a}
\begin{aligned}
(R^+_1)_{12}&={M^+_{12}}^2J^+_{21}+O(N^{-2-\ep}),\qquad
(R^+_1)_{21}=-{M^+_{22}}^2J^+_{21}+O(N^{-2-\ep}),\\
(R^+_1)_{11}&=-M^+_{12}M_{22}J^+_{21}+O(N^{-2-\ep}),\qquad
(R^+_1)_{22}=M^+_{12}M_{22}J^+_{21}+O(N^{-2-\ep}),
\end{aligned}
\end{equation}
where 
\begin{equation}\label{as18}
J^+_{21}=\sum_{j:\;\kappa_j\le 2}\underset{z=z_j}{\Res}\left[e^{-NG_N(z)}\right]
=ie^{iN\om}\sum_{j:\;\kappa_j\le 2}C_jN^{-\kappa_j}+O(N^{-2-\ep}),
\end{equation}
and $M^+_{ij}$ are the matrix elements of the matrix $M(+i0)$.
By applying (\ref{3.32}), we obtain that
\begin{equation}\label{as19}
(R^+_1)_{12}=(p-iq)^2J^+_{21}+O(N^{-2-\ep}),\qquad
(R^+_1)_{21}=-(p+iq)^2J^+_{21}+O(N^{-2-\ep}),
\end{equation}
Similarly, we evaluate the contributions from the 
contour $\Sigma_R^-$ as 
\begin{equation}\label{as20}
(R^-_1)_{12}=(p+iq)^2J^-_{21}+O(N^{-2-\ep}),\qquad
(R^-_1)_{21}=-(p-iq)^2J^-_{21}+O(N^{-2-\ep}),
\end{equation}
where 
\begin{equation}\label{as21}
J^-_{21}=-\sum_{j:\;\kappa_j\le 2}\underset{z=-z_j}{\Res}\left[e^{NG_N(z)}\right]
=ie^{-iN\om}\sum_{j:\;\kappa_j\le 2}C_jN^{-\kappa_j}+O(N^{-2-\ep}).
\end{equation}
By combining (\ref{as19}) and (\ref{as20}), we obtain that
\begin{equation}\label{as22}
\begin{aligned}
(R^+_1)_{12}+(R^-_1)_{12}&=i2\left[(p^2-q^2)
\cos(N\om)+2pq\sin(N\om)\right]\sum_{j:\;\kappa_j\le 2}C_jN^{-\kappa_j}+O(N^{-2-\ep}),\\
(R^+_1)_{21}+(R^-_1)_{21}&=-i2\left[(p^2-q^2)
\cos(N\om)-2pq\sin(N\om)\right]\sum_{j:\;\kappa_j\le 2}C_jN^{-\kappa_j}+O(N^{-2-\ep})\,.
\end{aligned}
\end{equation}
From (\ref{3.33}) we find that
\begin{equation}\label{as23}
p^2-q^2=\frac{1}{2}\,,\qquad
2pq=\frac{1}{2}\tan\frac{\pi\z}{2}\,,
\end{equation}
hence
\begin{equation}\label{as22a}
\begin{aligned}
(R^+_1)_{12}+(R^-_1)_{12}&=i\left[
\cos(N\om)+\tan\left(\frac{\pi\z}{2}\right)\sin(N\om)\right]
\sum_{j:\;\kappa_j\le 2}C_jN^{-\kappa_j}+O(N^{-2-\ep}),\\
(R^+_1)_{21}+(R^-_1)_{21}&=-i\left[
\cos(N\om)-\tan\left(\frac{\pi\z}{2}\right)\sin(N\om)\right]
\sum_{j:\;\kappa_j\le 2}C_jN^{-\kappa_j}+O(N^{-2-\ep})\,.
\end{aligned}
\end{equation}

{\bfseries Evaluation of $R_1^{\infty}$.} From (\ref{J11}) we obtain that
$R_1^{\infty}$ is exponentially small as $N\to\infty$,
\begin{equation}\label{as23b}
R_1^{\infty}=O(e^{-c_0N}).
\end{equation}

{\bfseries Summary for the linear term.}
The evaluation of the linear term can be summarized as follows:
\begin{equation}\label{as23a}
\begin{aligned}
(R_1^{(1)})_{12}&=-N^{-1}\frac{5i(\be-\al)}{96}
+i\left[
\cos(N\om)+\tan\left(\frac{\pi\z}{2}\right)\sin(N\om)\right]
\sum_{j:\;\kappa_j\le 2}C_jN^{-\kappa_j}\\
&+c^{(1)}_{12}N^{-2}+O(N^{-2-\ep}),\\
(R_1^{(1)})_{21}&=-N^{-1}\frac{5i(\be-\al)}{96}
-i\left[
\cos(N\om)-\tan\left(\frac{\pi\z}{2}\right)\sin(N\om)\right]
\sum_{j:\;\kappa_j\le 2}C_jN^{-\kappa_j}\\
&+c^{(1)}_{21}N^{-2}+O(N^{-2-\ep}),
\end{aligned}
\end{equation}
where $c^{(1)}_{12}$, $c^{(1)}_{21}$ are some constants.

{\bfseries Evaluation of the quadratic term.} We obtain from (\ref{J17a}),
(\ref{as6b}), that the quadratic term is equal to
\begin{equation}\label {qua1}
R_1^{(2)}=-\frac{1}{(2\pi)^2}\int_{\Sigma_R}
\int_{\Sigma_R} 
\frac {J_R^{\circ}(\mu)J_R^{\circ}(\la)}{\la_- -\mu}\,d\mu
d\la.
\end{equation} 
We can split it as
\begin{equation}\label {qua2}
R_1^{(2)}=\sum_{a,b\in A}R_1^{a,b},\qquad A=\{\al,\be,+,-,\infty\},
\end{equation}
where
\begin{equation}\label {qua3}
R_1^{a,b}=-\frac{1}{(2\pi)^2}\int_{\Sigma_R^a}
\int_{\Sigma_R^b} 
\frac {J_R^{\circ}(\mu)J_R^{\circ}(\la)}{\la_- -\mu}\,d\mu
d\la.
\end{equation} 
If $a\not= b$ then we can replace $\la_-$ by $\la$ and in this case
we obtain that 
\begin{equation}\label {qua4}
R_1^{a,b}=-\frac{1}{(2\pi)^2}\int_{\Sigma_R^a}
\int_{\Sigma_R^b} 
\frac {J_R^{\circ}(\mu)J_R^{\circ}(\la)}{\la -\mu}\,d\mu
d\la,\qquad a\not= b.
\end{equation}
It is tempting to say that $R_1^{b,a}=-R_1^{a,b}$, but in general
it is not true, because the matrices $J_R^{\circ}(\la)$ and 
$J_R^{\circ}(\mu)$ do not commute.
By (\ref{J10}) $J_R^{\circ}(z)$ is analytic
on $\Sigma_R^{\infty}$ and by (\ref{J11}) it is exponentially small
in $N|z|$, hence $R_1^{a,b}$ is exponentially small in $N$,
if at least one of $a,b$ is equal to $\infty$,
\begin{equation}\label {qua6}
|R_1^{a,b}|\le C_0e^{-c_0N},\qquad C_0,c_0>0; 
\qquad a=\infty \quad\textrm{or}\quad b=\infty.
\end{equation}
From (\ref{J2}) we obtain that
\begin{equation}\label {qua6a}
\frac{1}{2\pi i}\int_{\Sigma_R^{\be}}
\frac {J_R^{\circ}(\mu)}{\la_- -\mu}\,d\mu
=N^{-1}\underset{\mu=\be}{\Res}\left[J_R^1(\mu)\right]
\frac{1}{\la-\be}+O(N^{-2})\,,
\end{equation} 
hence
\begin{equation}\label {qua7}
R_1^{\be,\be}
=N^{-2}\underset{\mu=\be}{\Res}\left[J_R^1(\mu)\right]
\underset{\la=\be}{\Res}\left[\frac{J_R^1(\la)}{\la-\be}\right]
+O(N^{-3})\,.
\end{equation} 
Similarly,
\begin{equation}\label {qua8}
R_1^{\al,\al}
=N^{-2}\underset{\mu=\al}{\Res}\left[J_R^1(\mu)\right]
\underset{\la=\al}{\Res}\left[\frac{J_R^1(\la)}{\la-\al}\right]
+O(N^{-3})\,.
\end{equation}
The cross terms are evaluated as
\begin{equation}\label {qua8a}
\begin{aligned}
R_1^{\al,\be}
&=-N^{-2}\frac{1}{\be-\al}\underset{\mu=\be}{\Res}\left[J_R^1(\mu)\right]
\underset{\la=\al}{\Res}\left[J_R^1(\la)\right]
+O(N^{-3})\,,\\
R_1^{\be,\al}
&=N^{-2}\frac{1}{\be-\al}\underset{\mu=\al}{\Res}\left[J_R^1(\mu)\right]
\underset{\la=\be}{\Res}\left[J_R^1(\la)\right]
+O(N^{-3})\,.
\end{aligned}
\end{equation}
 Thus,
\begin{equation}\label {qua9}
R_1^{\al,\al}+R_1^{\be,\be}+R_1^{\al,\be}+R_1^{\be,\al}
=c_1N^{-2}+O(N^{-3}),
\end{equation} 
where
\begin{equation}\label {qua9a}
\begin{aligned}
c_1&=\underset{\la=\al}{\Res}\left[J_R^1(\la)\right]
\underset{\la=\al}{\Res}\left[\frac{J_R^1(\la)}{\la-\al}\right]
+\underset{\la=\be}{\Res}\left[J_R^1(\la)\right]
\underset{\la=\be}{\Res}\left[\frac{J_R^1(\la)}{\la-\be}\right]\\
&+\frac{1}{\be-\al}\underset{\la=\al}{\Res}\left[J_R^1(\la)\right]
\underset{\la=\be}{\Res}\left[J_R^1(\la)\right]
-\frac{1}{\be-\al}\underset{\la=\be}{\Res}\left[J_R^1(\la)\right]
\underset{\la=\al}{\Res}\left[J_R^1(\la)\right].
\end{aligned}
\end{equation} 
Let us evaluate $R_1^{+,+}$. Consider
\begin{equation}\label {qua10}
v_1^+(\la)\equiv -\frac{1}{2\pi i}\int_{\Sigma_R^{+}}
\frac {J_R^{\circ}(\mu)}{\la_- -\mu}\,d\mu.
\end{equation} 
By deforming the contour of integration up, we obtain the asymptotic
expansion of $v_1^+(\la)$ as $N\to\infty$,
\begin{equation}\label{qua11}
v_1^+(\la)
\sim -\sum_{j=1}^{\infty} \frac{1}{\la-z_j}\,
\underset{z=z_j}{\Res}\left[e^{-NG_N(z)}\right] M(z_j)\sg_+M(z_j)^{-1}.
\end{equation}
Now we substitute this asymptotic expansion into the formula,
\begin{equation}\label {qua12}
R_1^{+,+}=-\frac{1}{2\pi i}\int_{\Sigma_R^{+}}
v_1^+(\la)J_R^{\circ}(\la)\,d\la,
\end{equation} 
and move the contour of integration up. This gives the
asymptotic series,
\begin{equation}\label{qua13}
\begin{aligned}
R_1^{+,+}
&\sim \sum_{j,k=1;\; j\not=k}^\infty \frac{1}{z_k-z_j}\,
\underset{z=z_j}{\Res}\left[e^{-NG_N(z)}\right] 
\underset{z=z_k}{\Res}\left[e^{-NG_N(z)}\right] 
M(z_j)\sg_+ M(z_j)^{-1}M(z_k)\\
&\times \sg_+ M(z_k)^{-1}+\sum_{j=1}^\infty 
\underset{z=z_j}{\Res}\left[e^{-NG_N(z)}\right] 
\underset{z=z_j}{\Res}\left[\frac{e^{-NG_N(z)}}{z-z_j}\right] 
M(z_j)\sg_+^2 M(z_j)^{-1}.
\end{aligned}
\end{equation}
Observe that the last sum is equal to 0, because $\sg_+^2=0$.
Furthermore, since $M(z_j)=M(+i0)+O(N^{-1})$, we obtain that
\begin{equation}\label {qua14}
M(z_j)^{-1} M(z_k)=I+O(N^{-1}).
\end{equation} 
When we substitute $I$ for $M(z_j)^{-1} M(z_k)$ in the first
sum in (\ref{qua13}), we 
again get 0. When we substitute $O(N^{-1})$ for $M(z_j)^{-1} M(z_k)$,
we get a term of the order of $O(N^{-2\kappa_1})$. Thus,
\begin{equation}\label {qua15}
R_1^{+,+}=O(N^{-2\kappa_1}).
\end{equation} 
Similarly,
\begin{equation}\label {qua16}
R_1^{-,-}=O(N^{-2\kappa_1}).
\end{equation}
Observe that by (\ref{as14}),
\begin{equation}\label {qua16a}
\kappa_1=1+\frac{4\ga}{\pi-2\ga}>1.
\end{equation} 
Consider now
\begin{equation}\label {qua17}
R_1^{-,+}=-\frac{1}{2\pi i}\int_{\Sigma_R^{-}}
v_1^+(\la)J_R^{\circ}(\la)\,d\la.
\end{equation} 
When we substitute asymptotic expansion (\ref{qua11}) into this formula
and move the contour of integration, $\Sigma_R^-$, down, crossing the
poles of $J_R^{\circ}(\la)$, we obtain the asymptotic expansion,
\begin{equation}\label{qua18}
\begin{aligned}
R_1^{-,+}
\sim \sum_{j,k=1}^\infty \frac{1}{z_j+z_k}
&\underset{z=z_j}{\Res}\left[e^{-NG_N(z)}\right] 
\underset{z=-z_k}{\Res}\left[e^{NG_N(z)}\right] \\
&\times M(z_j)\sg_+ M(z_j)^{-1}M(-z_k)\sg_+ M(-z_k)^{-1}.
\end{aligned}
\end{equation}
Since
\begin{equation}\label {qua19}
M(z_j)=M(+i0)+O(N^{-1}),\qquad M(-z_k)=M(-i0)+O(N^{-1}),
\end{equation} 
we have that
\begin{equation}\label {qua20}
\begin{aligned}
&M(z_j)\sg_+ M(z_j)^{-1}M(-z_k)\sg_+ M(-z_k)^{-1}\\
&=M(+i0)\sg_+ M(+i0)^{-1}M(-i0)\sg_+ M(-i0)^{-1}+O(N^{-1}).
\end{aligned}
\end{equation} 
From (\ref{3.29}) we obtain, by a direct computation, that
\begin{equation}\label {qua21}
\sg_+ M(+i0)^{-1}M(-i0)\sg_+=-\sg_+,
\end{equation} 
and from (\ref{as13}), (\ref{as14a}), that
\begin{equation}\label {qua22}
\underset{z=z_j}{\Res}\left[e^{-NG_N(z)}\right] 
\underset{z=-z_k}{\Res}\left[e^{NG_N(z)}\right]
=C_jC_k N^{-\kappa_j-\kappa_k}(1+O(N^{-1})).
\end{equation} 
Hence
\begin{equation}\label{qua23}
R_1^{-,+}
\sim -\sum_{j,k=1}^\infty \frac{1}{z_j+z_k}
N^{-\kappa_j-\kappa_k}\left[ C_jC_k M(+i0)\sg_+  M(-i0)^{-1}+O(N^{-1})\right].
\end{equation}
Observe that
\begin{equation}\label {qua24}
\sg_+ M(-i0)^{-1}M(+i0)\sg_+=\sg_+,
\end{equation} 
and, therefore, a similar computation for $R_1^{+,-}$ gives that
\begin{equation}\label{qua25}
R_1^{+,-}
\sim \sum_{j,k=1}^\infty \frac{1}{z_j+z_k}
N^{-\kappa_j-\kappa_k}\left[ C_jC_k  M(-i0)\sg_+  M(+i0)^{-1}+O(N^{-1})\right].
\end{equation}
Since
\begin{equation}\label{qua26}
M(-i0)\sg_+  M(+i0)^{-1}-M(+i0)\sg_+  M(-i0)^{-1}=I,
\end{equation}
we obtain that
\begin{equation}\label{qua27}
R_1^{+,-}+R_1^{-,+}
\sim \sum_{j,k=1}^\infty \frac{1}{z_j+z_k}
 N^{-\kappa_j-\kappa_k}\left[ C_jC_k I+O(N^{-1})\right].
\end{equation}
If we restrict this matrix formula to the elements (12) and (21), then we
obtain that
\begin{equation}\label{qua28}
(R_1^{+,-})_{12}+(R_1^{-,+})_{12}=O(N^{-2\kappa_1}),\qquad
(R_1^{+,-})_{21}+(R_1^{-,+})_{21}=O(N^{-2\kappa_1}),
\end{equation}
because $\frac{1}{z_j+z_k}=O(N)$. Finally, the cross terms of
the form $R_1^{a,b}$, where $a=\pm$, $b=\al,\be$, or vice versa,
are estimated as
\begin{equation}\label{qua29}
R_1^{a,b}=O(N^{-1-\kappa_1}),\qquad
a=\pm,\;b=\al,\be,\quad
\textrm{or}\quad a=\al,\be,\; b=\pm.
\end{equation}

{\bfseries Summary for the quadratic term.}
By combining formulae (\ref{qua6}), (\ref{qua9}), (\ref{qua15}),
(\ref{qua16}), (\ref{qua28}), and (\ref{qua29}), we obtain that
\begin{equation}\label{qua29a}
(R_1^{(2)})_{12}=(c_1)_{12}N^{-2}+O(N^{-2-\ep}),\qquad
(R_1^{(2)})_{21}=(c_1)_{21}N^{-2}+O(N^{-2-\ep}),
\end{equation}
where the matrix $c_1$ is given in (\ref{qua9a}) and $\ep>0$.

{\bfseries Evaluation of the higher order terms.} The higher order
terms, $R_1^{(k)}$, $k\ge 3$, are evaluated in the same way 
as the quadratic terms, and we obtain that
\begin{equation}\label{qua30}
(R_1^{(k)})_{12},(R_1^{(k)})_{21}=O(N^{-2-\ep}),\qquad k\ge 3.
\end{equation}
Consider, for instance, the cubic term,
\begin{equation}\label{qua31}
R_1^{(3)}=-\frac{1}{2\pi i}\int_{\Sigma_R}v_2(\la)J_R^{\circ}(\la)d\la
=\left(-\frac{1}{2\pi i}\right)^3\int_{\Sigma_R}\int_{\Sigma_R}\int_{\Sigma_R}
\frac{J^{\circ}_R(\nu)J^{\circ}_R(\mu)J_R^{\circ}(\la)}
{(\la_- -\mu)(\mu_- -\nu)}\,d\nu d\mu d\la.
\end{equation}
As for the quadratic term, we split $R_1^{(3)}$ into a sum of terms
$R_1^{a,b,c}$, and the only nontrivial terms in regard to estimate 
(\ref{qua30}) are $R_1^{+,-,+}$ and $R_1^{-,+,-}$. We have that 
\begin{equation}\label{qua32}
R_1^{+,-,+}
=\left(-\frac{1}{2\pi i}\right)^3\int_{\Sigma_R^+}\int_{\Sigma_R^-}\int_{\Sigma_R^+}
\frac{J^{\circ}_R(\nu)J^{\circ}_R(\mu)J_R^{\circ}(\la)}
{(\la_- -\mu)(\mu_- -\nu)}\,d\nu d\mu d\la.
\end{equation}
We move the contour of integration $\Sigma_R^+$ up and the one $\Sigma_R^-$
down, and obtain the asymptotic series,
\begin{equation}\label{qua33}
\begin{aligned}
R_1^{+,-,+}
&\sim \sum_{j,k,l=1}^\infty \frac{1}{(z_j+z_k)(z_k+z_l)}
\underset{z=z_j}{\Res}\left[e^{-NG_N(z)}\right] 
\underset{z=-z_k}{\Res}\left[e^{NG_N(z)}\right] \underset{z=z_l}{\Res}\left[e^{-NG_N(z)}\right]\\
&\times M(z_j)\sg_+ M(z_j)^{-1}M(-z_k)\sg_+ M(-z_k)^{-1}M(z_l)\sg_+ M(z_l)^{-1}.
\end{aligned}
\end{equation}
By using (\ref{qua19}) and (\ref{qua21}), we obtain that
\begin{equation}\label{qua34}
R_1^{+,-,+}
\sim i\sum_{j,k,l=1}^\infty \frac{1}{(z_j+z_k)(z_k+z_l)}
N^{-\kappa_j-\kappa_k-\kappa_l}\left[ C_jC_kC_l M(+i0)\sg_+  M(-i0)^{-1}+O(N^{-1})\right].
\end{equation}
A similar computation for $R_1^{-,+,-}$ gives that
\begin{equation}\label{qua35}
R_1^{-,+,-}
\sim -i\sum_{j,k,l=1}^\infty \frac{1}{(z_j+z_k)(z_k+z_l)}
N^{-\kappa_j-\kappa_k-\kappa_l}\left[ C_jC_kC_l M(-i0)\sg_+  M(+i0)^{-1}+O(N^{-1})\right],
\end{equation}
and by using (\ref{qua26}), we obtain that
\begin{equation}\label{qua35a}
R_1^{+,-,+}+R_1^{-,+,-}
\sim -i\sum_{j,k,l=1}^\infty \frac{1}{(z_j+z_k)(z_k+z_l)}
N^{-\kappa_j-\kappa_k-\kappa_l}\left[ C_jC_kC_l I+O(N^{-1})\right],
\end{equation}
hence
\begin{equation}\label{qua36}
(R_1^{+,-,+})_{12}+(R_1^{-,+,-})_{12}=O(N^{-3\kappa_1+1}),\qquad
(R_1^{+,-,+})_{21}+(R_1^{-,+,-})_{21}=O(N^{-3\kappa_1+1}).
\end{equation}
Since $3\kappa_1-1>2$, we obtain estimate (\ref{qua30}) for $R_1^{+,-,+}+R_1^{-,+,-}$. 
It is straightforward to get the estimate,
\begin{equation}\label{qua37}
R_1^{a,b,c}=O(N^{-2-\ep}),
\end{equation}
for all other combinations of $a,b,c$ and hence (\ref{qua30}) follows.
The same argument holds for $k>3$.

{\bfseries Evaluation of $R_{NN}$.} Let us go back now to formula
(\ref{as5a}) and evaluate  the terms on the right in this
formula with an error term
of the order of $N^{-2-\ep}$. From (\ref{VN7}),
\begin{equation}\label{as24a}
\be_N-\al_N=\frac{2\pi}{\cos\frac{\pi\zeta}{2}}+N^{-2}
\frac{2\ga^2}{3(\pi-2\ga)\cos\frac{\pi\zeta}{2}}+O(N^{-3}),
\end{equation}
hence
\begin{equation}\label{as25}
\left(\frac{\be_N-\al_N}{4}\right)^2
=\left(\frac{\pi}{2\cos\frac{\pi\zeta}{2}}\right)^2+N^{-2}
\frac{\pi\ga^2}{6(\pi-2\ga)\cos^2\frac{\pi\zeta}{2}}+O(N^{-3}),
\end{equation}
Next, from (\ref{as23a}), (\ref{qua29}), and (\ref{qua30}) we obtain that
\begin{equation}\label{as30}
\begin{aligned}
\frac{\be-\al}{4i}\,[(R_1)_{12}-(R_1)_{21}]
&=\frac{\be-\al}{2}\cos(N\om)\sum_{j:\;\kappa_j\le 2}C_jN^{-\kappa_j}+cN^{-2}+O(N^{-2-\ep})\\
&=\cos(N\om)\sum_{j:\;\kappa_j\le 2}c_jN^{-\kappa_j}+c^0N^{-2}+O(N^{-2-\ep})\,,
\end{aligned}
\end{equation}
where
\begin{equation}\label{as31}
c_j=\frac{\be-\al}{2}C_j=\frac{2\ga e^{\f(y_j)}}{\cos\frac{\pi\z}{2}}
(-1)^j\sin\frac{\pi j}{1-\frac{2\ga}{\pi}}\,,
\end{equation}
and $c^0$ is a computable constant.
From (\ref{as5a}) we obtain now that
\begin{equation}\label{as32}
\begin{aligned}
R_{NN}&=\left(\frac{\pi}{2\cos\frac{\pi\zeta}{2}}\right)^2+
\cos(N\om)\sum_{j:\;\kappa_j\le 2}c_jN^{-\kappa_j}+cN^{-2}+O(N^{-2-\ep})\,,
\end{aligned}
\end{equation}
where
\begin{equation}\label{as33}
c=\frac{\pi\ga^2}{6(\pi-2\ga)\cos^2\frac{\pi\zeta}{2}}+c_0.
\end{equation}
Here the first term in the expression for $c$ comes from the difference 
$\left(\frac{\be_N-\al_N}{4}\right)^2
-\left(\frac{\be-\al}{4}\right)^2$, see (\ref{as25}),
while the second term, $c_0$, is  determined by
calculations of other terms of the order of $N^{-2}$ on the right in formula
(\ref{as5a}). The constant $c_0$ can be evaluated explicitly by
tracing down all the terms of the order of $N^{-2}$ in the above computations.
To avoid these somewhat tedious computations, we will use the fact
that we know the exact expression for $R_{NN}$ on the free fermion line.

Observe that $c_0$ is calculated in terms of
contour integrals around the turning points $\al_N$ and $\be_N$,
and it depends only on the limiting values of the end points, $\al,\be$.
The exact values of $\al,\be$ are given in (\ref{V4}) and they 
depend on the parameter $\z$ only. 
This implies that $c_0$ is a function of the parameter $\z$ as well, 
$c_0=c_0(\z)$,  and it is independent of $\ga$. To find an
exact value of $c_0(\z)$, consider 
the free fermion line $\gamma=\frac{\pi}{4}$. In this case $c=0$, which gives
\begin{equation}\label{as34}
c_0(\z)=-\frac{\pi^2}{48\cos^2\frac{\pi\zeta}{2}}\,.
\end{equation}
Thus,
\begin{equation}\label{as35}
c=\frac{\pi\ga^2}{6(\pi-2\ga)\cos^2\frac{\pi\zeta}{2}}
-\frac{\pi^2}{48\cos^2\frac{\pi\zeta}{2}}\,.
\end{equation}
This proves formula (\ref{resc8a}) and hence Theorem \ref{R_n}.

\section{Proof of Theorems \ref{F_N_t} and \ref{kappa}}
\label{Proofs}

We omit the proof of Theorem \ref{F_N}, because it follows from Theorem \ref{kappa}.

{\bf Proof of Theorem \ref{F_N_t}}. By (\ref{dph24}) and (\ref{dph19}),
\begin{equation}\label{fnt1}
\frac{\partial^2 F_N}{\partial t^2}=\frac{R_N}{N^2}
=\frac{1}{\ga^2}\left[R+\cos (N\om)\sum_{j:\;\kappa_j\le 2}c_j N^{-\kappa_j}+cN^{-2}+O(N^{-2-\ep})\right].
\end{equation}
It is easy to check that
\begin{equation}\label{fnt2}
\frac{\partial^2 F}{\partial t^2}
=\frac{R}{\ga^2}\,,
\end{equation}
hence
(\ref{dph25}) follows. Theorem \ref{F_N_t} is proved.

{\bf Proof of Theorem \ref{kappa}}. 
By (\ref{dph19}),
\begin{equation}\label{k1}
R_n=\frac{n^2R}{\ga^2}e^{b_n},
\end{equation}
where
\begin{equation}\label{k2}
b_n=\cos(n\om)\sum_{j:\;\kappa_j\le 2} 
d_j n^{-\kappa_j}-\kappa n^{-2}+O(n^{-2-\ep}),
\end{equation}
and $d_j=\frac{c_j}{R}\,$, $\kappa=-\frac{c}{R}$. From (\ref{dph20}),
(\ref{dph20e}) we obtain that
\begin{equation}\label{k2a}
\kappa=-\frac{c}{R}=\frac{1}{12}-\frac{2\ga^2}{3\pi(\pi-2\ga)}\,.
\end{equation}
From (\ref{dph18}) and (\ref{k1}) we obtain that
\begin{equation}\label{k3}
\tau_N=h_0^N \left(\frac{R}{\ga^2}\right)^{\frac{N(N-1)}{2}}
\left(\prod_{n=0}^{N-1}n!\right)^2 e^{B_N},
\end{equation}
where 
\begin{equation}\label{k4}
B_N=(N-1)b_1+(N-2)b_2+\dots+b_{N-1},
\end{equation}
hence by (\ref{dph21}),
\begin{equation}\label{k5}
\begin{aligned}
F_N&=N^{-2}\ln \frac{\tau_N}{\left(\prod_{n=0}^{N-1} n!\right)^2}
=N^{-2}\left[ N\ln h_0+\frac{N(N-1)}{2}\ln\frac{R}{\ga^2}+B_N\right]\\
&=\frac 12 \ln \frac{R}{\ga^2}+C_0N^{-1}+N^{-2}B_N,
\end{aligned}
\end{equation}
where $C_0$ is a constant. Let us evaluate $B_N$. We have that
\begin{equation}\label{k6}
B_N=N(b_1+b_2+\dots+b_N)-b_1-2b_2-\dots-Nb_N,
\end{equation}
and
\begin{equation}\label{k7}
b_1+b_2+\dots+b_N=B-\sum_{n=N+1}^\infty b_n,
\end{equation}
where
\begin{equation}\label{k8}
B=\sum_{n=1}^\infty b_n.
\end{equation}
It follows from (\ref{k2}), that
\begin{equation}\label{k9a}
\sum_{n=N+1}^\infty b_n=-\kappa N^{-1}+O(N^{-1-\ep}),
\end{equation}
because
\begin{equation}\label{k9b}
\sum_{n=N+1}^{\infty} n^{-\kappa_j}\cos(n\om)=
O(N^{-\kappa_j}),\qquad 0<\om<2\pi.
\end{equation}
It also follows from (\ref{k2}), that
\begin{equation}\label{k10}
\sum_{n=1}^N nb_n=-\kappa\ln N+C_1+O(N^{-\ep}),
\end{equation}
where $C_1$ is a constant, because
\begin{equation}\label{k11}
\sum_{n=1}^N n^{-\kappa_j+1}\cos(n\om)=C(\kappa_j)+O(N^{-\kappa_j+1}),\qquad 0<\om<2\pi.
\end{equation}
Thus,
\begin{equation}\label{k12}
B_N=C_2N+\kappa\ln N +C_3+O(N^{-\ep}),
\end{equation}
where $C_2,C_3$ are some constants, hence from (\ref{k5}) we obtain that
\begin{equation}\label{k13}
F_N=F+c_0N^{-1}+\kappa N^{-2}\ln N+C_3N^{-2}+O(N^{-2-\ep}),
\end{equation}
where $c_0$ is a constant. This implies that
\begin{equation}\label{k13a}
Z_N=Ce^{N^2f+Nc_0}N^{\kappa}\left(1+O(N^{-\ep})\right),
\end{equation}
where $C=e^{C_3}$. To finish the proof of Theorem \ref{kappa}, 
it remains to prove  the following lemma.

\begin{lem} \label{C=0} $c_0=0$.
\end{lem}

{\it Proof.} By (\ref{resc5}),
\begin{equation}\label{k16}
h_n=\left(\frac{n}{\ga}\right)^{2n+1}h_{nn},
\end{equation}
and by (\ref{as1}),
\begin{equation}\label{k17}
h_{nn}=-2\pi i e^{nl_n}(S_1)_{12}.
\end{equation}
Observe that by (\ref{lN6}),
\begin{equation}\label{k18}
l_n=l+O(n^{-2}),\qquad l=2\ln(\be-\al)-2-4\ln 2,
\end{equation}
and by (\ref{as3}), (\ref{as5}), 
\begin{equation}\label{k19}
(S_1)_{12}=(M_1)_{12}+(R_1)_{12}=-\frac{\be-\al}{4i}\left(1+O(n^{-1})\right).
\end{equation}
Therefore,
\begin{equation}\label{k20}
h_n=\left(\frac{n}{\ga}\right)^{2n+1}\frac{\pi(\be-\al)}{2}\exp\left(nl+O(n^{-1})\right),
\end{equation}
and 
\begin{equation}\label{k21}
\begin{aligned}
\tau_N=\prod_{n=0}^{N-1}h_n&=h_0\left(\prod_{n=1}^{N-1} n^{2n+1}\right)\ga^{-N^2}
\left(\frac{\pi(\be-\al)}{2}\right)^{N-1}\\
&\times\exp\left(\frac{N(N-1)}{2}l+O(\ln N)\right).
\end{aligned}
\end{equation}
By applying (\ref{k18}), (\ref{dph23}), and (\ref{V4a}), we obtain that 
\begin{equation}\label{k21a}
\tau_N=C_N\exp\left(N^2F+O(\ln N)\right),
\end{equation}
where $C_N$ does not depend on $\ga$ and $t$. By (\ref{pf7}) and
(\ref{exact1}), this implies that
\begin{equation}\label{k22}
Z_N=\tilde C_N\exp\left(N^2f+O(\ln N)\right),
\end{equation}
where $\tilde C_N$ also does not depend on $\ga$ and $t$.
Since on the free fermion line, $Z_N=1$ and $f=0$, we obtain that
\begin{equation}\label{k23}
\ln \tilde C_N=O(\ln N),
\end{equation}
hence
\begin{equation}\label{k24}
Z_N=\exp\left(N^2f+O(\ln N)\right),
\end{equation}
so that $c_0=0$. Lemma \ref{C=0} is proved.

\appendix

\section{Large $N$ asymptotics of $A(N)$ and $A(N;3)$}\label{AN}

{\bf Large $N$ asymptotics of $A(N)$.}
We will find in this appendix the large $N$ asymptotics of
\begin{equation} \label{AN1}
A(N)=\prod_{n=0}^{N-1}\frac{(3n+1)!n!}{(2n)!(2n+1)!}\,.
\end{equation}
We start with the asymptotics of 
\begin{equation} \label{AN2}
a(N)=\prod_{n=1}^{N-1}n!.
\end{equation}
We have that
\begin{equation} \label{AN3}
\ln a(N)=\sum_{n=1}^N (N-n)\ln n=N^2\sum_{n=1}^N \left(1-\frac{n}{N}\right)
\left(\ln \frac{n}{N}\right)N^{-1}+\sum_{n=1}^N (N-n)\ln N.
\end{equation}
In addition,
\begin{equation} \label{AN4}
\sum_{n=1}^N \left(1-\frac{n}{N}\right)
\left(\ln \frac{n}{N}\right)N^{-1}
=-\frac{3}{4}+\frac{\ln N}{2N}+\frac{\ln(2\pi)}{2N}
-\frac{\ln N}{12 N^2}+\frac{\z'(-1)}{N^2}-\frac{1}{240 N^4}+\dots,
\end{equation}
where $\z(s)$ is the Riemann zeta-function. This gives 
\begin{equation} \label{AN5}
\ln a(N)
=\frac{N^2\ln N}{2}-\frac{3N^2}{4}+\frac{N\ln(2\pi)}{2}
-\frac{\ln N}{12}+\z'(-1)-\frac{1}{240 N^2}+\dots,
\end{equation}
so that
\begin{equation} \label{AN6}
a(N)=\prod_{n=1}^{N-1}n!=N^{\frac{N^2}{2}}e^{-\frac{3}{4}N^2}(2\pi)^{\frac{N}{2}}
N^{-\frac{1}{12}}e^{\z'(-1)-\frac{1}{240N^2}+\dots}.
\end{equation}
Consider now
\begin{equation} \label{AN7}
a_{31}(N)=\prod_{n=1}^{N-1}(3n+1)!.
\end{equation}
We have that
\begin{equation} \label{AN8}
\ln a_{31}(N)=b_1(N)+b_0(N)+b_{-1}(N)
\end{equation}
where
\begin{equation} \label{AN9}
b_j(N)=\sum_{n=1}^N (N-n)\ln (3n+j),\qquad j=1,0,-1.
\end{equation}
Observe that 
\begin{equation} \label{AN10}
b_0(N)=\sum_{n=1}^N (N-n)\ln (3n)=\frac{(\ln 3)N(N-1)}{2}+\ln a(N),
\end{equation}
hence by (\ref{AN5}),
\begin{equation} \label{AN11}
b_0(N)
=\frac{(\ln 3)N(N-1)}{2}+\frac{N^2\ln N}{2}-\frac{3N^2}{4}+\frac{N\ln(2\pi)}{2}
-\frac{\ln N}{12}+\z'(-1)-\frac{1}{240 N^2}+\dots,
\end{equation}
Now,
\begin{equation} \label{AN12}
\begin{aligned}
b_1(N)+b_{-1}(N)-2b_0(N)&=\sum_{n=1}^N (N-n)\ln \left(1-\frac{1}{9n^2}\right)\\
&=N\ln\left(\frac{3\sqrt 3}{2\pi}\right)+\frac{\ln N}{9}+\ga_0+\frac{2}{243 N^2}+\dots,
\end{aligned}
\end{equation}
where $\ga_0$ is a constant,
\begin{equation} \label{AN12a}
\ga_0=\lim_{N\to\infty}\left[-\sum_{n=1}^N n\ln\left(1-\frac{1}{9n^2}\right)
-\frac{\ln N}{9}\right]\,.
\end{equation}
Therefore,
\begin{equation} \label{AN13}
\begin{aligned}
\ln a_{31}(N)&=\frac{3(\ln 3)N(N-1)}{2}+\frac{3N^2\ln N}{2}-\frac{9N^2}{4}+\frac{3N\ln(2\pi)}{2}
-\frac{\ln N}{4}+3\z'(-1)\\
&-\frac{1}{80 N^2}+\dots
+N\ln\left(\frac{3\sqrt 3}{2\pi}\right)+\frac{\ln N}{9}+\ga_0+\frac{2}{243 N^2}+\dots,
\end{aligned}
\end{equation}
and 
\begin{equation} \label{AN14}
a_{31}(N)=
N^{\frac{3N^2}{2}}3^{\frac{3N^2}{2}}e^{-\frac{9}{4}N^2}
(2\pi)^{\frac{N}{2}}
N^{-\frac{5}{36}}e^{3\z'(-1)+\ga_0-\frac{83}{19440N^2}+\dots}.
\end{equation}
Finally,
\begin{equation} \label{AN15}
\prod_{n=0}^{N-1}[(2n)!(2n+1)!]=\prod_{n=0}^{2N-1} n!=a(2N).
\end{equation}
By (\ref{AN6}),
\begin{equation} \label{AN16}
a(2N)=(2N)^{2N^2}e^{-3N^2}(2\pi)^{N}
(2N)^{-\frac{1}{12}}e^{\z'(-1)-\frac{1}{960N^2}+\dots}.
\end{equation}
Thus, (\ref{AN1}) reduces to
\begin{equation} \label{AN17}
\begin{aligned}
A(N)&=\frac{a_{31}(N)a(N)}{a(2N)}
=\frac{N^{\frac{3N^2}{2}}3^{\frac{3N^2}{2}}e^{-\frac{9}{4}N^2}
(2\pi)^{\frac{N}{2}}
N^{-\frac{5}{36}}
N^{\frac{N^2}{2}}e^{-\frac{3}{4}N^2}(2\pi)^{\frac{N}{2}}
N^{-\frac{1}{12}}}
{(2N)^{2N^2}e^{-3N^2}(2\pi)^{N}
(2N)^{-\frac{1}{12}}}\\
&\times \frac{e^{3\z'(-1)+\ga_0-\frac{83}{19440N^2}+\dots}
e^{\z'(-1)-\frac{1}{240N^2}+\dots}}
{e^{\z'(-1)-\frac{1}{960N^2}+\dots}}\,.
\end{aligned}
\end{equation}
By simplifying, we obtain that
\begin{equation} \label{AN18}
A(N)=C\left(\frac{3\sqrt 3}{4}\right)^{N^2}N^{-\frac{5}{36}}
\left(1-\frac{115}{15552N^2}+O(N^{-3}\right),
\end{equation}
where
\begin{equation} \label{AN19}
C=2^{\frac{1}{12}}e^{3\z'(-1)+\ga_0}\,.
\end{equation}

{\bf Large $N$ asymptotics of $A(N;3)$.} From (\ref{exact8})
we have that
\begin{equation} \label{AN3:1}
\left\{
\begin{aligned}
A(2m;3)&=3^{m^2}\frac{m!}{(3m)!}\prod_{k=0}^{m-1}
\left[\frac{(3k+2)!}{(m+k)!}\right]^2,\\
A(2m+1;3)&=3^{m^2+m}\prod_{k=0}^{m-1}
\left[\frac{(3k+2)!}{(m+k+1)!}\right]^2,
\end{aligned}\right.
\end{equation}
cf. \cite{CP}. Let us start with $A(2m;3)$. We can rewrite it as
\begin{equation} \label{AN3:2}
A(2m;3)=3^{m^2}\frac{m!}{(3m)!}\left[\frac{a_{32}(m)a(m)}{a(2m)}\right]^2,
\end{equation}
where
\begin{equation} \label{AN3:3}
a_{32}(m)=\prod_{k=0}^{m-1}(3k+2)!.
\end{equation}
Observe that
\begin{equation} \label{AN3:4}
a_{32}(m)=a_{31}(m)\prod_{k=0}^{m-1}(3k+2)=a_{31}(m)3^m
\frac{\Gamma\left(m+\frac{2}{3}\right)}{\Gamma\left(\frac{2}{3}\right)},
\end{equation}
hence from (\ref{AN3:2}) and (\ref{AN17}) we obtain that
\begin{equation} \label{AN3:5}
A(2m;3)=3^{m^2}\frac{m!}{(3m)!}\left[\frac{3^m\Gamma\left(m+\frac{2}{3}\right)A(m)}
{\Gamma\left(\frac{2}{3}\right)}\right]^2,
\end{equation}
We have that
\begin{equation} \label{AN3:6}
\frac{(m!)^3}{(3m)!}=3^{-3m}\frac{2\pi m}{\sqrt 3}e^{\frac{2}{9m}+O(m^{-3})}
\end{equation}
and
\begin{equation} \label{AN3:7}
\frac{\Gamma\left(m+\frac{2}{3}\right)}{m!}
=m^{-\frac13}e^{-\frac{1}{9m}+\frac{1}{162m^2}+\dots}.
\end{equation}
By combining this with asymptotics (\ref{AN18}), we obtain that
\begin{equation} \label{AN3:8}
A(2m;3)=C_3\left(\frac{3}{2}\right)^{4m^2}3^{-m}
(2m)^{\frac{1}{18}}\left(1+\frac{77}{7776m^2}+O(m^{-3})\right),
\end{equation}
where
\begin{equation} \label{AN3:9}
C_3=\frac{2^{\frac{10}{9}}\pi}
{\left[\Gamma\left(\frac{2}{3}\right)\right]^2\sqrt 3}\,e^{6\z'(-1)+2\ga_0}.
\end{equation}
Consider now $A(2m+1;3)$. From (\ref{AN3:1}),
\begin{equation} \label{AN3:10}
A(2m+1;3)=3^m\frac{(3m)!m!}{[(2m)!]^2}A(2m;3).
\end{equation}
By using the Stirling formula we obtain that
\begin{equation} \label{AN3:11}
3^m\frac{(3m)!m!}{[(2m)!]^2}
=\left(\frac{3}{2}\right)^{4m}\frac{\sqrt 3}{2}e^{\frac{1}{36m}+O(m^{-3})}.
\end{equation}
Also,
\begin{equation} \label{AN3:11a}
\left(\frac{2m}{2m+1}\right)^{\frac{1}{18}}=e^{-\frac{1}{36m}+\frac{1}{144m^2}+O(m^{-3})}
\end{equation}
By combining these formulae with (\ref{AN3:8}), we get 
\begin{equation} \label{AN3:12}
A(2m+1;3)=
C_3\left(\frac{3}{2}\right)^{(2m+1)^2}(\sqrt 3)^{-(2m+1)}
(2m+1)^{\frac{1}{18}}\left(1+\frac{131}{7776m^2}+O(m^{-3})\right).
\end{equation}

\section {Proof of formula (\ref{V11})}\label{A}

We have:
\begin{equation}
\int\om(z)\,dz=z\om(z)-\int z\om'(z) dz.
\end{equation}
From (\ref{V3}),
\begin{equation}
\begin{aligned}
\om'(z)&=\frac{2}{i\pi}\left[
\frac{\frac{\sqrt{\be}}{2\sqrt{z-\al}}
-\frac{i\sqrt{-\al}}{2\sqrt{z-\be}}}
{\sqrt{\be(z-\al)}-i\sqrt{-\al(z-\be)}}
-\frac{1}{2z}\right]\\
&=\frac{1}{i\pi}\left[
\frac{\left(\frac{\sqrt{\be}}{\sqrt{z-\al}}
-\frac{i\sqrt{-\al}}{\sqrt{z-\be}}\right)
\left(\sqrt{\be(z-\al)}+i\sqrt{-\al(z-\be)}\right)}
{(\be-\al)z}
-\frac{1}{z}\right]\\
&=\frac{1}{\pi}
\frac{\sqrt{\be(-\al)}}{(\be-\al)z}\left(\sqrt{\frac{z-\be}{z-\al}}
-\sqrt{\frac{z-\al}{z-\be}}\right)
=-\frac{\sqrt{\be(-\al)}}{\pi z\sqrt{(z-\al)(z-\be)}},
\end{aligned}
\end{equation}
hence
\begin{equation}
\begin{aligned}
\int\om(z)\,dz&=z\om(z)+\frac{\sqrt{\be(-\al)}}{\pi}
\int\frac{dz}{\sqrt{(z-\al)(z-\be)}}\\
&=z\om(z)+\frac{2\sqrt{\be(-\al)}}{\pi}
\log\left(\sqrt{z-\al}+\sqrt{z-\be}\right).
\end{aligned}
\end{equation}
From (\ref{V4a}), $\sqrt{\be(-\al)}=\pi$, hence 
\begin{equation}
g(z)=z\om(z)+2
\log\left(\sqrt{z-\al}+\sqrt{z-\be}\right)+C.
\end{equation}
As $z\to \infty$,
\begin{equation}
g(z)=\log z+O(z^{-1})=z[z^{-1}+O(z^{-2})]+2[\log (2\sqrt z)+O(z^{-1})]
+C,
\end{equation}
hence $C=-1-2\ln 2$, and (\ref{V11}) follows.

\section {Proof of Proposition \ref{alN-beN}} \label{AppD}

From (\ref{VN5}), (\ref{VN6}) we have that 
\begin{equation}\label{16} 
F_N(\al_N,\be_N)\equiv \frac{1}{2\pi}\int_{\al_N}^{\be_N}\frac{V'_N(x)}{\sqrt{(x-\al_N)(\be_N-x)}}dx=0,
\end{equation}
and
\begin{equation}\label{17}
G_N(\al_N,\be_N)\equiv\frac{1}{2\pi}
\int_{\al_N}^{\be_N}\frac{xV'_N(x)}{\sqrt{(x-\al_N)(\be_N-x)}}dx=1.
\end{equation}
In (\ref{16}), (\ref{17}) we can rewrite the integrals as the contour integrals, 
\begin{equation}\label{17a} 
\begin{aligned}
F_N(\al_N,\be_N)&=\frac{1}{4\pi i}\oint_{\Ga_\ep}\frac{V'_N(z)}{\sqrt{(z-\al_N)(z-\be_N)}}dz,\\
G_N(\al_N,\be_N)&=\frac{1}{4\pi i}\oint_{\Ga_\ep}\frac{zV'_N(z)}{\sqrt{(z-\al_N)(z-\be_N)}}dz,
\end{aligned}
\end{equation}
where the function $\sqrt{(z-\al_N)(z-\be_N)}$ is considered on the principal sheet, with a cut
on $[\al_N,\be_N]$, and
$\Ga_\ep$ is a positively oriented contour on the complex plane around $[\al_N,\be_N]$,
which consists of
the two circles, $\{|z-\al|=\ep\}$ and $\{|z-\be|=\ep\}$, and the two intervals,
$[\al+\ep,\be-\ep]$, along the lower shore of the cut, and $[\be-\ep,\al+\ep]$,
along the upper shore, see Fig. 8. It follows from representation (\ref{17a}) that both $F_N$ and $G_N$ 
are analytic functions of $\al_N,\be_N$.

\begin{center}
 \begin{figure}[h]\label{figure8}
\begin{center}
   \scalebox{0.5}{\includegraphics{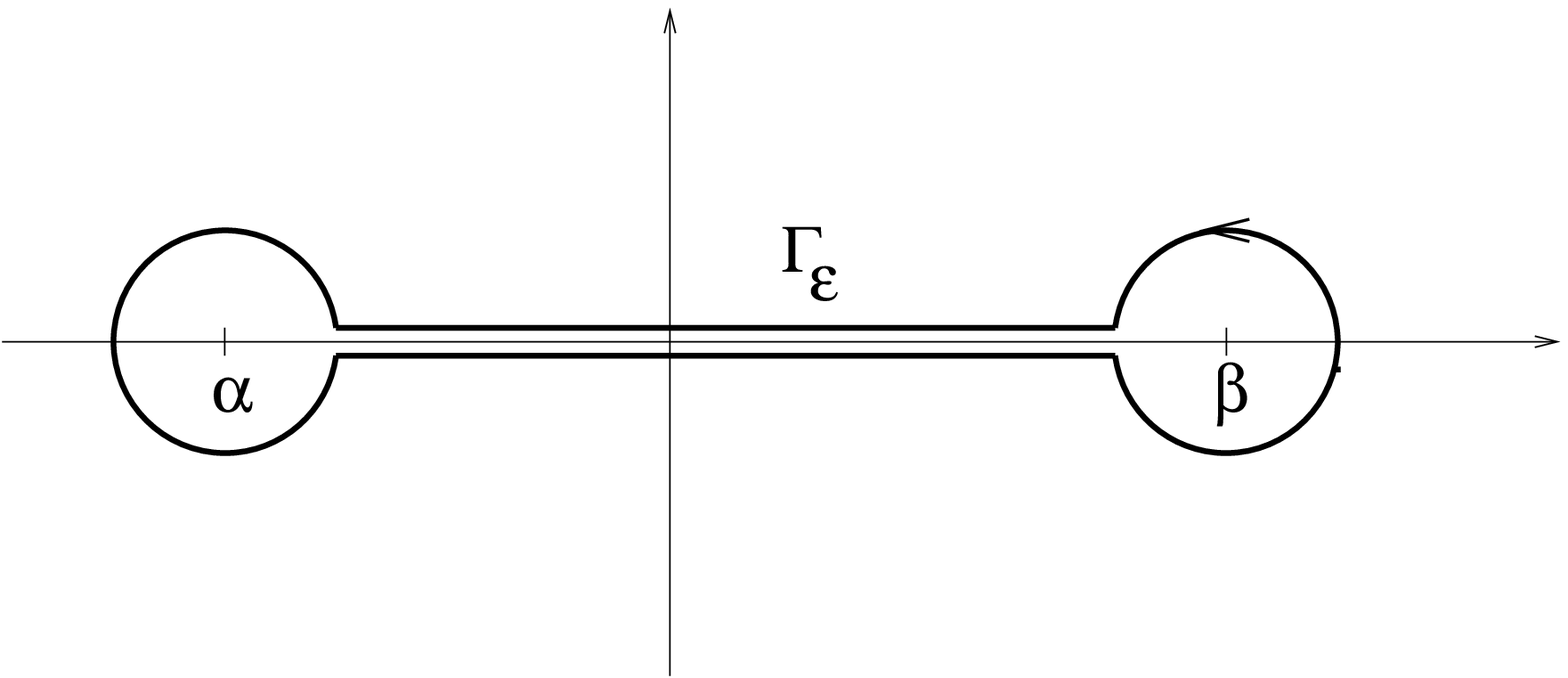}}
\end{center}
        \caption{The contour $\Gamma_{\ep}$.}
    \end{figure}
\end{center}

By (\ref{VNa}),
\begin{equation}\label{17b}
V'_N(z)=V'(z)+f(Nz),
\end{equation}
where
\begin{equation}\label{17c}
V(z)=z\,\sign\Re z-\z z,\qquad V'(z)=\sign\Re z -\z,
\end{equation}
and
\begin{equation}\label {17d}
f(z)=\frac{\pi}{2\ga}\coth z\frac{\pi}{2\ga}
-\left(\frac{\pi}{2\ga}-1\right)
\coth z\left(\frac{\pi}{2\ga}-1\right)
-\sign \Re z.
\end{equation}
Therefore, we can rewrite equations (\ref{16}), (\ref{17}) as
\begin{equation}\label{17e}
\left\{ 
\begin{aligned}
F(\al_N,\be_N)&\equiv \frac{1}{4\pi i}\oint_{\Ga_\ep}\frac{V'(z)}{\sqrt{(z-\al_N)(z-\be_N)}}dz
=-\frac{1}{4\pi i}\oint_{\Ga_\ep}\frac{f(Nz)}{\sqrt{(z-\al_N)(z-\be_N)}}dz,\\
G(\al_N,\be_N)&\equiv \frac{1}{4\pi i}\oint_{\Ga_\ep}\frac{zV'(z)}{\sqrt{(z-\al_N)(z-\be_N)}}dz
=1-\frac{1}{4\pi i}\oint_{\Ga_\ep}\frac{zf(Nz)}{\sqrt{(z-\al_N)(z-\be_N)}}dz.
\end{aligned}
\right.
\end{equation}
We will assume that $\al_N-\al=O(N^{-2})$ and $\be_N-\be=O(N^{-2})$ as $N\to\infty$,
where $\al$ and $\be$ solve the system
\begin{equation}\label{17f}
\left\{ 
\begin{aligned}
F(\al,\be)&=0,\\
G(\al,\be)&=1,
\end{aligned}
\right.
\end{equation}
and we will prove the existence of $\al_N,\be_N$ by using the implicit function theorem.
Observe that $\al$ and $\be$ are given by formulae (\ref{V4}). 

The function $f(z)$ is exponentially decaying as $|\Re z|\to\infty$, and this allows us to evaluate the 
integrals on the right in (\ref{17e}) asymptotically,  as $N\to\infty$. Namely,
\begin{equation}\label{17g}
\begin{aligned}
\frac{1}{4\pi i}\oint_{\Ga_\ep}&\frac{f(Nz)}{\sqrt{(z-\al_N)(z-\be_N)}}dz
=\frac{1}{2\pi N}\int_{-N\al_N}^{N\be_N}\frac{f(x)}{\sqrt{(N^{-1}x-\al_N)(\be_N-N^{-1}x)}}dx\\
&=\frac{1}{2\pi N\sqrt{(-\al_N)\be_N}}\int_{-\infty}^{\infty}f(x)\left[
1 +\frac{x(\al_N+\be_N)}{2N(-\al_N)\be_N}\right]dx+O(N^{-3}).
\end{aligned}
\end{equation}
Observe that that $f(-x)=-f(x)$, hence
\begin{equation}\label{18}
\int_{-\infty}^{\infty}f(x)dx=0
\end{equation}
and
\begin{equation}\label{19}
\begin{aligned}
\int_{-\infty}^{\infty}xf(x)dx=2\int_{0}^{\infty}x\left(\frac{\pi}{\ga}\frac{1}{e^{x\frac{\pi}{\ga}}-1}-
\left(\frac{\pi}{\ga}-2\right)\frac{1}{e^{x(\frac{\pi}{\ga}-2)}-1}\right)dx
\\=2\left(\frac{\ga}{\pi}-\frac{\ga}{\pi-2\ga}\right)
\int_{0}^{\infty}\frac{u}{e^u-1}du=-\frac{2\ga^2\pi}{3(\pi-2\ga)}.
\end{aligned}
\end{equation}
Also we can replace $\al_N,\be_N$ for $\al,\be$ in (\ref{17g}) and use 
formulae (\ref{V4a}). This gives us that
\begin{equation}\label{20}
\frac{1}{4\pi i}\oint_{\Ga_\ep}\frac{f(Nz)}{\sqrt{(z-\al_N)(z-\be_N)}}dz
=-N^{-2}\frac{\ga^2\tan\frac{\pi\z}{2}}{ 3\pi^2(\pi-2\ga)}+O(N^{-3})\,.
\end{equation}
Similarly we obtain that
\begin{equation}\label{21}
\frac{1}{4\pi i}\oint_{\Ga_\ep}\frac{zf(Nz)}{\sqrt{(z-\al_N)(z-\be_N)}}dz
=-N^{-2}\frac{\ga^2}{ 3\pi(\pi-2\ga)}+O(N^{-3})\,.
\end{equation}
Thus, system (\ref{17e}) reduces to the following one:
\begin{equation}\label{22}
\left\{ 
\begin{aligned}
F(\al_N,\be_N)&=N^{-2}\frac{\ga^2\tan\frac{\pi\z}{2}}{ 3\pi^2(\pi-2\ga)}+O(N^{-3}),\\
G(\al_N,\be_N)&=1+N^{-2}\frac{\ga^2}{ 3\pi(\pi-2\ga)}+O(N^{-3}).
\end{aligned}
\right.
\end{equation}
In the linear approximation the latter system reads
\begin{equation}\label{23}
\left\{
\begin{aligned}
&(\al_N-\al)F_{\al_N}(\al,\be)+(\be_N-\be)F_{\be_N}(\al,\be)
=N^{-2}\frac{\ga^2\tan\frac{\pi\z}{2}}{ 3\pi^2(\pi-2\ga)}+O(N^{-3}),\\
&(\al_N-\al)G_{\al_N}(\al,\be)+(\be_N-\be)G_{\be_N}(\al,\be)
=N^{-2}\frac{\ga^2}{ 3\pi(\pi-2\ga)}+O(N^{-3})\,.
\end{aligned}
\right.
\end{equation}
The coefficients of this linear system can be evaluated explicitly.
Namely, we have that
\begin{equation}\label{24}
\begin{aligned}
&F(\al_N,\be_N)=-\frac{\z}{2}+\frac{1}{\pi}\arcsin\frac{\be_N+\al_N}{\be_N-\al_N}\,,\\
&G(\al_N,\be_N)=-\frac{\z(\be_N+\al_N)}{4}+\frac{\sqrt{\be_N(-\al_N)}}{\pi}
+\frac{\be_N+\al_N}{2\pi}\arcsin\frac{\be_N+\al_N}{\be_N-\al_N}\,,
\end{aligned}
\end{equation}
which gives that
\begin{equation}\label{25}
\begin{aligned}
&F_{\al_N}(\al,\be)=\frac{1}{2\pi^2}\left(1+\sin\frac{\pi\zeta}{2}\right),
\qquad
F_{\be_N}(\al,\be)=\frac{1}{2\pi^2}\left(1-\sin\frac{\pi\zeta}{2}\right),
\\
&G_{\al_N}(\al,\be)=-\frac{1}{2\pi}\cos\frac{\pi\zeta}{2}\,,
\qquad
G_{\be_N}(\al,\be)=\frac{1}{2\pi}\cos\frac{\pi\zeta}{2}\,.
\end{aligned}
\end{equation}
By solving system (\ref{23}), we obtain that
\begin{equation}\label{26} 
\begin{aligned} 
&\al_N=\al+N^{-2}\frac{\ga^2\left(2\sin\frac{\pi\zeta}{2}-1\right)}{3(\pi-2\ga)\cos\frac{\pi\zeta}{2}}
+O(N^{-3}),\\
&\be_N=\be+N^{-2}\frac{\ga^2\left(2\sin\frac{\pi\zeta}{2}+1\right)}{3(\pi-2\ga)\cos\frac{\pi\zeta}{2}}
+O(N^{-3}).
\end{aligned}
\end{equation}
The determinant of system (\ref{23}) is not equal to zero, and this guarantees,
by the implicit function theorem, that there exists a solution to (\ref{16}), (\ref{17}),
which has the same asymptotics (\ref{26}).
Proposition \ref{alN-beN} is proved.

\section{Proof of Proposition \ref{rhoN1}}\label{E}

To prove (\ref{rhoN1_1}), we would like to replace
$r_N(\mu)$ and $r_N(x)$ in (\ref{VNd}) by $r_N(0)$ and
to estimate the error term as $O(N^{-2})$. Fix any 
$0<r<\frac 12\min\{-\al,\be\}$ .

{\it Case 1, $\mu\in [\al+r,\be-r]$.}
From (\ref{VNd}) we have that
\begin{equation}\label{E1}
\begin{aligned}
\rho_N^1(\mu)+\frac{1}{2\pi^2}k(N\mu)
&=-\frac{1}{2\pi^2} P.V.\int_{\al_N}^{\be_N}
\left[\sqrt{\frac{r_N(\mu)}{r_N(x)}}-1\right]
\frac{f(Nx)dx}{\mu-x}\\
&+\frac{1}{2\pi^2} \int_{\R^1\setminus[\al_N,\be_N]}
\frac{f(Nx)dx}{\mu-x}\,.
\end{aligned}
\end{equation}
Due to estimate (\ref{VNg}), the second integral is exponentially
small as $N\to \infty$, hence we can drop it. 
In the first integral we can drop the sign of the principal value,
because the function under the integral is smooth,
and we can restrict the limits of integration to $(\al+\frac r2)$
and $(\be-\frac r2)$ plus an exponentially small term. Finally,
the function
\[
\left[\sqrt{\frac{r_N(\mu)}{r_N(x)}}-1\right]
\frac{1}{\mu-x},
\]
is a uniformly bounded analytic function in a fixed complex
neighborhood of $(x,\mu)\in [\al+\frac r2,\be-\frac r2]\times 
[\al+r,\be-r]$, hence
\begin{equation}\label{E2}
\int_{\al+\frac r2}^{\be-\frac r2}
\left[\sqrt{\frac{r_N(\mu)}{r_N(x)}}-1\right]
\frac{f(Nx)dx}{\mu-x}=O(N^{-2}),
\end{equation}
because $f$ is an odd exponentially decaying function.
This proves Proposition \ref{rhoN1} for $\mu\in [\al+r,\be-r]$.

{\it Case 2, $\mu\in [\al_N,\be_N]\setminus [\al+r,\be-r]$.}
Suppose $\mu\in [\be-r,\be_N]$. From (\ref{VNd}),
\begin{equation}\label{E3}
\begin{aligned}
\rho_N^1(\mu)=&
-\frac{\sqrt{r_N(\mu)}}{2\pi^2}\int_{\al_N}^{\al+2r}
\frac{f(Nx)dx}{(\mu-x)\sqrt{r_N(x)}}-\frac{\sqrt{r_N(\mu)}}{2\pi^2}\int_{\al+2r}^{\be-2r}
\frac{f(Nx)dx}{(\mu-x)\sqrt{r_N(x)}}\\
&-\frac{\sqrt{r_N(\mu)}}{2\pi^2}P.V.\int_{\be-2r}^{\be_N}
\frac{f(Nx)dx}{(\mu-x)\sqrt{r_N(x)}}\,.
\end{aligned}
\end{equation}
The first term is exponentially small as $N\to\infty$
(because $f$ is exponentially decaying),
and the second one is $O(N^{-2})$ (because $f$ is odd and the
integration is with respect to a smooth kernel). Let us consider the
third term. We can rewrite it as
\begin{equation}\label{E4}
-\frac{\sqrt{r_N(\mu)}}{2\pi^2}P.V.\int_{\be-2r}^{\be_N}
\frac{[f(Nx)-f(N\be_N)]dx}{(\mu-x)\sqrt{r_N(x)}}
-\frac{\sqrt{r_N(\mu)}}{2\pi^2}P.V.\int_{\be-2r}^{\be_N}
\frac{f(N\be_N)dx}{(\mu-x)\sqrt{r_N(x)}}\,.
\end{equation}
The second term is evaluated explicitly as const.$f(N\be_N)$,
and it is exponentially small as $N\to\infty$. We can represent the
first term as a half-sum of contour integrals over two contours,
$\Ga_{\pm}$, where $\Ga_+$ ($\Ga_-$) goes from $\be-2r$ to $\mu-\de$, where
$\de=\frac 13(\be_N-\mu)$, then along the upper (respectively, lower) semicircle of
radius $\de$ centered at $\mu$, and then from $\mu+\de$ to $\be_N$. 
The both integrals are exponentially small as $N\to\infty$,
hence the third term in (\ref{E3}) is exponentially small, and 
$\rho_N^1(\mu)=O(N^{-2})$ when $\mu\in [\be-r,\be_N]$. 
From (\ref{kmu2}) we obtain that $k(N\mu)=O(N^{-2})$ when $\mu\in [\be-r,\be_N]$.
This proves (\ref{rhoN1_1}) for $\mu\in [\be-r,\be_N]$. 
Similarly, it holds for $\mu\in [\al_N,\al+r]$.  Proposition
\ref{rhoN1} is proved.

\end{document}